  \documentclass{emulateapj}
\shorttitle{Shallow Decay Segment of the {\em Swift}/XRT Data}
\shortauthors{Liang et
al.}
\slugcomment{Submitted to ApJ on April 29, 2007}
\begin{document}

\title{A Comprehensive Analysis of the {\em Swift}/XRT Data: \\
II. Diverse Physical Origins of the Shallow Decay Segment}
\author{En-Wei Liang\altaffilmark{1,2}, Bin-Bin Zhang\altaffilmark{1,3},
Bing Zhang\altaffilmark{1}
} \altaffiltext{1}{Department of Physics and Astronomy, University of Nevada, Las Vegas,
NV 89154, USA; lew@physics.unlv.edu; zbb@physics.unlv.edu;bzhang@physics.unlv.edu}
\altaffiltext{2}{Department of Physics, Guangxi University, Nanning 530004, China}
\altaffiltext{3}{National Astronomical Observatories/Yunnan Observatory, CAS, Kunming
650011, China}

\begin{abstract}
The origin of the shallow decay segment in the {\em Swift}/XRT light curves is
still a puzzle. We analyze the properties of this segment with a sample of 53
long {\em Swift} GRBs detected before Feb., 2007. We show that the distributions
of its characteristics are log-normal or normal, and its isotropic X-ray energy
($E_{\rm iso,X}$) is linearly correlated with the prompt gamma-ray energy, but
with a steeper photon spectrum except for some X-ray flashes. No significant
spectral evolution is observed from this phase to the follow-up phase, and the
follow-up phase is usually consistent with the external shock models, implying
that this shallow decay phase is also of external shock origin, likely due to a
refreshed external shock. Within the refreshed shock model, the data are
generally consistent with a roughly constant injection luminosity up to the end
of this phase $t_b$. A positive correlation between $E_{\rm iso, X}$ and $t_b$
also favors the energy injection scenario. Among the 13 bursts that have
well-sampled optical light curves, 6 have an optical break around $t_b$ and the
breaks are consistent with being achromatic. However, the other 7 bursts either
do not show an optical break or have a break at a different epoch than $t_b$.
This raises a concern to the energy injection scenario, suggesting that the
optical and X-ray emissions may not be the same component at least for a fraction
of bursts. There are 4 significant outliers in the sample, GRBs 060413, 060522,
060607A, and 070110. The shallow decay phase in these bursts is immediately
followed by a very steep decay after $t_b$, which is inconsistent with any
external shock model. The optical data of these bursts evolve independently from
the X-ray data. These X-ray plateaus likely have an internal origin and demand
continuous operation of a long-term GRB central engine. We conclude that the
observed shallow decay phase likely has diverse physical origins.
\end{abstract}


\section{Introduction\label{sec:intro}}
The observations for the gamma-ray burst (GRB) phenomenon with the {\em Swift}
satellite (Gehrels et al. 2004) have revolutionized our understanding on this
phenomenon in many aspects (see recent reviews by M\'{e}sz\'{a}ros 2006; Zhang
2007). In its first two years of operation, the on-board X-ray telescope (XRT;
Burrows et al. 2004) has accumulated a large set of well-sampled X-ray light
curves from tens of seconds to days (even months) since the GRB triggers.

The generally accepted GRB models are the relativistic fireball models (Rees \&
M\'esz\'aros 1994; M\'esz\'aros \& Rees 1997a; Sari et al. 1998; see reviews by
M\'{e}sz\'{a}ros 2002; Zhang \& M\'{e}sz\'{a}ros 2004; Piran 2005). This model
invokes a fireball powered by a GRB central engine that ejects an intermittent,
relativistic outflow. Internal shocks from stochastic collisions within the
ejecta power the observed prompt gamma-rays, and deceleration of the fireball by
the ambient medium excites a long term external forward shock that powers the
broad band afterglow (M\'esz\'aros \& Rees 1997a; Sari et al. 1998). Swift data
suggest possible late internal shocks that are the origin of the erratic late
X-ray flares seen in XRT light curves (Burrows et al. 2005; Zhang et al. 2006;
Dai et al. 2006; Fan \& Wei 2005; King et al. 2005; Proga \& Zhang 2006; Perna et
al. 2006). The XRT light curves generally begin with a rapidly decaying segment
(Tagliaferri et al. 2005; O'Brien et al. 2006b), which is explained as the prompt
emission tail due to the so-called ``curvature effect'' (Kumar \& Panaitescu
2000b; Zhang et al. 2006; Liang et al. 2006; Yamazaki et al. 2006)\footnote{Other
suggestions to interpret this segment include cooling of a hot cocoon surrounding
the GRB jet (Pe'er et al. 2006) or a highly radiative blast wave (Dermer 2007).}.
The broadband afterglows, which usually decay as a power-law with an index of
$\alpha\sim -1$ (normal decay phase), are believed to be related to the external
shock. If the external shocks are refreshed by continuous energy injection into
the blastwave, a shallow decay phase prior of the normal decay phase could be
observed (Rees \& M\'{e}z\'{a}ros 1998; Dai \& Lu 1998a,b; Panaitescu et al.
1998; Sari \& M{\'e}sz{\'a}ros 2000 ; Zhang \& M\'{e}sz\'{a}ros et al. 2001; Wang
\& Dai 2001; Dai 2004; Granot \& Kumar 2006; Panaitescu 2007; Yu \& Dai 2007; see
Zhang 2007 for a review).

As the fireball is decelerated by the ambient medium, the normal decay phase
transits to a jet-like decay phase (with a decay index $\alpha\sim -2$), when the
bulk Lorentz factor is degraded to $\Gamma\sim \theta^{-1}_{j}$, where
$\theta_{j}$ is the opening angle of a conical jet (Rhoads 1997; Sari et al.
1999). Therefore, four successive emission episodes are invoked in the framework
of the fireball models, i.e., prompt gamma-ray phase with a tail, shallow decay
phase, normal decay phase, and jet-like decay phase. These power law segments,
together with erratic X-ray flares, composes a canonical X-ray afterglow
lightcurve revealed by Swift (Zhang et al. 2006; Nousek et al. 2006; O'Brien et
al. 2006b). The physical origins of these segments have been discussed in the
literature (Zhang et al. 2006; Nousek et al. 2006; Panaitescu et al. 2006a).
Empirically,  O'Brien et al. (2006a,b) and Willingale et al. (2007) show that the
data can be fitted by the superposition of a prompt emission component and an
afterglow component.

In order to explore the physical origin of this canonical afterglow lightcurve,
we perform a systematic analysis of the Swift XRT data. In the first paper of
this series (Zhang, Liang, \& Zhang 2007, Paper I of this series), we have
studied the steep decay phase for a sample of bright tails, and revealed an
apparent hard-to-soft spectral evolution for some bursts (see also Campana et al.
2006; Mangano et al. 2007; Butler \& Kocevski 2007). This paper will focus on the
shallow decay phase and the subsequent phase. This is motivated by some puzzling
facts related to the shallow decay phase. For example, simultaneous X-ray/optical
observations suggest that the break between the shallow and the normal decay
segments in the X-ray band for some GRBs is chromatic (Panaitescu et al. 2006b;
Fan \& Piran 2006). This is inconsistent with the simplest energy injection
model. One fundamental question is whether X-ray and optical afterglows have the
same physical origin. Another interesting fact is that the XRT light curve of GRB
070110 shows a long-lived plateau followed by an abrupt fall-off (the decay slope
$\sim -9$ with zero time at the trigger time). This feature is hard to interpret
within the external shock models, and it likely indicates a long-lasting central
engine emission component (Troja et al. 2007).

Theoretically, several models have been proposed to interpret the shallow decay
phase (e.g. Zhang 2007 for a review). Besides the energy injection models (Zhang
et al. 2006; Nousek et al. 2006; Panaitescu et al. 2006a), other models include
the combination of the GRB tail with the delayed onset of the afterglow emission
(Kobayashi \& Zhang 2007); off-beam jet model (Toma et al. 2006; Eichler \&
Granot 2006); pre-cursor model (Ioka et al. 2006); two-component jet (Granot et
al. 2006; Jin et al. 2007), varying microphysics parameter model (Ioka et al.
2006; Panaitescu et al. 2006b; Fan \& Piran 2006; Granot et al. 2006), etc. The
chromaticity of some X-ray shallow-to-normal breaks drives several ideas that go
beyond the traditional external forward shock model. For example, Shao \& Dai
(2007) interpret the X-ray lightcurve as due to dust scattering of some prompt
X-rays, so that it has nothing to do with the external shock. Uhm \& Beloborodov
(2007) and Genet, Daigne \& Mochkovitch (2007) interpret both X-ray and optical
afterglow as emission from a long-lived reverse shock. Ghisellini et al. (2007)
even suggested that the shallow-to-normal transition X-ray afterglows may be
produced by late internal shocks, and the end of this phase is due to the jet
effect in the prompt ejecta (see also Nava et al. 2007).

The observational puzzles and theoretical chaos call for a systematic
understanding of the shallow decay phase data for a large sample of GRBs. In
particular, it is desirable to find out how bad the standard external forward
shock model is when confronted with the data, e.g. what fraction of bursts
actually call for models beyond the standard external forward shock model. This
is the primary goal of this paper. Data reduction and sample selection are
presented in \S 2. The characteristics of the shallow decay segment and their
relations with the prompt gamma-ray phase are explored in \S 3. In \S4, we test
the external origin of the power law segment following the shallow decay phase
and explore whether or not the shallow decay segment is also of external origin.
Assuming an energy injection model for shallow decay phase, we further analyze
the the energy injection model parameters of these bursts in \S 5. The relation
among the isotropic X-ray energy ($E_{\rm iso, X}$), peak energy of the prompt
gamma-ray $\nu f_\nu$ spectrum ($E_{\rm p}$), and $t_{\rm b}$ is investigated in
\S 6. The results are summarized in \S 7 with some discussion. Throughout the
paper the cosmological parameters $H_0 = 71$ km s$^{-1}$ Mpc$^{-1}$,
$\Omega_M=0.3$, and $\Omega_\Lambda=0.7$ have been adopted.

\section{Data Reduction and Sample Selection \label{sec:data}}
The XRT data are taken from the Swift data archive. We developed a script to
automatically download and maintain all the XRT data. The {\em Heasoft} packages,
including {\em Xspec}, {\em Xselect}, {\em Ximage}, and {\em Swift} data analysis
tools, are used for the data reduction. We have developed an IDL code to
automatically process the XRT data for a given burst in any user-specified time
interval. Our procedure is described as follows. The details of our code have
been presented in Paper I.

Our code first runs the XRT tool {\em xrtpipeline} to reproduce the XRT clean
event data, and then makes pile-up corrections with the same methods as discussed
in Romano et al. (2006) (for the Window Timing [WT] mode data) and Vaughan et al.
(2006) (for the Photon Counting [PC] mode data). Both the source and background
regions are annuli (for PC) or rectangular annuli (for WT). The inner radius of
the (rectangular) annuli are dynamically determined by adjusting the inner radius
of the annuli through fitting the source brightness profiles with the King's
point source function (for PC) or determined by the photon flux using the method
described in Romano et al 2006 (for WT). If the pipe-up effect is not
significant, the source regions are in the shape of a circle with radius $R=20$
pixels (for PC) or of a 40$\times$20 pixels rectangle (for WT) centered at the
bursts' positions. The background regions have the same size as the source
region, but has a distance of 20 pixels away from the source regions. The
exposure correction is also made with an exposure map created by XRT tools {\em
xrtexpomap}. By considering these corrections, the code extracts the
background-subtracted light curve and spectrum for the whole XRT data set. The
signal-to-noise ratio is normally taken as 3 $\sigma$, but it is not rigidly
fixed at this value and may be flexibly adjusted depending on the source
brightness.

With our code we process all the XRT data observed between Feb., 2005 and Jan.,
2007. We inspect all the light curves to identify the beginning ($t_1$) of the
shallow decay segment and the end ($t_2$) of the decay phase following the
shallow phase (which usually is the normal decay phase, but in some cases the
decay slope could be much steeper). Please note that the selection of $t_1$ and
$t_2$ is guided by eye without a rigid criterion. Generally,  $t_1$ is taken
as the end of the steep decay segment or the beginning of the observation time,
unless significant flares or high level emission bumps following the GRB
tails were observed. The ending time $t_2$ is taken as the end of the observation
time except for GRBs 050416A, 050803, 060413, 060908, 060522, 061121,and 070110,
which have an additional break at later times, and $t_2$ is chosen as that
break time. For example, GRBs 060522
and 070110 have a distinct ``normal-decay'' emission component following the sharp
decay segment, and $t_2$ is taken the end of the sharp decay. The last
data points of GRB 050416A, 050803, 060413, 060908, and
061121 show a flattening feature, which significantly deviates from the power
law decay trend post $t_b$. We thus do not include those data points.

Physically, the temporal break of an external shock origin should be smooth
(due to the equal-arrival-time effect of a relativistic shell of conical geometry).
Therefore, a smoothed broken power law is used to fit the light
curve in the time interval [$t_1,t_2$],
\begin{equation}
F=F_0\left[\left(\frac{t}{t_{\rm b}}\right)^{\omega {\alpha_1}}+\left(\frac{t}{t_{\rm
b}}\right)^{\omega {\alpha_2}}\right]^{-1/\omega},
\end{equation}
where $\omega$ describes the sharpness of the break. The larger the $\omega$,
the sharper the break. In order to constrain $\omega$ it is required that the time
interval covers a range from $t_1\ll t_b$ to $t_2\gg t_b$, and that the light curve
around $t_b$ is well-sampled. The parameter $t_b$ is not significantly affected
by $\omega$, but both $\alpha_1$ and $\alpha_2$ are. We show the comparison of
the fitting results with $\omega=1$ and $\omega=3$ for the bursts in our sample
(see below) in Fig. 1. We find that systematically, $t_b^{\omega=1}\sim
t_b^{\omega=3}$, $\alpha_1^{\omega=1}<\alpha_1^{\omega=3}$, and
$\alpha_2^{\omega=1}>\alpha_2^{\omega=3}$. We notice that Willingale et al. (2007)
fit the XRT light curves with a superposition model of both the prompt and afterglow
emission components. The derived $\alpha_2$ from our fitting with $\omega=3$ is
generally consistent with their results. We therefore fix $\omega=3$ in this
analysis, except for GRBs 060413, 060522, 060607A, and 070710. The XRT light
curves of these bursts abruptly drop at $t_{\rm b}$, and we take $\omega=10$. We
then create a time filter array that contains two time intervals of [$t_1, t_{\rm
b}$] and [$t_{\rm b}, t_2$] for each burst. By specifying the time filter array
we run our code again to extract the spectra in the two time intervals and derive
their photon indices, $\Gamma_{\rm X, 1}$ and $\Gamma_{\rm X, 2}$, by fitting the
spectra with a power law model incorporating with absorptions by both the Milky Way
Galaxy and the host galaxy, wabs$^{\rm Gal}\times$zwabs$^{\rm host}\times$ power law
(when the redshift is unknown, zwabs$^{\rm host}$ is replaced with the model of
wabs). The $N_H^{\rm host}$ value in the time-resolved spectral analysis is fixed
to the value obtained from fitting the time-integrated spectrum during the whole
time span of each burst.

The $t_{\rm b}$ is roughly considered as the duration of the shallow decay phase.
As suggested by Lazzati \& Begelman (2006) and Kobayashi \& Zhang (2007), the
zero time of the external-origin power-law segments should be roughly the BAT
trigger time. In our calculation, in order to account for the onset of the
afterglow we take a $t_0$ as 10 seconds after the GRB trigger. The X-ray fluence
($S_X$) of the shallow decay phase is derived by integrating the fitting light
curve from 10 seconds post the GRB trigger to $t_{\rm b}$ without considering the
contributions of both early X-ray flares and the GRB tail emissions. Since the
shallow decay phase has a temporal decay index shallower than -1, the results are
not sensitive to the choice of $t_0$. We estimate the uncertainty of $S_X$ with a
boostrap method based on the errors of the fitting parameters, assuming that the
errors of the fitting parameters, $\sigma_{\log F_0}$, $\sigma_{\log t_{\rm b}}$,
$\sigma_{ \alpha_1}$, and $\sigma_{ \alpha_2}$, are of Gaussian distributions. We
generate $5\times 10^3$ parameter sets of ($F_0$, $t_{\rm b}$, $\alpha_1$,
$\alpha_2$) from the distributions of these parameters for each burst, and then
calculate $S_{\rm X}$ for each parameter set. We make a Gaussian fit to the
distribution of $\log S_{\rm X}$ and derive the central value of $\log S_{\rm X}$
and its error $\sigma_{\log S_{\rm X}}$. In our fittings, $\alpha_1$ and/or $t_b$
are fixed for GRBs 050801 and 060607A. We do not calculate the errors for the two
bursts.

We use the following criteria to select our sample. First, the XRT light curves
have a shallow decay segment following the GRB tails. Since the decay slope of
the ``normal'' decay phase predicted by the external GRB models is generally
steeper than 0.75, we require that the so-called shallow decay segment has a
slope $\alpha_{\rm X,1}<0.75$ at $1\sigma$ error. Second, both the shallow decay
segment and the follow-up segment are bright enough to perform spectral analysis.
Systematically going through all the {\em Swift} XRT data before Feb. 2007 we use
the above criteria to compile a sample of 53 bursts. Please note that the
apparently long GRB 060614 is also included in our sample, although it may belong
to the short-type bursts (Gehrels et al. 2006; Zhang et al. 2007b; Zhang 2006).
The XRT light curves and the fitting results are shown in Fig. 2, and the data
are summarized in Table 1. We collect the BAT observations of these bursts from
GCN circular reports, and report them in Table 2. We search the optical afterglow
data of these bursts from published papers and GCN circular reports\footnote{The
{\rm GRBlog} web page (http://grad40.as.utexas.edu/grblog.php) has been used.}.
We identify a burst as optically bright, if three or more detections in the
UV-optical bands were made. We find that 30 out of the 53 bursts are optically
bright, but only 15 bursts have an optical light curve with good temporal
coverage. We make the Galactic extinction correction and convert the observed
magnitudes to energy fluxes. We fit these light curves with a simple power law or
the smooth broken power law ($\omega$ is also fixed as 3). The fitting results
are summarized in Table 3. We directly compare the optical data with the XRT data
in Fig. 2 in order to perform a quick visual check of achromaticity of these
light curves. If multi-wavelength optical light curves are available, we show
only the one that was observed around the X-ray shallow decay phase with the best
sampling. Notice that the contribution from the host galaxy to the optical light
curve of GRB 060614 has been removed.

Twenty-seven out of the 53 GRBs in our sample have redshift measurements. Table 4
reports the properties of these bursts in the burst rest frame, including the
durations ($T_{90}^{'}$ and $t_{\rm b}^{'}$) and the equivalent-isotropic
radiation energies ($E_{\rm iso, \gamma}$ and $E_{\rm iso, X}$) in the prompt
phase and in the shallow decay phase, and the peak energy of the $\nu f_\nu$
spectrum ($E_{\rm p}^{'}$). The $E_{\rm iso, \gamma}$ and $E_{\rm iso, X}$ are
calculated by
\begin{equation}
E_{\rm iso,(\gamma,X)}=\frac{4\pi D_L^2 S_{(\gamma,X)}}{1+z},
\end{equation}
where $S_\gamma$ is the gamma-ray fluence in the BAT band and $S_X$ is the X-ray
fluence in the shallow decay phase in the XRT band, and $D_L$ is the
luminosity distance of the source. Due to the narrowness of the BAT band, the BAT
data cannot well constrain the spectral parameters of GRBs (Zhang et al. 2007a).
Generally the BAT spectrum can be fitted by a simple power law, and the power
law index $\Gamma$ is correlated with $E_{\rm p}$ (Zhang et al. 2007b; see also
Sakamoto et al. 2007; Cabrera et al.2007)\footnote{We should point out that
this empirical relation is for BAT observations only. The origin of this relation
is due to the narrowness of the BAT instrument. It can be robustly used for those
bursts whose $E_{\rm p}$ are roughly within the BAT band.}, i.e.,
\begin{equation}\label{E_p-Gamma-Relation}
\log E_p=(2.76\pm 0.07)-(3.61\pm 0.26)\log \Gamma.
\end{equation}
We estimate $E_{\rm p}$ with this relation if it is not constrained by the BAT
data. We then calculate the {\em bolometric} energy $E^b_{\rm iso,\gamma}$ in
the $1-10^4$ keV band with the $k$-correction method used by Bloom et al. (2001),
assuming that the photon indices are -1 and -2.3 before and after $E_{\rm p}$,
respectively (Preece et al. 2000). Both $E_{\rm p}^{'}$ and $E^{b}_{\rm iso,
\gamma}$ are listed in Table 4.

\section{The Characteristics of the Shallow Decay Phase and its Relations
to the Prompt Gamma-Ray Phase
\label{sec:Shallow-Prompt}}

We display the
distributions of the characteristics of the shallow decay phase in Fig. 3. It is
found that these distributions are consistent with being normal/lognormal, i.e.
$\log t_{\rm b} /s=4.09\pm 0.61 $, $\log S_X / {\rm erg\ cm}^{-2}=-6.52\pm 0.69$,
$\Gamma_{\rm X,1}=2.09\pm 0.21$, and $\alpha_1=0.35\pm 0.35$. Quoted errors are at
$1\sigma$ confidence level.

We investigate the relation of the shallow decay phase to the prompt gamma-ray
phase. Figure 3 shows $t_{\rm b}$, $S_X$, $\Gamma_{\rm X,1}$, and $E_{\rm iso,
X}$ as functions of $T_{90}$, $S_\gamma$, $\Gamma_\gamma$, and $E_{\rm iso,
\gamma}$, respectively. No correlation between $\Gamma_\gamma$ and $\Gamma_{\rm
X,1}$ is observed. However, $\Gamma_{\rm X,1}$ is larger then
$\Gamma_\gamma$, except for some X-ray flashes (XRFs), indicating that
the photon spectrum of the shallow decay phase is generally steeper than that of the
prompt gamma-ray phase for typical GRBs. It is interesting to note that in contrast
to $\Gamma_\gamma$ $\Gamma_{\rm X,1}$ is narrowly clustered around 2.1 (see also
O'Brien et al. 2006a), hinting a possible common microscopic mechanism during the
shallow decay phase.

From Fig. 3 we find tentative correlations of durations, energy fluences,
and isotropic energies between the gamma-ray and X-ray phases. The best fits yield
$\log t_{\rm b}=(0.61\pm 0.16)\log T_{90}+(3.00\pm 0.27)$ ($r=0.48$ and $p=0.003$ for
$N=53$), $\log S_X=(0.76\pm 0.11)\log S_{\gamma}+(-2.33\pm 0.60)$ ($r=0.70$ and
$p<10^{-4}$ for $N=53$), and $\log E_{\rm iso, X}=(1.00\pm 0.16) \log E_{\rm iso,
\gamma}+(-0.50\pm 8.10)$ ($r=0.79$ and $p<10^{-4}$ for $N=27$). It is found that
$t_{\rm b}$ weakly depends on $T_{90}$. However, X-ray fluence and
isotropic energy are almost linearly correlated with gamma-ray fluence and
gamma-ray energy, respectively. $E_{iso, \gamma}$ is greater than
$E_{X,iso}$ for most of the bursts, but for a few cases $E_{iso, X}$ is even
larger than $E_{\gamma,iso}$. In order to reveal possible linear correlations for
the quantities in the two phases, we define a $2\sigma$ linear correlation
regions with $y=x+(A\pm 2\times \sigma_A)$, where $y$ and $x$ are the two quantities
in question, and
$A$ and $\sigma_A$ are the mean and its
$1\sigma$ standard error of the $y-x$ correlation, respectively.
The $1\sigma$ regions of the correlations are shown with
dashed lines in Fig. 4.  These results indicate that radiation during the
shallow decay phase is correlated with that in the prompt gamma-ray phase.

\section{Testing the physical origin of the shallow decay segment using the properties
of the follow-up segment \label{sec:Shallow-Prompt}}

The leading scenario of the shallow decay phase is a refreshed forward shock due to
either a long-term central engine or a spread of the ejecta Lorentz factor (Zhang
et al. 2006; Nousek et al. 2006; Rees \& M\'esz\'aros 1998; Dai \& Lu 1998a,b;
Zhang \& M\'esz\'aros 2001; Granot \& Kumar 2006; Yu \& Dai 2007). Within such
a scenario, the shallow decay phase ends at $t_{\rm b}$ and
transits to a ``normal'' decay phase consistent with the standard external forward
shock models. Three criteria are required to claim
an energy injection break $t_b$. First, there should be no spectral evolution
across $t_b$ since energy injection is a pure hydrodynamical effect. Second, due
to the same reason, the break at $t_b$ should be achromatic. Third,
the power-law decay phase after $t_b$ should comply with
the standard external shock models. In this section, we test whether all three
criteria are satisfied with the data.

Figure 4(a) shows $\Gamma_{\rm X,2}$ as a function of $\Gamma_{\rm X,1}$. The
solid line is $\Gamma_{\rm X,2}=\Gamma_{\rm X,1}$, and the dashed lines marks
the $3\sigma$ region of the equality, which is defined with $\Gamma_{\rm
X,2}=\Gamma_{\rm X,1}+(\mathcal{G}\pm 3\delta_\mathcal{G})$, where $\mathcal{G} $
and $\delta_\mathcal{G}$ are the mean and statistical uncertainty ($1\sigma$
level) of the difference $\Gamma_{\rm X,2}-\Gamma_{\rm X,1}$. Please note that
$\delta_\mathcal{G}$ does not include the observational uncertainty. It
statistically describes the scatter of $\mathcal{G}$ for the bursts in our
sample. We find that only one burst GRB 061202 is out of the region. The
comparison between the distributions of $\Gamma_{\rm X,2}$ and $\Gamma_{\rm X,1}$
is shown in Fig. 4(b). Excluding GRB 061202, the two distributions are
consistent. The Kolmogorov-Smirnov  test suggests that the significance level of
this consistency is 0.96. These results indicate that $\Gamma_{\rm X,1}$ and
$\Gamma_{\rm X,2}$ for the bursts in our sample are globally consistent with
each other. In order to verify this consistency within observed uncertainty
for individual bursts, Fig. 4(c)
shows the distribution of the ratio $\mu=\mathcal{G}/\sigma$, where
$\sigma^2={\delta \Gamma^2_{X,1}}+{\delta \Gamma^2_{X,2}}$ is the observed
uncertainty of $\mathcal{G}$. A positive value of $\mu$ would indicate a hard-to-soft
spectral evolution. This ratio indicates the significance level of the difference
between $\Gamma_{\rm X,1}$ and $\Gamma_{\rm X,2}$ for individual bursts within the
observational uncertainties of the two quantities. As shown in Fig. 4(c), most of
the bursts ($\sim 90\%$) have $\mu\lesssim 1$, and only one burst (GRB 061202) has
$\mu> 3$. These results prove that no significant spectral evolution between the
two phases with a confidence level above $3\sigma$ is observed for the bursts in
our sample within the observational error, except for GRB 061202. This is
consistent with the expectation of the refreshed shock afterglow models. Please
note that GRB 061202 shows significant hard-to-soft spectral evolution from the
shallow to the normal decay phases, i.e., from $\Gamma_{\rm X,1}=2.25\pm 0.07$ to
$\Gamma_{\rm X,2}=3.55\pm 0.44$. One caveat for this spectral evolution is that
there is a long observational gap between the first epoch in the shallow decay
phase ($4\times 10^3$ to $2\times 10^4$ seconds) and the second epoch in normal
decay phase ($1\times 10^5\sim 5\times 10^5$ seconds) when the spectral indices
are measured. Without detecting the break itself, it may be dangerous to draw the
conclusion that spectral variation is clearly seen across $t_b$. We cannot rule
out the possibility that the plateau extends further and drops dramatically
before landing onto a normal decay segment as is seen in GRBs 060522 and 070110
(see discussion below).

Although the mechanism of energy injection into the forward shock could vary
(e.g. Rees \& M\'{e}z\'{a}ros 1998; Dai \& Lu 1998a,b; Zhang \& M\'{e}sz\'{a}ros
et al. 2001; Yu \& Dai 2007), the kinetic energy of the fireball after the energy
injection is over should be constant and this ``normal'' decay phase should be
explained with the standard external shock models. Without broadband afterglow
modeling, the ``closure relations'' between the observed spectral index $\beta$
and temporal decay index $\alpha$ present a simple test to the models. In Fig.5,
we present $\alpha_{\rm X,2}$ as a function of spectral index $\beta_{\rm X,2}$,
where $\beta_{\rm X,2}=\Gamma_{\rm X,2}-1$. The closure correlations of the
external shock afterglow models for different spectral regimes, different cooling
schemes, different ambient medium properties, and different electron
distributions (the spectral index $p>2$ and $p<2$) are shown in Fig. 5 (see Table
1 of Zhang \& M\'{e}sz\'{a}ros 2004 and reference therein, in particular Sari et
al. 1998; Chevalier \& Li 2000; Dai \& Cheng 2001). The fact that the observed
$\beta_{X,2}$ is greater than 0.5 for the bursts in our sample suggests that
these X-rays are in the spectral regime $\nu_X>\max(\nu_m,\nu_c)$ (Regime I) or
$\nu_m<\nu_X<\nu_c$ (Regime II), where $\nu_m$ and $\nu_c$ are the characteristic
frequency and cooling frequency of synchrotron radiation. The relation between
$\alpha$ and $\beta$ for the spectral Regime I is $\alpha=(3\beta-1)/2$
regardless of the type of the medium (ISM or wind medium). If the X-ray band is
in the Regime II, we have $\alpha=3\beta/2$ (for ISM) and $\alpha=(3\beta+1)/2$
(for wind). We define
\begin{eqnarray}
D=|\alpha^{\rm obs}-\alpha(\beta^{\rm obs})|,\nonumber \\
\delta_D=\sqrt{(\delta \alpha^{\rm obs})^2+[\delta \alpha(\beta^{\rm obs})]^2},
\end{eqnarray}
where $\alpha^{\rm obs}(\delta \alpha^{\rm obs})$ and $\alpha(\beta^{\rm obs}) $
are the temporal decay slopes (errors) from the observations and from the closure
relations. The ratio $\phi=D/\delta_D$ reflects the ``nearness'' of the data point
to the model predictions within the error scope. If $\phi <1$, we consider that
the data point goes cross the corresponding closure relation line. If $\phi<3$,
we regard that the
model cannot be excluded within a 3$\sigma$ significance level. Those bursts that
have large uncertainties on both $\alpha$ and $\beta$ may be interpreted with
more than one models. In this case, we compare $\phi$ values derived from these
models and take the model that gives the smallest $\phi$.

As shown in Fig. 5, 24 out of the 53 bursts distribute around the line for the
spectral Regime I. A group of bursts have a decay slope shallower
than the model prediction, but they are slightly below and almost keep abreast
with the Regime I model line (see also Fig. 5 of Willingale et al.
2007). At the $3\sigma$ confidence level, this model cannot be excluded for these
bursts. Eighteen bursts are consistent with the ISM external shock afterglow model
in the spectral regime II.

Six bursts (GRBs 050315, 050318, 050803, 060614, 051008, and 060906) agree with
both the regime I ISM jet model and the regime II wind model. The observed
$\beta$ of these six bursts are $\sim 1$. The two models are almost degenerate at
$\beta\sim 1$. We therefore use the spectral and temporal behaviors of the prior
segment to distinguish the two models. Since the observed $\beta>0.5$ in our
sample, the decay slope of the light curves before a jet break should be steeper
than $(3\beta)/2\sim 0.75$. From Table 1 we can see that the $\alpha_1$ values
are $0.66\pm 0.03$, $0.90\pm 0.23$, $0.25\pm 0.03$, $0.18\pm 0.06$, $0.78\pm
0.11$, and $0.35\pm 0.10$, respectively, for the six bursts. So there is no
confident evidence to claim a jet break within the uncertainty of the decay slope
for these bursts\footnote{The possibility that a jet break is temporarily
coincident with an energy injection break is however not ruled out. The normal
decay phase could be missed in data fitting, if the normal decay segment is
short, the data is sparse, or the jet break is not significant(e.g. Wei \& Lu
2000; Kumar \& Panaitescu 2000a; Gou et al. 2001). One example of this scenario
is GRB 060614. Our best fit with a smooth broken power law does not reveal a
jet-like break from a normal decay phase. However, Mangano et al. (2007) suggest
a normal decay phase between $3.66\times 10^4$ and $1.04\times 10^5$ seconds by
fitting the light curve with a joint-power-law model (the breaks are guided by
eye). They showed that the decay slope during the period is $1.03\pm 0.02$. This
normal decay phase thus satisfies a closure relation of the standard forward
shock models. For a detailed study of jet breaks please see our paper III in the
series, E.-W. Liang et al. 2007, in preparation.} Since the energy injection
model the shallow decay slope depends on a free parameter $q$,  we tentatively
suggest that these six cases can be explained with a wind afterglow model in the
spectral regime II. GRB 060108 is also consistent with this model according to
our criterion.

As shown above, the spectral index and temporal decay slope of the normal decay
phase for most bursts in our sample (49 out of 53 bursts) are roughly consistent
with the closure relations of the external shock models. This further favors the
idea that the shallow decay segment is also of external shock origin, and
probably is related to a long-term energy injection effect. In this scenario, the
energy injection break should be achromatic, if the multi-wavelength radiations
are all from the same emission region, presumably the forward shock. We therefore
inspect the optical light curves of these bursts to examine whether the breaks
observed in the XRT light curves are achromatic. Among these 13 bursts have
well-sampled optical light curves, as shown in Fig. 2 and Table 3. The optical
light curves of GRBs 050801, 051109A, 060614, 060714, 060729, and 061121 show a
break around $t_b$, indicating that the breaks in both the X-ray and the optical
bands are consistent with being achromatic. However, the optical light curves of
GRBs 050318, 050319, 050802, 060124, and 050401 do not have a break around $t_b$
(see also Panaitescu et al. 2006b). They can be well fitted by a simple power law
model. GRBs 060210 and 060526 have an optical break, but the breaks are not
around $t_b$.

GRBs 060413, 060522, 060607A, and 070110 have a plateau with a step-like sharp
drop ($\omega=10$ is required in our data fitting). Except for GRB 060522, the
other three bursts deviate significantly from any external shock afterglow models
at $3\sigma$ significance level. Although the sharp drop segment of GRB 060522 is
consistent with the regime II, wind-jet model, the plateau convincingly rules out
this model since it cannot be explained as the pre-jet segment within the same
model. These results suggest that the sharp drop segment and its prior plateau in
these bursts are very likely not of external shock origin. A direct support to
this speculation is that the optical lightcurves of these bursts, if available,
are all evolve independently with respect to the X-ray lightcurves. For example,
The optical light curve of GRB 060607A rapidly increases (with $F\propto t^{3}$)
up to a maximum at $t=160$ seconds post the GRB trigger, and then continuously
decays with an index of $-1.18\pm 0.02$ (Molinari et al. 2007). The X-ray
lightcurve, on the other hand, shows significant flares before 600 seconds, and a
plateau lasting from 600 seconds to $\sim 1.2\times 10^4$ seconds after the GRB
trigger. At the end of the plateau, the XRT light curve drops sharply with
$\alpha_2=3.35\pm 0.09$. During the plateau in the XRT light curve, the optical
light curve ``normally'' decays until a significant flare around 2000 seconds.
The optical light curve is consistent with a external forward shock, and the peak
is consistent with onset of the afterglow (Molinari et al. 2007). The plateau and
the sharp drop in the XRT light curve of GRB 070110 is similar to that of GRB
060607A, but an additional ``normal''-decay component post the steep fall-off was
also observed (Troja et al. 2007). The decay slope of this late X-ray emission
component is similar to that of the optical light curve and is likely of external
shock origin (see also GRB 060522). This reinforces the suggestion that the early
X-ray plateau is of internal origin and is connected to a long-lasting central
engine (Troja et al. 2007). A common signature of these internal-origin plateaus
is that the flux almost keeps constant on the plateau but with significant
flickering. Although it may not be unreasonable to interpret it as late internal
shocks (which usually give rise to erratic collisions within the ejecta and may
power X-ray flares), another possibility is that the plateau is powered by
tapping the spindown energy of the central engine, as suggested by Troja et al.
(2007).

\section{Energy Injection Behavior}

As shown above, the normal decay phase for most of the bursts in our sample (49 out
of 53) are consistent with the external shock models. This suggests that,
in general, the observed shallow decay phase is also of the external origin and
may be related to continuous energy injection into the fireball. In this section,
we assume the standard energy injection model and infer from the data the
parameters of the long-lived central engine.

We describe the energy injection behavior as $L(t)\propto t^{-q}$ (e.g. Zhang \&
M\'esz\'aros 2001)\footnote{Another injection scenario invoking a distribution of
the Lorentz factor of the ejecta (Rees \& M\'esz\'aros 1998) can be effectively
represented by a long-term central engine (Zhang et al. 2006). The
internal-origin plateaus discussed above suggest that at least for some GRBs, a
long-lived central engine is indeed in operation.}. The difference between the
decay slopes before and after $t_{\rm b}$ depends on the observed spectral regime
and the type of abient medium, which can be summarized as (derived from Table 2
of Zhang et al. 2006),

\begin{equation}
\label{q} \Delta \alpha=\cases{ \frac{(p+2)(1-q)}{4},
            & spectal regime I, ISM and Wind, \cr
 \frac{(p+3)(1-q)}{4},
            & spectal regime II, ISM,\cr
 \frac{(p+1)(1-q)}{4},
            & spectral regime II, Wind, \cr
            }
\end{equation}
where $p$ is the power law index of the electron distribution. The $p$ value is
derived from the observed spectral index, depending on the observed spectral
regime. We identify the spectral regime for these bursts by comparing the
observed $\alpha_{\rm X,2}$ and $\beta_{X,2}$ with the closure correlations, and
then derive their $q$ values from Eqs. \ref{q}. The distributions of theses GRBs
in the two dimensional $q-\Delta \alpha$ and $q-p$ planes are shown in Fig. 6, along
with the contours of constant $p$ and $\Delta \alpha$ lines derived from the models
(Eq.[\ref{q}]). No correlation between $\alpha_1$ and $\alpha_2$ is found. The
steepening index $\Delta\alpha$ is found to vary among bursts, with an average of
$1.11\pm 0.39$.  The $p$ values range
from $2\sim 3.5$ without evidence of clustering (see also Shen et al. 2006). The
$q$ values for most bursts are around $-0.75$ to $0.55$, with an average of $\sim
-0.07\pm 0.35$. It is worth commenting that a specific energy injection model
invoking a spindown pulsar predicts a $q$ value of 0 (Dai \& Lu 1998a; Zhang
\& M\'esz\'aros 2001). The average $q$ value is close to this model prediction.

\section{Empirical Relation among $E_{\rm iso}-E_{\rm p}-t_{\rm b}$}
An empirical relation among $E_{\rm iso}-E_{\rm p}^{'}-t'_{b,opt}$ was discovered
with pre-Swift GRBs, where $t'_{b,opt}$ is the temporal break of the optical
afterglow light curve in the rest frame of the burst (Liang-Zhang relation; Liang
\& Zhang 2005)\footnote{If $t'_{b,opt}$ is interpreted as a jet break, then the
relation is similar to the Ghirlanda-relation (Ghirlanda et al. 2004). However,
the more empirical Liang-Zhang relation allows more freedom to understand the
origin of the breaks.}. Willingale et al. (2007) found a similar correlation to
the Ghirlanda relation by assuming that $t_b$ in the X-ray band are jet breaks.
Such a relation may be interpreted as an effective $E_{\rm iso}-E_{\rm
p}^{'}-t_{b}{'}$ relation similar to the Liang-Zhang relation. In this section we
investigate the relations among the energies $E_{\rm iso, \gamma}^{b}$, $E_{\rm
iso,X}$ and the other two parameters $E_{\rm p}^{'}$ and $t_{\rm b}^{'}$. With
the data reported in Table 4 (for bursts with redshift measurements), we use a
multi-variable regression analysis method to search for possible dependences of
$E_{\rm iso, X}$ and $E_{\rm iso, \gamma}^{b}$ on both $E_{\rm p}^{'}$ and
$t_{\rm b}^{'}$. Our sample is limited to those bursts whose $t_b$ can be
explained as an energy injection break (without considering the achromaticity of
the break). Among the 49 bursts 27 have redshift measurements. Since only two
bursts in the internal-origin plateau sample have redshift measurements, we
cannot make an analysis to them. Our regression model reads
\begin{equation}\label{MVR}
\log \hat{E}_{{\rm{iso}}}=\kappa_{0}+ \kappa_{1}\log E_{\rm p}^{'}+\kappa_{2}\log
t_{b}^{'}
\end{equation}
where $E_{\rm p}^{'}=E_{\rm {p}}(1+z)$ and $t_{b}^{'}=t_{\rm{b}}/(1+z)$. We
measure the significance level of the dependences of each variable on the model
by the probability of a t-test ($p_t$). The significance of the global regression
is measured by a F-test (with a chance probability $p_F$). Statistically, a
robust statistical analysis requires the chance probability to be less than
$10^{-4}$. Our multiple regression analysis to $E_{\rm iso, X}(E_{\rm p}^{'},
t_{\rm b}^{'})$ shows $\kappa_0=44.0\pm 1.1$ (with $p_t<10^{-4}$),
$\kappa_1=1.82\pm 0.33$ (with $p_t<10^{-4}$), and $\kappa_2=0.61\pm 0.18$ (with
$p_t=3\times 10^{-3}$). The $p_F$ is $<10^{-4}$. These results suggest a strong
correlation between $E_{\rm iso, X}$ and $E_{\rm p}^{'}$ and a tentative
correlation between $E_{\rm iso, X}$ and $t_{\rm b}^{'}$. On the other hand, our
multiple regression analysis to $E_{\rm iso,\gamma}^{b} (E_{\rm p}^{'}, t_{\rm
b}^{'})$ gives $\kappa_0=48.3\pm 0.8$ (with $p_t<10^{-4}$), $\kappa_1=1.70\pm
0.25$ (with $p_t<10^{-4}$), and $\kappa_2=0.07\pm 0.13$ (with $p_t=0.486$). The
$p_F$ is $<10^{-4}$. It is found that the correlation between $E_{\rm iso,
\gamma}^{b}$ and $E_{\rm p}^{'}$ is significant, but statistically no correlation
between $E_{\rm iso, \gamma}^{b}$ and $t_{\rm b}^{'}$ can be claimed.

With the relation discovered by Willingale et al.(2007), one would naively expect
a multi-variable correlation among $E_{\rm iso, \gamma}^{b}-E_p^{'}-t_b^{'}$.
According to our results, a significant correlation among $E_{\rm iso,
\gamma}^{b}-E_p^{'}-t_b^{'}$ could be indeed claimed with a chance probability
$p_F<10^{-4}$. However, this correlation is dominated by the correlation of
$E_{\rm iso, \gamma}^{b}-E_p^{'}$ only (with a $p_t<10^{-4}$), which is
essentially the Amati-relation (Amati et al. 2002). The $p_t$ of the dependence
between $E_{\rm iso, \gamma}^{b}$ and $t_b^{'}$ is 0.486. This strongly rules out
such a dependence. Therefore, we suspect that the apparent relation found by
Willingale et al. (2007) would be intrinsically a manifestation of the
Amati-relation. A similar conclusion has been also achieved by Nava et al.
(2007). The $t_b$ essentially did not enter the problem, since the distribution
of $t_b$ is narrower than that of $t_{b,opt}$ as discussed in Liang \& Zhang
(2005).

It is interesting to note the dependence $E_{\rm iso, X}\propto t_{\rm
b}^{'0.61\pm 0.18}$. This is in sharp contrast to the Liang-Zhang relation, in
which $E_{\rm iso, \gamma}\propto t_{\rm b,opt}^{'-1.24}$ was discovered. In
order to compare the $E_{\rm iso, X}-E_{\rm p}^{'}-t_{\rm b}^{'}$ correlation
with the Liang-Zhang relation in a 2-dimensional plane, we define $\Sigma=\log
E_{\rm iso}-\kappa_{2}\log t'_{\rm b}$, and show $\Sigma_X$ and $\Sigma_\gamma$
as a function of $\log E_{\rm p}^{'}$ in Fig. 8. We observe that the $E_{\rm iso,
X}-E_{\rm p}^{'}-t_{\rm b}^{'}$ correlation is significant, but it has a larger
scatter than the Liang-Zhang relation. Although we can not rule out the
possibility that the large dispersion is intrinsic, the observational
uncertainties of both $E_{X,iso}$ and $E_{\rm p}$ could make such a dispersion.
Figure 8 evidently shows that the $E_{\rm iso, X}-E_{\rm p}-t_{\rm b}$
correlation is different from the Liang-Zhang relation. This suggests that
$t_{\rm b}$ and $t_{\rm b,opt}$ may have distinct physical origins. The positive
correlation between  $E_{\rm iso, X}$ and $t_{\rm b}$ is consistent with energy
injection origin of $t_{\rm b}$, namely, a longer injection episode gives more
energy. The negative correlation between $E_{\rm iso, \gamma}$ and $t_{\rm
b,opt}$ may step back to the standard energy reservoir argument of Frail et al.
(2001), which suggests a connection between $t_{\rm b,opt}$ and the jet opening
angle of the outflow.

\section{Conclusions and Discussion}
We have presented a comprehensive analysis of the {\em Swift} XRT light curves of
long GRBs, focusing on the properties of the shallow decay phase and its relation
with the follow-up decay phase. Our sample includes 53 bursts whose X-ray
emissions are bright enough to perform spectral and temporal analyses for both
phases. We summarize our results as follows.

(1) We find that the distributions of the characteristic properties of the
shallow decay phase are log-normal or normal, i.e., $\log t_{\rm b} /s=4.09\pm
0.61 $, $\log S_X / {\rm erg~cm}^{-2}=-6.52\pm 0.69$, $\Gamma_{\rm
X,1}=2.09\pm 0.21$, and $\alpha_1=0.35\pm 0.35$ (quoted errors are at $1\sigma$
confidence level).

(2) The $E_{\rm iso,X}$ of the shallow decay phase is linearly correlated with
the prompt gamma-ray phase, i.e., $\log E_{X,iso}=(1.00\pm 0.16)\log
E_{iso,\gamma}-(0.5\pm 8.12)$ (with a Spearman correlation coefficient $r=0.79$
and a chance probability $p<10^{-4}$). The spectrum of the shallow decay phase is
softer than the prompt gamma-ray phases, except for some typical XRFs.

(3) Except for GRB 061202, no spectral evolution is observed during the
transition from the shallow decay to the follow-up decay phases. The post break
phase in most bursts is consistent with the closure relations of the external
shock models. Six out of the 13 bursts with well-sampled optical light curves
show an achromatic break in both X-ray and optical bands, but the other 7 cases
either do not show any break or have a break at a different epoch in the optical
band.  This poses an issue to explain $t_b$ of these bursts as the end of the
energy injection phase.

(4) Four bursts (GRBs 060413, 060522, 060607A, and 070110) in our sample have a
post-break phase significantly deviate from the external shock models. The decay
indices are much steeper than model requirements. The optical light curves of the
latter two bursts evolve distinctly from the X-ray light curves. We suggest that
the X-ray and optical emissions of these bursts are from different emission
sites, and the X-ray plateaus are of internal origin and demand a long-live
emission component from the central engine.

(5) Within the scenario of the refreshed external shocks, the average energy
injection index $q\sim 0$, suggesting a roughly constant injection luminosity
from the central engine.

(6) With a sub-sample of 27 bursts with known redshifts that satisfy the closure
relations of the standard external fireball models, we discover an empirical
multi-variable relation among $E_{\rm iso, X}$, $E_{\rm p}^{'}$, and $t'_{\rm b}$
(Eq.[7]), which is distinctly different from the $E_{\rm iso,\gamma}-E_{\rm
p}^{'} -t'_{\rm b,opt}$ relation discussed by Liang \& Zhang (2005).

(7) There is no significant correlation between $t'_b$ and the
other parameters $E_{\rm iso,\gamma}$ and $E_{\rm p}^{'}$ (unlike $t'_{\rm
b,opt}$). This suggests that the apparent $E_{\rm j, \gamma}-E_p^{'}$ relation by
assuming a jet origin of $t_b$ (Willingale et al. 2007) is likely a manifestation
of the Amati-relation.

These results suggest that the shallow decay segment observed in most bursts is
consistent with having an external forward shock origin, probably due to a
continuous energy injection into the forward shock from a long-lived central
engine. Therefore, the scenarios that
completely abandon the external shock models (e.g. Ghisellini et al. 2007;
Genet et al. 2007; Uhm \& Beloborodov 2007; Shao \& Dai 2007) may
not be demanded by the data, and these models need to explain the apparent
consistency of the $\alpha-\beta$ data with the simple closure relations of
the forward shock models.

Since the energy injection break is due to a hydrodynamic effect, achromatism is
one of a key feature of the model. Although about half cases satisfy such a
constraint, at least some X-ray breaks are chromatic (even if the post break
segment is well consistent with the standard afterglow model). This poses a great
issue to argue that these X-ray breaks are energy injection breaks. Invoking
different emission regions (e.g. Zhang \& M\'esz\'aros 2002) may solve the
problem, although more detailed modelling is needed. Crossing of a cooling break
would also result in a temporal break, but it would also lead to a change of the
spectral index by $\sim 0.5$. From Table 1 we find that the changes of the X-ray
spectral indices across the break of these bursts are $0.01\pm 0.10$ (050318),
$0.04\pm 0.10$ (050319), $0.08\pm 0.12$(050401), and $0.03\pm 0.09$ (050802).
These results confidently rule out such a possibility. Genet et al. (2007)
account for these chromatic breaks as due to a long-live reverse shock in which
only a small fraction of electrons are accelerated. The big issue of such an
interpretation is how to ``hide'' the emission from the forward shock, which
carries most of the energy.

Assuming a simple energy injection law $L(t) \propto t^{-q}$, we find that
averagely speaking the injection luminosity could be almost a constant. This
places some constraints on the physical models of the energy injection models.
The constant injection luminosity agree with the expectation of the energy
injection model from a central pulsar (Dai \& Lu 1998a; Zhang \& M\'esz\'aros
2001), suggesting that the pulsar injection model may be consistent with the data
at least for some GRBs (see also Grupe et al. 2007; Fan \& Xu 2006; De Pasquale
et al. 2007; Yu et al. 2007).

The temporal decay slopes of some bursts following the shallow decay phase are
shallower than the model predictions [Fig. 5, see also Fig. 5(a) in Willingale et
al. 2007]. This discrepancy may be alleviated by different ways. First, as shown
in Fig. 1, $\alpha_2$ could be systematically steeper if a smoother broken power
law model (with smaller $\omega$) is adopted. Second, theoretically the temporal
breaks involving external shocks are usually not sharp. Other effects, such as
delay of transfer of the fireball energy to the forward shock (Kobayashi \& Zhang
2007), and the structured jet effect (Zhang et al. 2004; Kumar et al. 2004;
Yamazaki et al. 2006) would modify the simplest closure relations to make a band
rather than a line in the $\alpha-\beta$ plane.

One interesting conclusion from this study is that at least for a small fraction
of bursts (e.g. GRBs 060413, 060522, 060607A, and 070110), the observed shallow
decay phase is likely of internal origin. This is another component other than
X-ray flares that are possibly of internal origin. Contrary to the
erratic X-ray flares, this component has a smoother light curve with flickering,
likely due to a steady component from the central engine. A possible energy
source for such a component would be the spin energy from the central engine, and
an internal dissipation of the spindown power may be the origin (e.g. Troja et
al. 2007). Tapping of the rotation energy is likely through magnetic fields, either
through dipolar spindown for a central millisecond pulsar (Usov 1992; Dai \& Lu
1998a,b; Zhang \& M\'esz\'aros 2001) or by the Blandford-Znajek mechanism for a black
hole central engine (Blandford \& Znajek 1977; M\'esz\'aros \& Rees 1997b; Li
2000). If one accepts that such a component is common among bursts, one can
speculate that the observed early X-ray emission is the sum of different emission
components. The competition among these components shapes the variety of the X-ray
light curves one observes. Depending on the relative importance of the internal and
external components, the shallow decay segment could be possibly dominated by
either the radiation from the refreshed shocks or by the steady radiation
component from the internal dissipation of the central engine. In the former
scenario, the shallow decay phase transits to a normal decay phase that is
consistent with the external shock models. In the later scenario, the emission
level of the underlying afterglow component is weaker than that from the emission
component of the central engine, so that one needs a steep dropoff from the
plateau to land onto the external shock emission component.

\acknowledgments We acknowledge the use of the public data from the Swift data
archive. We appreciate valuable comments and suggestions from the anonymous
referee and P. O'Brien, Z. G. Dai, X. Dai, D. M. Wei, K. Ioka, L. Nava, X. Y.
Wang, X. F. Wu, F. W. Zhang. This work is supported by NASA under grants
NNG06GH62G, NNG05GB67G, NNX07AJ64G, NNX07AJ66G, and the National Natural Science
Foundation of China under grant No. 10463001 (EWL) and 10640420144.

\begin{deluxetable}{lllllllllllllllllllll}


\tablewidth{450pt} \tabletypesize{\tiny}

\tablecaption{XRT observations and the fitting results of our sample}

\tablenum{1}

\tablehead{\colhead{GRB}&\colhead{$t_1$(ks)\tablenotemark{a}}&\colhead{$t_2$(ks)\tablenotemark{a}}&\colhead{$t_{b}$(ks)\tablenotemark{b}}&\colhead{$\alpha_{\rm
X,1}$\tablenotemark{b}}&\colhead{$\alpha_{\rm
X,2}$\tablenotemark{b}}&\colhead{$\chi^2$(dof)\tablenotemark{b}}&\colhead{$S_X$\tablenotemark{c}}&\colhead{$\Gamma_{\rm
X,1}$\tablenotemark{d}}&\colhead{$\Gamma_{\rm X,2}$\tablenotemark{d}}}

\startdata

050128 & 0.25 & 70.72 & 2.76(0.62) & 0.49(0.07) & 1.26(0.03) & 40(48) & 3.70(1.07) & 1.87(0.14) & 1.95(0.06)\\
050315 & 5.40 & 450.87 & 224.64(38.68) & 0.66(0.03) & 1.90(0.28) & 42(52) & 10.88(2.56) & 2.06(0.11) & 2.18(0.08)\\
050318 & 3.34 & 45.19 & 10.64(4.97) & 0.90(0.23) & 1.84(0.19) & 27(20) & 5.92(6.32) & 2.09(0.08) & 2.02(0.06)\\
050319 & 6.11 & 84.79 & 11.20(13.26) & 0.23(0.59) & 0.99(0.25) & 9(9) & 1.26(1.42) & 2.00(0.06) & 2.04(0.07)\\
050401 & 0.14 & 801.04 & 5.86(0.78) & 0.57(0.02) & 1.37(0.06) & 106(92) & 9.32(1.31) & 1.91(0.05) & 1.99(0.11)\\
050416A & 0.25 & 261.69 & 1.74(1.12) & 0.43(0.12) & 0.90(0.04) & 36(38) & 0.62(0.38) & 2.18(0.31) & 2.15(0.10)\\
050505 & 3.07 & 97.19 & 7.87(1.57) & 0.15(0.19) & 1.30(0.06) & 26(45) & 2.34(0.68) & 2.00(0.07) & 2.03(0.04)\\
050713B & 0.79 & 478.50 & 10.80(1.59) & -0.00(0.07) & 0.94(0.04) & 40(63) & 3.28(0.35) & 1.85(0.10) & 1.94(0.09)\\
050726 & 0.42 & 17.05 & 1.17(0.33) & 0.08(0.33) & 1.31(0.09) & 13(21) & 1.17(0.53) & 2.25(0.07) & 2.07(0.06)\\
050801 & 0.07 & 46.10 & 0.25(fixed) & 0.00(fixed) & 1.10(0.03) & 44(45) & 0.16(0.01) & 1.70(0.19) & 1.91(0.12)\\
050802 & 0.51 & 83.83 & 4.09(0.61) & 0.32(0.10) & 1.61(0.04) & 58(72) & 3.66(0.94) & 1.91(0.06) & 1.89(0.07)\\
050803 & 0.50 & 368.89 & 13.71(0.90) & 0.25(0.03) & 2.01(0.07) & 94(57) & 5.96(0.51) & 1.76(0.14) & 2.00(0.08)\\
050822 & 6.41 & 523.32 & 66.99(44.38) & 0.60(0.10) & 1.25(0.19) & 29(44) & 4.05(3.12) & 2.29(0.13) & 2.36(0.11)\\
051008 & 3.09 & 43.77 & 14.67(3.82) & 0.78(0.11) & 1.96(0.21) & 17(19) & 6.87(3.43) & 2.00(0.11) & 2.06(0.07)\\
051016B & 4.78 & 150.47 & 66.40(23.09) & 0.71(0.08) & 1.84(0.46) & 15(16) & 2.18(1.10) & 2.15(0.13) & 2.19(0.13)\\
051109A & 3.73 & 639.16 & 27.28(7.90) & 0.79(0.07) & 1.53(0.08) & 39(48) & 10.59(4.71) & 1.91(0.07) & 1.90(0.07)\\
060105 & 0.10 & 360.83 & 2.31(0.14) & 0.84(0.01) & 1.72(0.02) & 653(754) & 42.98(3.84) & 2.23(0.05) & 2.15(0.03)\\
060108 & 0.77 & 165.26 & 22.08(7.38) & 0.26(0.09) & 1.43(0.17) & 7(7) & 0.53(0.17) & 2.17(0.32) & 1.75(0.15)\\
060109 & 0.74 & 48.01 & 4.89(1.10) & -0.17(0.14) & 1.32(0.09) & 19(13) & 0.91(0.20) & 2.32(0.15) & 2.34(0.14)\\
060124 & 13.30 & 664.01 & 52.65(10.33) & 0.78(0.10) & 1.65(0.05) & 165(132) & 29.65(12.09) & 2.10(0.03) & 2.08(0.06)\\
060204B & 4.06 & 98.80 & 5.55(0.66) & -0.59(0.72) & 1.45(0.07) & 21(34) & 0.87(0.36) & 2.54(0.14) & 2.77(0.18)\\
060210 & 3.90 & 861.94 & 24.24(5.01) & 0.63(0.05) & 1.38(0.05) & 144(133) & 10.41(2.90) & 2.06(0.03) & 2.12(0.09)\\
060306 & 0.25 & 124.39 & 4.67(2.91) & 0.40(0.11) & 1.05(0.07) & 30(32) & 1.58(0.98) & 2.09(0.16) & 2.21(0.10)\\
060323 & 0.33 & 16.28 & 1.29(0.32) & -0.11(0.23) & 1.55(0.16) & 4(7) & 0.27(0.08) & 1.99(0.16) & 2.02(0.13)\\
060413 & 1.20 & 253.52 & 26.43(1.12) & 0.18(0.03) & 3.42(0.21) & 78(71) & 13.77(0.82) & 1.60(0.08) & 1.50(0.10)\\
060428A & 0.23 & 271.10 & 11.04(6.58) & 0.27(0.09) & 0.88(0.08) & 25(21) & 3.79(1.74) & 2.11(0.24) & 2.05(0.14)\\
060502A & 0.24 & 593.06 & 72.57(15.05) & 0.53(0.03) & 1.68(0.15) & 11(26) & 5.09(1.19) & 2.20(0.12) & 2.15(0.13)\\
060507 & 3.00 & 86.09 & 6.95(1.68) & -0.37(0.48) & 1.25(0.09) & 2(8) & 0.40(0.16) & 2.15(0.19) & 2.13(0.12)\\
060510A & 0.16 & 343.41 & 9.18(0.67) & 0.10(0.03) & 1.51(0.03) & 93(142) & 17.28(1.65) & 1.91(0.09) & 1.96(0.06)\\
060522 & 0.20 & 0.90 & 0.53(0.06) & 0.14(0.36) & 3.15(0.79) & 11(11) & 0.26(0.12) & 2.03(0.16) & 2.13(0.30)\\
060526 & 1.09 & 322.75 & 10.02(4.55) & 0.30(0.12) & 1.50(0.23) & 34(48) & 0.79(0.32) & 2.08(0.09) & 2.08(0.16)\\
060604 & 3.52 & 403.81 & 11.37(6.80) & 0.19(0.48) & 1.17(0.08) & 34(41) & 0.79(0.67) & 2.44(0.15) & 2.43(0.17)\\
060607A & 1.52 & 39.52 & 12.34(0.19) & 0.00(fixed) & 3.35(0.09) & 132(139) & 8.45(0.17) & 1.44(0.06) & 1.64(0.05)\\
060614 & 5.03 & 451.71 & 49.84(3.62) & 0.18(0.06) & 1.90(0.07) & 70(54) & 4.35(0.49) & 2.02(0.02) & 1.93(0.06)\\
060707 & 5.32 & 813.53 & 22.21(54.08) & 0.37(0.96) & 1.09(0.17) & 8(11) & 0.64(2.01) & 1.88(0.09) & 2.06(0.20)\\
060708 & 3.81 & 439.09 & 6.66(3.84) & 0.49(0.54) & 1.30(0.09) & 39(34) & 0.96(1.06) & 2.41(0.17) & 2.28(0.12)\\
060714 & 0.32 & 331.97 & 3.70(0.97) & 0.34(0.10) & 1.27(0.05) & 53(73) & 1.48(0.46) & 2.15(0.08) & 2.04(0.11)\\
060719 & 0.28 & 182.15 & 9.57(2.70) & 0.40(0.06) & 1.31(0.10) & 19(26) & 1.30(0.37) & 2.35(0.13) & 2.38(0.26)\\
060729 & 0.42 & 2221.24 & 72.97(3.02) & 0.21(0.01) & 1.42(0.02) & 459(459) & 19.58(0.83) & 3.35(0.04) & 2.26(0.05)\\
060804 & 0.18 & 122.07 & 0.86(0.22) & -0.09(0.15) & 1.12(0.07) & 18(24) & 0.97(0.18) & 2.04(0.23) & 2.14(0.15)\\
060805A & 0.23 & 75.91 & 1.30(0.70) & -0.17(0.41) & 0.97(0.13) & 11(17) & 0.06(0.03) & 2.10(0.10) & 1.97(0.37)\\
060807 & 0.28 & 166.22 & 8.04(0.35) & 0.06(0.03) & 1.73(0.05) & 67(36)\tablenotemark{*} & 1.94(0.11) & 2.30(0.28) & 2.22(0.08)\\
060813 & 0.09 & 74.25 & 1.77(0.27) & 0.55(0.03) & 1.25(0.03) & 86(75) & 7.31(1.36) & 2.09(0.16) & 2.04(0.04)\\
060814 & 0.57 & 399.37 & 17.45(1.71) & 0.54(0.02) & 1.59(0.05) & 81(57) & 6.93(0.87) & 2.11(0.09) & 2.30(0.05)\\
060906 & 1.32 & 36.69 & 13.66(3.29) & 0.35(0.10) & 1.97(0.36) & 3(7) & 0.96(0.29) & 2.28(0.37) & 2.12(0.17)\\
060908 & 0.08 & 363.07 & 0.95(0.34) & 0.70(0.07) & 1.49(0.09) & 98(59)\tablenotemark{*}  & 1.28(0.61) & 2.41(0.21) & 2.00(0.08)\\
060912 & 0.42 & 86.80 & 1.13(0.31) & 0.13(0.30) & 1.19(0.08) & 8(26) & 0.37(0.15) & 2.08(0.11) & 1.95(0.13)\\
061021 & 0.30 & 594.16 & 9.59(2.17) & 0.52(0.03) & 1.08(0.03) & 94(87) & 3.59(0.87) & 1.81(0.04) & 1.70(0.13)\\
061121 & 4.89 & 353.10 & 24.32(4.38) & 0.75(0.06) & 1.63(0.05) & 121(147) & 19.89(6.14) & 2.00(0.04) & 1.93(0.05)\\
061202 & 0.93 & 357.04 & 41.65(5.36) & 0.10(0.04) & 2.20(0.18) & 55(49) & 13.80(1.12) & 2.15(0.09) & 3.55(0.44)\\
061222A & 22.78 & 724.64 & 32.73(2.17) & -0.61(0.45) & 1.75(0.04) & 102(59)\tablenotemark{*}  & 6.62(1.89) & 2.46(0.07) & 2.22(0.12)\\
070110 & 4.10 & 28.72 & 20.40(0.44) & 0.11(0.05) & 8.70(0.88) & 43(66) & 3.59(0.23) & 2.16(0.11) & 2.21(0.09)\\
070129 & 1.32 & 546.36 & 20.12(3.14) & 0.15(0.07) & 1.31(0.06) & 42(70) & 1.47(0.24) & 2.25(0.07) & 2.30(0.10)\\

\enddata

\tablenotetext{a}{The starting and ending time of our lightcurve fitting}

\tablenotetext{b}{The break time and the decay slopes before and after the break,
and the fitting $\chi^2$ (degrees of freedom).}

\tablenotetext{c}{The X-ray fluence (in units of $10^{-7}$ erg cm$^{-2}$) of the
shallow decay phase calculated by integrating the fitting light curve from 10
seconds post the GRB trigger to $t_b$.}

\tablenotetext{d}{The X-ray photon indices before and after $t_b$.}

\tablenotetext{*}{The fitting results of these bursts have an unaccepted reduced
$\chi^2$ due to significant flicking.}

\end{deluxetable}

\begin{deluxetable}{lllllllll}


\tablewidth{400pt} \tabletypesize{\tiny} \tablecaption{BAT observations and
redshifts of our sample}

\tablenum{2}

\tablehead{\colhead{GRB}& \colhead{$T_{90}$(s)}&
\colhead{$S_\gamma$\tablenotemark{a}}&
\colhead{$\Gamma_{\gamma,1}$\tablenotemark{b}}&
\colhead{$\Gamma_{\gamma,2}$\tablenotemark{b}}&\colhead{$E_{\rm
p}$(keV)\tablenotemark{b}}&\colhead{ref$_{\rm
BAT}$\tablenotemark{c}}&\colhead{$z$}&\colhead{ref$_{z}$ \tablenotemark{c}}}
\startdata
050128 & 13.8(2.0) & 45.00(5.00) & 1.50(0.05) &  & 133.1(30.2) & 2992 &  & \\
050315 & 96.0(10.0) & 28.00(3.00) & 1.28(0.00) & 2.20 & 37.0(8.0) & 3105 & 1.95 & 3101\\
050318 & 32.0(2.0) & 21.00(2.00) & 2.10(0.11) &  & 39.5(12.4) & 3134 & 1.44 &3122  \\
050319 & 10.0(2.0) & 8.00(0.80) & 1.25(0.00) & 2.15 & 28.0(6.0) & 3119, Z07 & 3.24 & 3136\\
050401 & 33.0(2.0) & 140.00(14.00) & 1.15(0.00) & 2.65 & 132.0(16.0) & 3173,Z07 & 2.90 &3176 \\
050416A & 2.4(0.2) & 3.80(0.40) & 1.00(0.00) & 3.22 & 16.0(3.0) &  3273 & 0.65 & 3542\\
050505 & 60.0(2.0) & 41.00(4.00) & 1.50(0.10) &  & 133.1(41.0) &  3364, Z07& 4.27 & 3368\\
050713B & 75.0(7.5) & 82.00(10.00) & 1.00(0.13) &  & 109.0(32.0) &3600, Z07  &  & \\
050726 & 30.0(3.0) & 43.00(7.00) & 1.00(0.16) &  & 984.0(200.0) &  3682, Z07& & \\
050801 & 20.0(3.0) & 4.40(1.00) & 1.40(0.00) & 2.00 & 33.0(7.0) & 3730, Z07 & & \\
050802 & 13.0(2.0) & 28.00(3.00) & 1.12(0.00) & 2.48 & 118.0(77.0) & 3737, Z07 & 1.71 &3749 \\
050803 & 110.0(11.0) & 39.00(3.00) & 1.05(0.10) & & 150.0(68.0) & 3757, Z07 & 0.42 &3758 \\
050822 & 102.0(2.0) & 34.00(3.00) & 1.00(0.00) & 2.48 & 36.0(7.0) & 3856 & & \\
051008 & 280.0(28.0) & 540.00(10.00) & 0.98(0.09) &  & 865.0(178.0) & 4077, Z07 & & \\
051016B & 4.0(0.1) & 1.70(0.20) & 2.38(0.23) &  & 25.2(11.2) & 4104 & 0.94 & 4186\\
051109A & 36.0(2.0) & 21.00(3.00) & 1.50(0.20) &  & 133.1(69.0) & 4217 & 2.35 & 4221\\
060105 & 55.0(5.0) & 182.00(4.00) & 1.11(0.03) &  & 394.8(75.2) & 4435 &  & \\
060108 & 14.4(1.0) & 3.70(0.40) & 2.01(0.17) &  & 46.3(18.1) & 4445 & 2.03 &4539 \\
060109 & 116.0(3.0) & 6.40(1.00) & 1.96(0.25) &  & 50.7(26.3) & 4476 &  & \\
060124 & 800.0(10.0) & 110.00(10.00) & 1.17(0.27) &  & 326.5(277.3) & 4601 & 2.30 & 4592\\
060204B & 134.0(5.0) & 30.00(2.00) & 0.82(0.40) &  & 96.8(41.0) & 4671 & & \\
060210 & 255.0(10.0) & 77.00(4.00) & 1.52(0.09) &  & 126.9(36.7) & 4748 & 3.91 &4729 \\
060306 & 61.0(2.0) & 22.00(1.00) & 1.85(0.10) &  & 62.4(18.7) & 4851 &  & \\
060323 & 18.0(2.0) & 5.70(0.60) & 1.53(0.17) &  & 124.0(55.3) & 4912 &  & \\
060413 & 150.0(10.0) & 36.00(1.00) & 1.67(0.08) &  & 90.4(24.5) & 4961 & & \\
060428A & 39.4(2.0) & 14.00(1.00) & 2.04(0.11) &  & 43.9(13.8) & 5022 & & \\
060502A & 33.0(5.0) & 22.00(1.00) & 1.43(0.08) &  & 158.2(43.4) & 5053 & 1.51 & 5052\\
060507 & 185.0(5.0) & 45.00(2.00) & 1.83(0.10) &  & 64.9(19.3) & 5091 & & \\
060510A & 21.0(3.0) & 98.00(5.00) & 1.55(0.10) &  & 118.3(36.1) & 5108 & & \\
060522 & 69.0(5.0) & 11.00(1.00) & 1.59(0.15) &  & 107.9(42.7) & 5153 & 5.11 & 5155\\
060526 & 13.8(2.0) & 4.90(0.60) & 1.66(0.20) &  & 92.3(44.5) & 5174 & 3.21 & 5170\\
060604 & 10.0(3.0) & 1.30(0.30) & 1.90(0.41) &  & 56.7(46.1) & 5214 & 2.68 & 5218\\
060607A & 100.0(5.0) & 26.00(1.00) & 1.45(0.07) &  & 150.5(38.6) &5242  & 3.08 &5237 \\
060614 & 102.0(5.0) & 217.00(4.00) & 2.13(0.04) &  & 37.5(9.9) & 5256 & 0.125 &5275 \\
060707 & 68.0(5.0) & 17.00(2.00) & 0.66(0.63) & & 66.0(25.0) & 5289 & 3.43 & 5298\\
060708 & 9.8(1.0) & 5.00(0.40) & 1.68(0.12) & & 88.4(29.4) & 5395 &  & \\
060714 & 115.0(5.0) & 30.00(2.00) & 1.99(0.10) & & 48.0(14.5) & 5334 & 2.71 & J06\\
060719 & 55.0(5.0) & 16.00(1.00) & 2.00(0.11) & & 47.1(14.7) & 5349 &  & \\
060729 & 116.0(10.0) & 27.00(2.00) & 1.86(0.14) & & 61.2(21.7) & 5370 & 0.54 & 5373\\
060804 & 16.0(2.0) & 5.10(0.90) & 1.78(0.28) & & 71.8(43.7) & 5395 &  & \\
060805A & 5.4(0.5) & 0.74(0.20) & 2.23(0.42) & & 31.8(23.2) & 5421 & & \\
060807 & 34.0(4.0) & 7.30(0.90) & 1.57(0.20) & & 112.9(56.6) & 5403 &  & \\
060813 & 14.9(0.5) & 55.00(1.00) & 0.53(0.15) & & 192.0(19.0) & 5443,5446 &  & \\
060814 & 146.0(10.0) & 150.00(2.00) & 1.56(0.03) & & 115.6(24.3) & 5459 &  & \\
060906 & 43.6(1.0) & 22.10(1.40) & 2.02(0.11) & & 45.5(14.2) & 5538 & 3.68 & 5535\\
060908 & 19.3(0.3) & 29.00(1.00) & 1.33(0.07) & & 205.5(53.4) & 5551 & 2.43 & 5555\\
060912 & 5.0(0.5) & 13.00(1.00) & 1.74(0.09) & & 77.9(22.3) & 5561 & 0.94 & 5617\\
061021 & 46.0(1.0) & 30.00(1.00) & 1.31(0.06) & & 217.1(52.4) & 5744 & & \\
061121 & 81.0(5.0) & 137.00(2.00) & 1.41(0.03) & & 166.5(33.2) & 5831 & 1.31 & 5826\\
061202 & 91.0(5.0) & 35.00(1.00) & 1.63(0.07) & & 98.6(25.4) &  5887&  & \\
061222A & 72.0(3.0) & 83.00(2.00) & 1.39(0.04) & & 175.3(36.8) &  5964& & \\
070110 & 85.0(5.0) & 16.00(1.00) & 1.57(0.12) & & 112.9(38.4) & 6007 & 2.35 & 6010\\
070129 & 460.0(20.0) & 31.00(3.00) & 2.05(0.16) & & 43.1(16.1) & 6058 & & \\
\enddata
\tablenotetext{a}{The observed gamma-ray fluence and its error in 15-150 keV
band, in units of $10^{-7}$ erg cm$^{-2}$.}

\tablenotetext{b} {The spectrum of the prompt gamma-rays is generally fitted by a
simple power-law, $f_\nu\propto \nu^{-\Gamma_{\gamma,1}}$. We estimate the $E_p$
for these bursts with the empirical relation between $\Gamma_{\gamma_1}$ and
$E_p$ for the BAT observations. Few cases are fitted with a cut-off power-law or
the Band function (the photon indices before and after the break energy is
$\Gamma_{\gamma,1}$ and $\Gamma_{\gamma,2}$, respectively).}

\tablenotetext{c} {References for BAT data and redshift data.}

\tablerefs{ J06: Jakobsson et al. 2006(c); Z07: Zhang et al. 2007a; 2992:
Cummings et al. (2005) ; 3101: Kelson and  Berger (2005); 3105: Krimm et al.
(2005c); 3119: Krimm et al. (2005a); 3122: Berger and Mulchaey (2005); 3134:
Krimm et al. (2005b); 3136: Fynbo et al. (2005a); 3173: Sakamoto et al. (2005a);
3176: Fynbo et al. (2005c); 3273: Sakamoto et al. (2005c); 3364: Hullinger et al.
(2005a); 3368: Berger et al. (2005b); 3542: Cenko et al. (2005); 3600: Parsons et
al. (2005a); 3682: Barthelmy et al. (2005a); 3730: Sakamoto et al. (2005b); 3737:
Palmer et al. (2005); 3749: Fynbo et al. (2005b); 3757: Parsons (2005b); 3758:
Bloom et al. (2005); 3856: Hullinger et al. (2006b); 4077: Barthelmy et al.
(2005b); 4104: Barbier et al. (2005); 4186: Soderberg et al. (2005); 4217:
Fenimore et al. (2005); 4221: Quimby et al. (2005); 4435: Markwardt et al.
(2006a); 4445: Sakamoto et al. (2006b); 4476: Palmer et al. (2006a); 4539:
Melandri et al. (2006); 4592: Cenko et al. (2006a); 4601: Lamb et al. (2006);
4671: Markwardt et al. (2006c); 4729: Cucchiara et al. (2006a); 4748: Sakamoto et
al. (2006a); 4851: Hullinger et al. (2006); 4912: Parsons et al. (2006b); 4961:
Barbier et al. (2006a); 5022: Markwardt et al.(2006b); 5052: Cucchiara et al.
(2006b); 5053: Parsons et al. (2006c); 5091:Barbier et al. (2006b); 5108: Barbier
et al. (2006c); 5153: Krimm et al.(2006c); 5155: Cenko et al. (2006a); 5162:
Berger and Gladders (2006a); 5170:Berger and  Gladders (2006b); 5174: Markwardt
et al. (2006d); 5214: Parsons et al.(2006e); 5218: Castro-Tirado et al. (2006);
5237: Ledoux et al. (2006); 5242:Tueller et al. (2006b); 5256: Barthelmy, S. et
al. (2006a); 5275: Price et al.(2006); 5289: Stamatikos et al. (2006b); 5295:
Fenimore et al. (2006a); 5298: Jakobsson et al. (2006a); 5320: Jakobsson et al.
(2006c); 5334: Krimm et al. (2006d); 5349: Sakamoto et al. (2006c); 5370: Parsons
et al. (2006a); 5373: Thoene et al. (2006); 5395: Tueller et al. (2006c); 5403:
Barbier et al. (2006d); 5421: Barthelmy et al. (2006b); 5443: Cummings et al.
(2006); 5446: Golenetskii et al. (2006); 5459: Stamatikos et al. (2006a); 5535:
Vreeswijk (2006); 5538: Sato et al. (2006); 5551: Palmer et al. (2006c); 5555:
Rol et al. (2006); 5561: Parsons et al. (2006d); 5617: Jakobsson (2006d); 5744:
Palmer et al. (2006b); 5826: Bloom (2006); 5831:Fenimore  et al. (2006b); 5887:
Sakamoto et al. (2006d); 5964: Tueller et al. (2006a); 6007: Cummings et al.
(2007); 6010: Jaunsen et al. (2007); 6058: Krimm et al. (2006e).}
\end{deluxetable}

\begin{deluxetable}{llllllll}


\tablewidth{330pt} \tabletypesize{\tiny} \tablecaption{The optical observations
and our fitting results}

\tablenum{3}

\tablehead{\colhead{GRB}& \colhead{$t_1$(ks)\tablenotemark{a}}&
\colhead{$t_2$(ks)\tablenotemark{a}}& \colhead{$t_{b,O}$(ks)\tablenotemark{b}}&
\colhead{$\alpha_{O,1}$\tablenotemark{b}}&
\colhead{$\alpha_{O,2}\tablenotemark{b}$}&
\colhead{$\chi^2$/(dof)\tablenotemark{b}}&\colhead{ref}} \startdata
050318 & 3.23 & 22.83 & - & 0.84(0.22)  & -& 0.5(1)&(1)\\
050319 & 2.00 & 204.74 & - & 0.42(0.02) & - & 11(16)&(2)-(4)\\
050401 & 0.06 & 1231.18 & - & 0.80(0.01) & - & 43(12)&(5)-(7)\\
050801 & 0.02 & 9.49 & 0.19(0.02) & -0.02(0.07) & 1.10(0.02) & 26(42)&(8)\\
050802 & 0.34 & 127.68 & - & 0.85(0.02) & - & 50(10)&(9)-(11)\\
051109A & 0.04 & 20170.00 & 21.80(10.95) & 0.66(0.02) & 1.10(0.08) & 106(42)&(12)\\
060124 & 3.34 & 1979.30 & - & 0.85(0.02) & - & 11(19)&(13)-(14)\\
060210 & 0.09 & 7.19 & 0.70(0.18) & 0.01(0.24) & 1.23(0.08) & 5(12)&(15)-(16)\\
060526 & 0.06 & 893.55 & 84.45(5.88) & 0.67(0.02) & 1.80(0.04) & 116(56)&(17)\\
060607A & 0.07 & 13.73 & 0.16(fixed) & -3.07(0.25) & 1.18(0.02) & 92(35)&(18)\\
060614 & 1.54 & 934.36 & 39.09(1.71) & -0.40(0.05) & 2.16(0.03) & 114(24)&(19)-(21)\\
060714 & 3.86 & 285.87 & 1.00 & 0.01(fixed) & 1.41(0.03) & 35(11)&(22)-(26)\\
060729 & 20.00 & 662.39 & 43.29(5.15) & -0.37(0.34) & 1.34(0.06) & 36(27)&(27)\\
061121 & 0.26 & 334.65 & 1.70(0.73) & 0.17(fixed) & 0.99(0.05) & 18(23)&(28)\\
070110 & 0.66 & 34.76 & - & 0.43(0.08) &  & 1(4)&29\\
\enddata
\tablenotetext{a}{The time interval concerned in our fitting}

\tablenotetext{b}{For a smooth broken power law fit, $t_{b,O}$, $\alpha_{O,1}$,
$\alpha_{O,2}$ are the break time and the decay slopes before and after the
break. For a simple power law fit, the decay index and its error are shown in column
$\alpha_{O,1}$. In order to make the fittings more reasonable, we assume an error
of 0.1 mag for those data points without observational error available.}

\tablerefs{(1)Still et al. (2005); (2) Quimby et al. (2006); (3) Huang et al.
(2007); (4) Mason et al. (2006); (5) De Pasquale (2006); (6) Rykoff et al.
(2005); (7) Watson et al. (2006); (8) Rykoff et al. (2006); (9) McGowan et
al.(2005a); (10)Pavlenko et al.(2005); (11: McGowan et al.(2005b);(12) Yost et
al. (2007); (13) Misra et al. (2007); (14): Romano et al. 2006; (15) Curran et
al.(2007);(16) Stanek et al.(2007); (17) Dai et al. 2007; (18) Molinari et
al.(2007); (19)Fynbo et al.(2006); (20)Della Valle et al.(2006); (21)Gal-Yam et
al.(2006); (22) Asfandyarov et al.(2006); (23)Rumyantsev et al.(2006); (24)
Pavlenko et al.(2006a); (25) Jakobsson et al.(2006c); (26)Pavlenko et al.(2006b);
(27) Grupe et al.(2007); (28) Page et al.(2007); (29) Troja et al.(2007).}

\end{deluxetable}

\begin{deluxetable}{lllllll}


\tablewidth{400pt} \tabletypesize{\tiny} \tablecaption{Rest-frame properties of
the bursts with known redshifts in our sample}

\tablenum{4}

\tablehead{\colhead{GRB}& \colhead{$\log T_{90}^{'}$(s)}& \colhead{$\log E_{\rm
p}^{'}$(keV)}& \colhead{$\log E_{\rm iso, \gamma}$(erg)}& \colhead{$\log E_{\rm
iso, \gamma}^{p}$(erg)}& \colhead{$\log E_{\rm iso, X}$(erg)}& \colhead{$\log
t_{\rm b}^{'}$(s)}}

\startdata
050315 & 1.51(0.05) & 2.0(0.1) & 52.41(0.05) & 52.84 & 51.94(0.10) & 4.83(0.07)\\
050318 & 1.12(0.03) & 2.0(0.1) & 52.04(0.04) & 52.38 & 51.88(0.46) & 4.03(0.20)\\
050319 & 0.37(0.09) & 2.1(0.1) & 52.24(0.04) & 52.69 & 52.03(0.49) & 4.01(0.51)\\
050401 & 0.93(0.03) & 2.7(0.1) & 53.41(0.04) & 53.69 & 52.82(0.06) & 3.77(0.06)\\
050416A & 0.16(0.04) & 1.4(0.1) & 50.62(0.05) & 51.00 & 49.69(0.26) & 2.88(0.28)\\
050505 & 1.06(0.01) & 2.8(0.1) & 53.14(0.04) & 53.51 & 52.62(0.13) & 3.90(0.09)\\
050802 & 0.68(0.07) & 2.5(0.3) & 52.31(0.05) & 52.62 & 51.86(0.11) & 3.61(0.06)\\
050803 & 1.89(0.04) & 2.3(0.2) & 51.24(0.03) & 51.67 & 50.29(0.04) & 3.85(0.03)\\
051016B & 0.32(0.01) & 1.7(0.2) & 50.59(0.05) & 50.95 & 50.99(0.22) & 4.82(0.15)\\
051109A & 1.03(0.02) & 2.6(0.2) & 52.43(0.06) & 52.82 & 52.65(0.19) & 4.44(0.13)\\
060108 & 0.68(0.03) & 2.1(0.2) & 51.56(0.05) & 51.89 & 51.20(0.14) & 4.34(0.15)\\
060124 & 2.38(0.01) & 3.0(0.4) & 53.13(0.04) & 53.72 & 53.08(0.18) & 4.72(0.09)\\
060210 & 1.72(0.02) & 2.8(0.1) & 53.36(0.02) & 53.72 & 53.18(0.12) & 4.38(0.09)\\
060502A & 1.12(0.07) & 2.6(0.1) & 52.10(0.02) & 52.54 & 51.82(0.10) & 4.81(0.09)\\
060522 & 1.05(0.03) & 2.8(0.2) & 52.69(0.04) & 53.03 & 51.29(0.19) & 2.16(0.05)\\
060526 & 0.52(0.06) & 2.6(0.2) & 52.02(0.05) & 52.36 & 51.21(0.26) & 3.38(0.25)\\
060604 & 0.43(0.13) & 2.3(0.4) & 51.32(0.10) & 51.64 & 50.88(0.37) & 3.27(0.26)\\
060607A & 1.39(0.02) & 2.8(0.1) & 52.72(0.02) & 53.12 & 52.84(0.01) & 4.09(0.01)\\
060614 & 1.96(0.02) & 1.6(0.1) & 50.90(0.01) & 51.24 & 49.25(0.05) & 4.70(0.03)\\
060707 & 1.19(0.03) & 2.5(0.2) & 52.61(0.05) & 52.98 & 51.44(1.36) & 3.95(1.06)\\
060714 & 1.49(0.02) & 2.3(0.1) & 52.69(0.03) & 53.01 & 51.95(0.14) & 3.57(0.11)\\
060729 & 1.88(0.04) & 2.0(0.2) & 51.31(0.03) & 51.65 & 51.35(0.02) & 4.86(0.02)\\
060906 & 0.97(0.01) & 2.3(0.1) & 52.78(0.03) & 53.09 & 51.57(0.13) & 3.62(0.10)\\
060908 & 0.75(0.01) & 2.8(0.1) & 52.59(0.01) & 53.07 & 51.12(0.21) & 2.32(0.16)\\
060912 & 0.41(0.04) & 2.2(0.1) & 51.47(0.03) & 51.83 & 50.21(0.17) & 3.05(0.12)\\
061121 & 1.54(0.03) & 2.6(0.1) & 52.78(0.01) & 53.23 & 51.83(0.13) & 3.90(0.08)\\
070110 & 1.40(0.03) & 2.6(0.1) & 52.31(0.03) & 52.68 & 52.19(0.03) & 4.31(0.01)\\
\enddata
\end{deluxetable}

\clearpage

\begin{figure}
\epsscale{0.8} \plotone{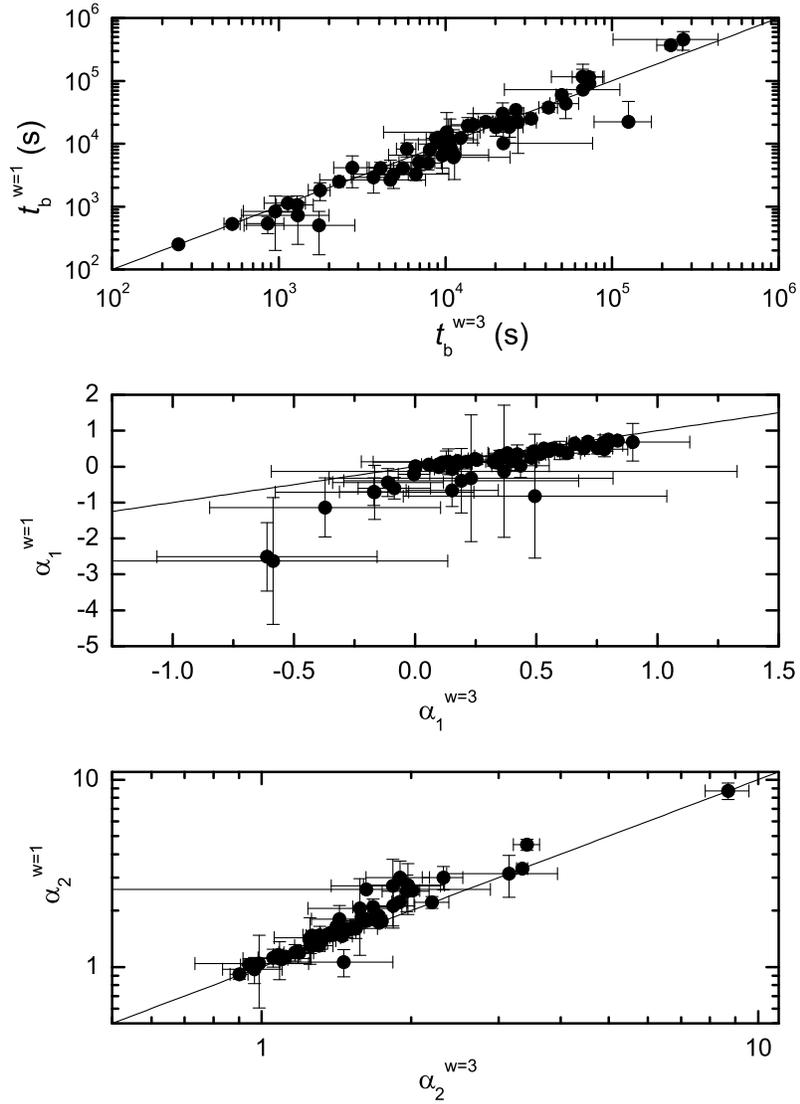} \caption{Comparison of the fitting results of
smooth broken power law models with different sharpness parameters: $\omega=1$
and $\omega=3$. The solid lines stand for the equality of the two quantities.}
\label{Fig_fittingmodel}
\end{figure}

\begin{figure*}
\includegraphics[angle=0,scale=0.40]{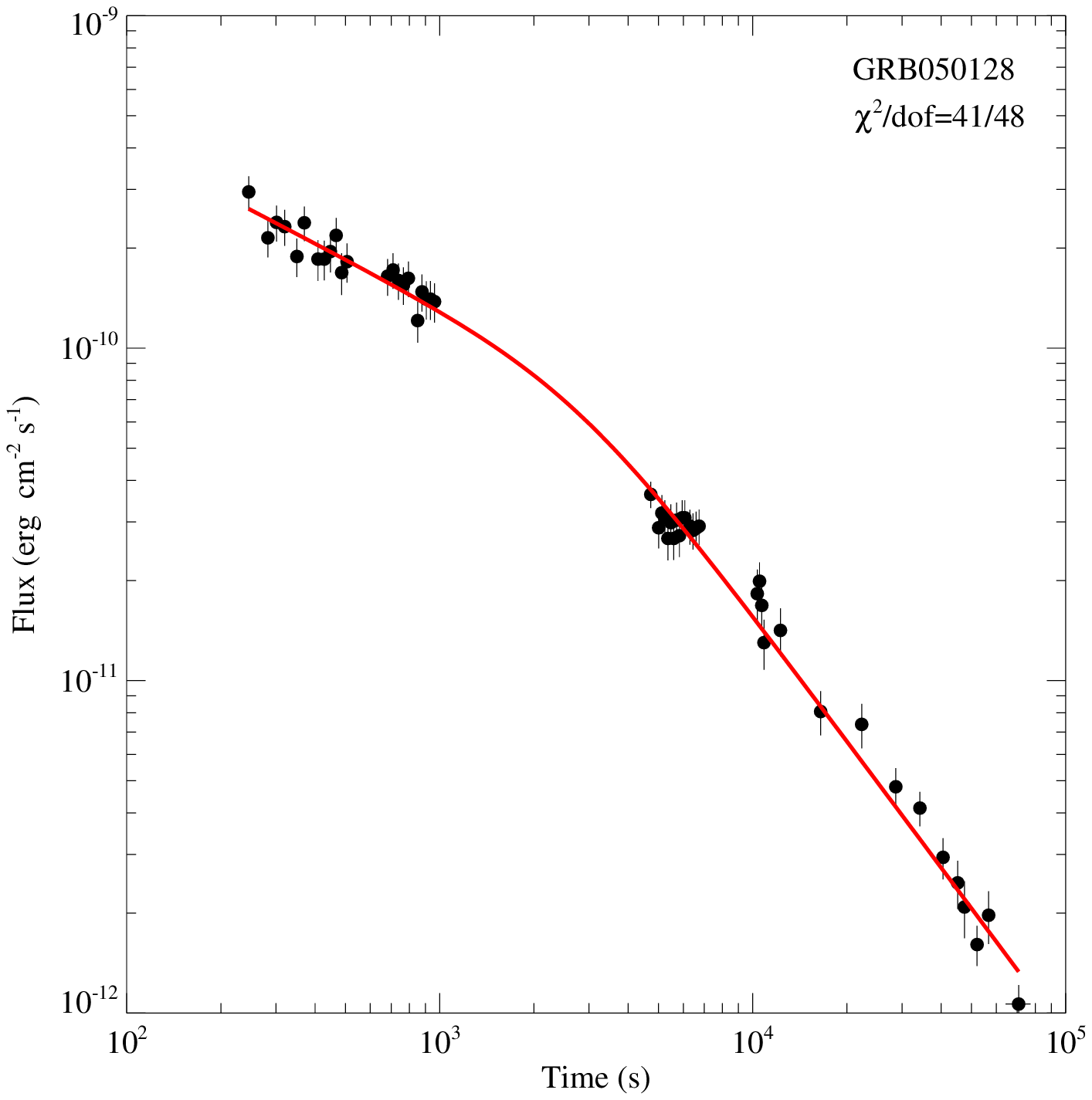}
\includegraphics[angle=0,scale=0.40]{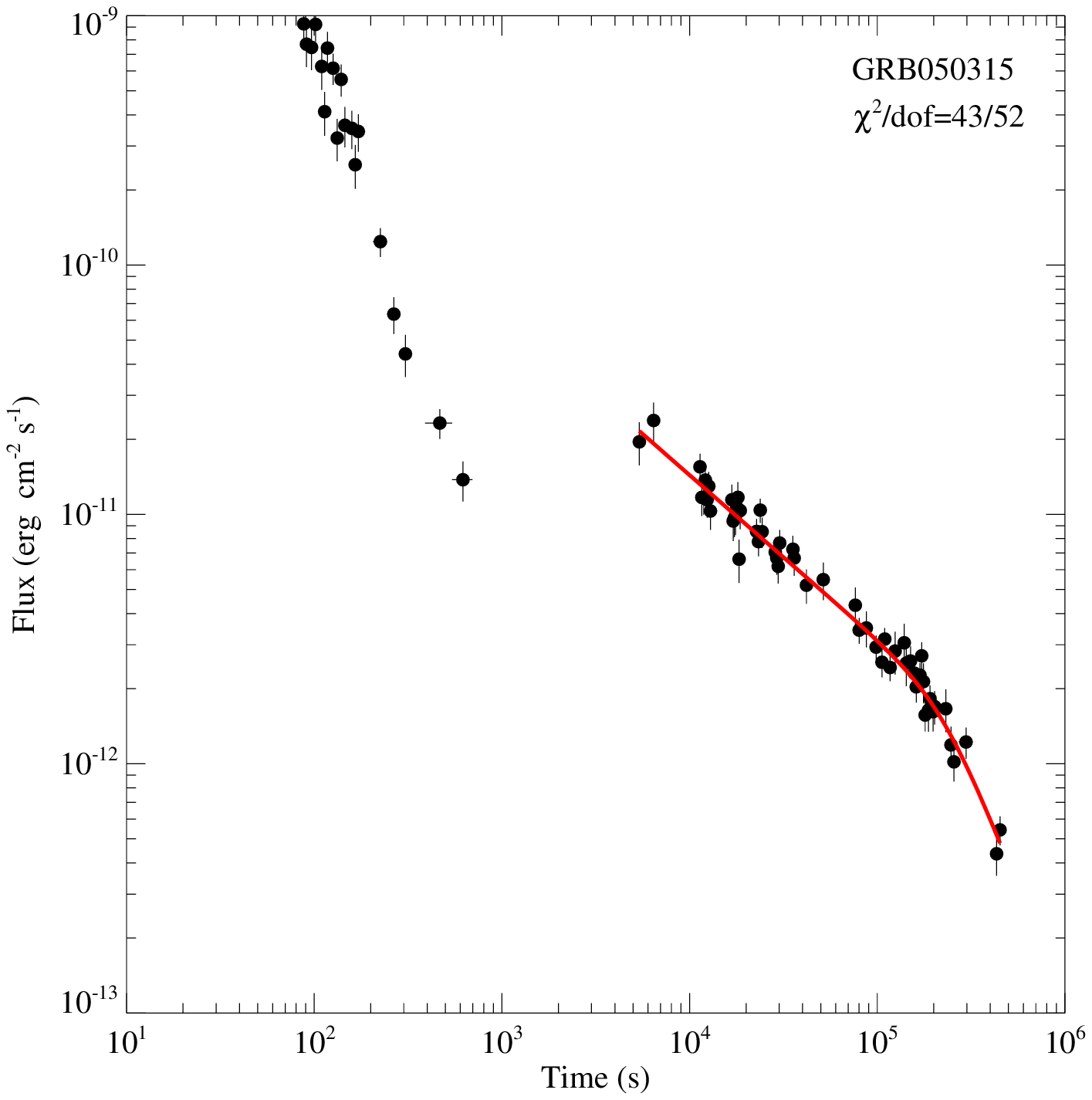}
\includegraphics[angle=0,scale=0.40]{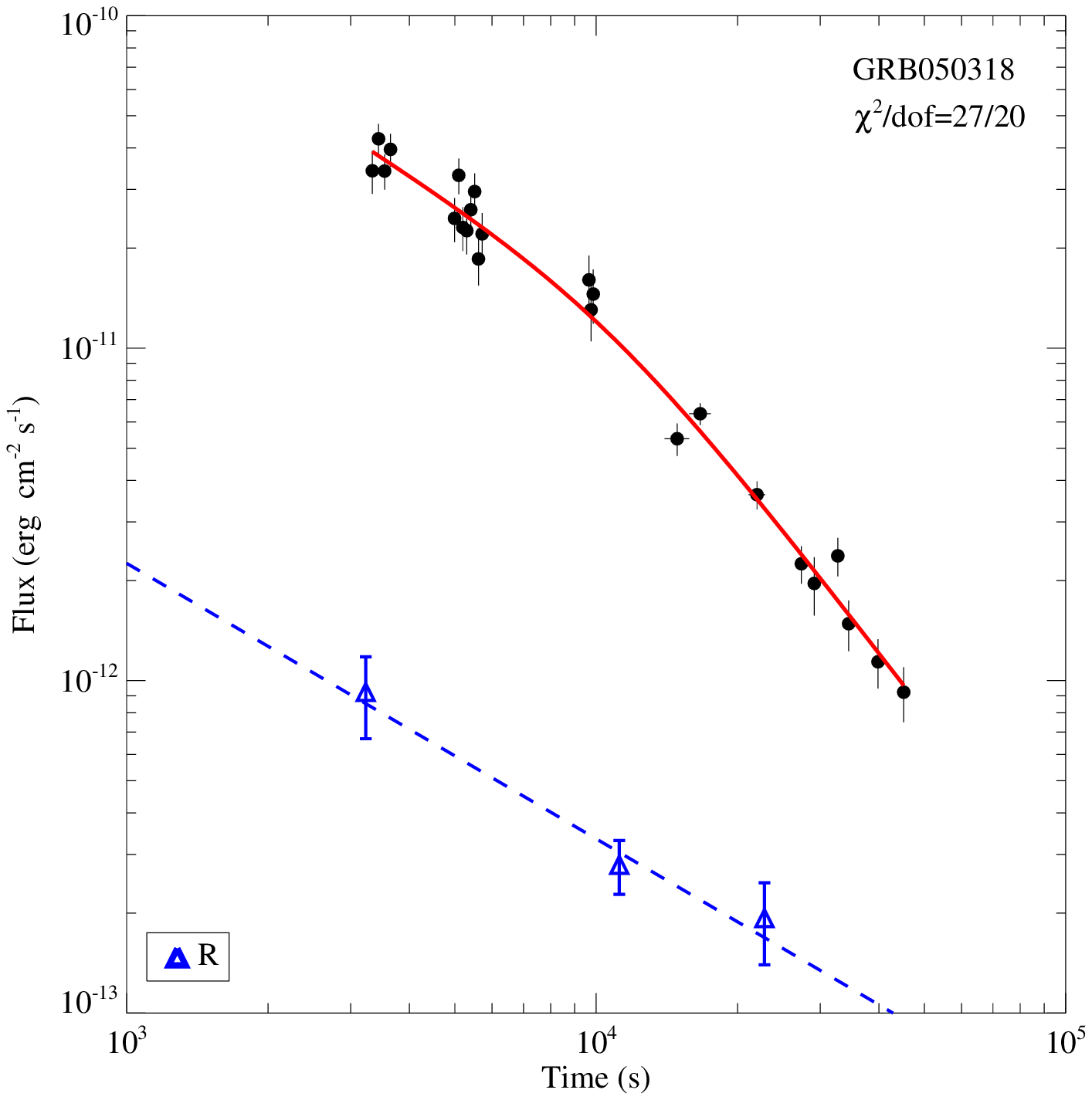}
\includegraphics[angle=0,scale=0.40]{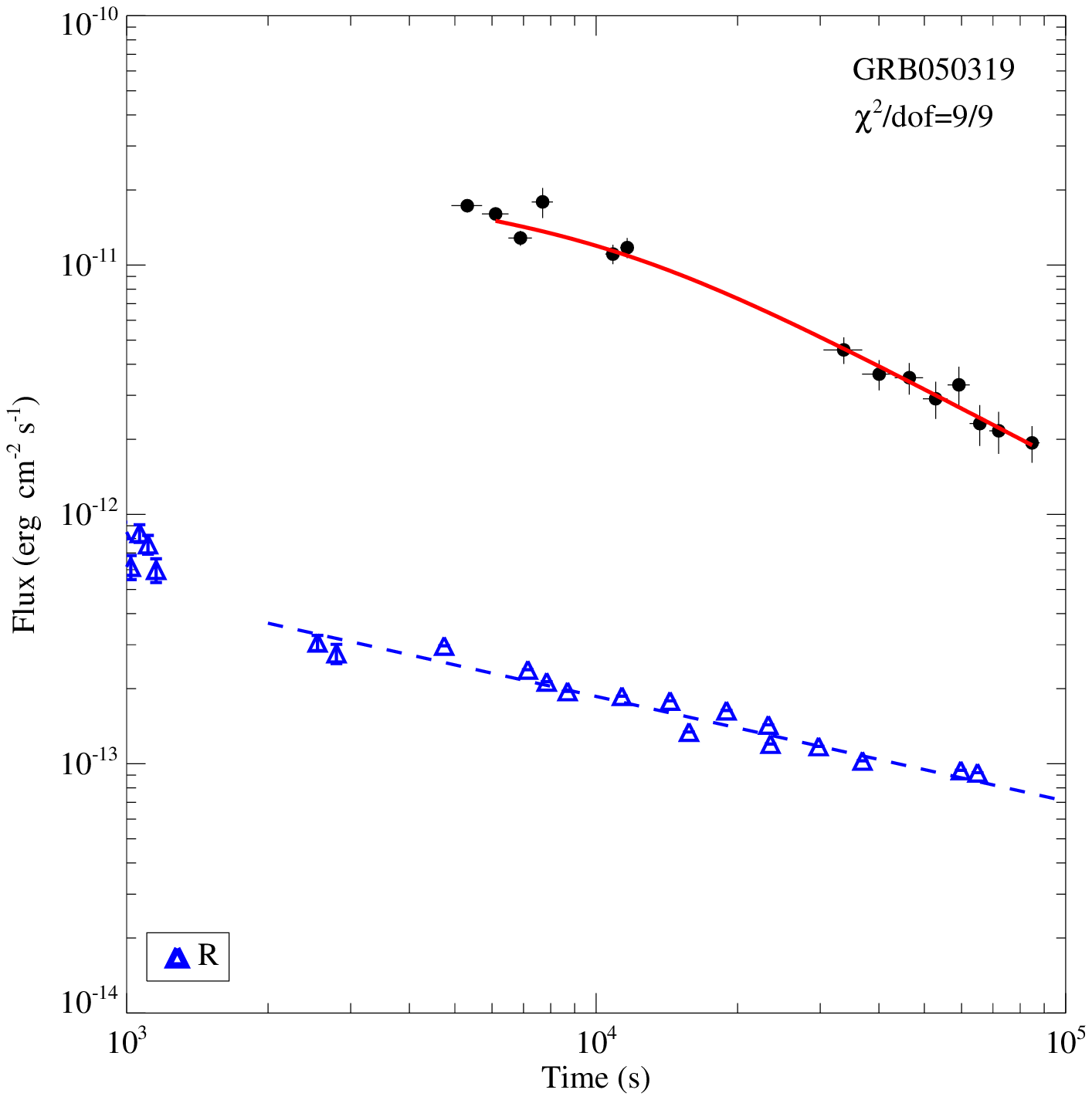}
\includegraphics[angle=0,scale=0.40]{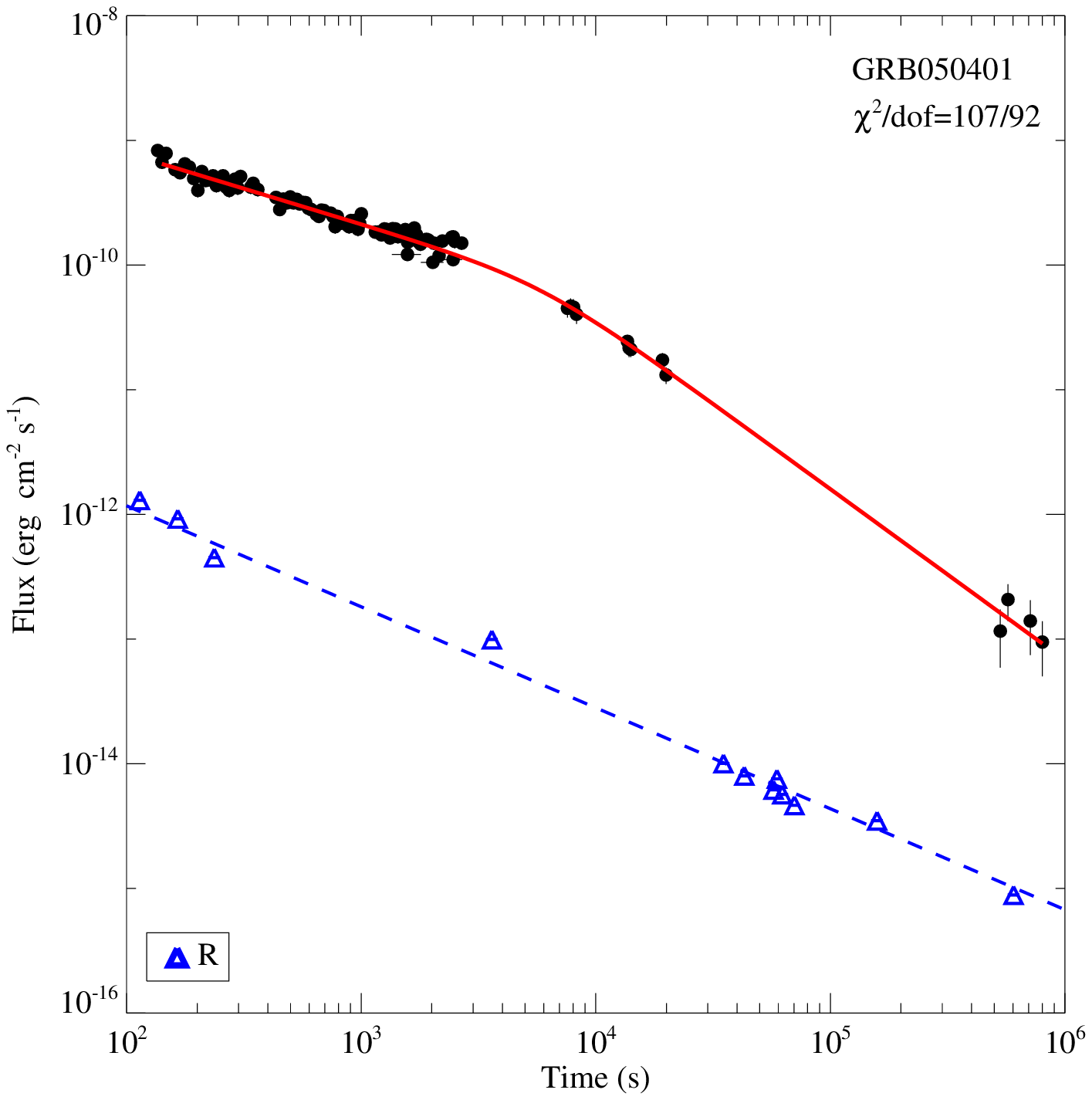}
\hfill
\includegraphics[angle=0,scale=0.40]{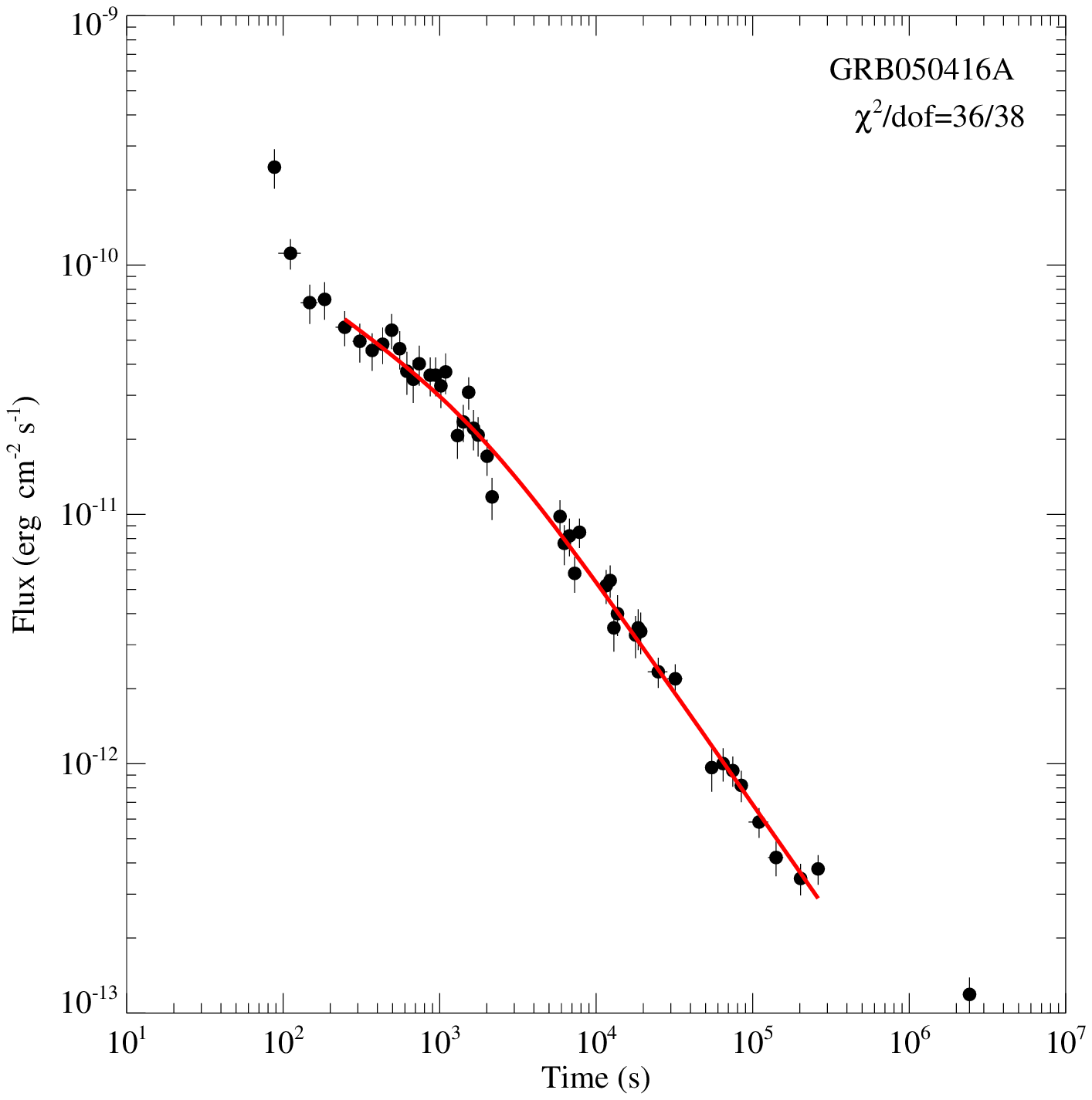}
\center{Fig.2  (continued)} \caption{The XRT light curves (dots) for the bursts
in our sample. The {\em solid} lines are the best fits with the smooth broken
power law for the shallow decay phase and its follow-up decay phase (usually the
``normal'' decay phase). The fitting $\chi^2$ and degrees of freedom are shown in
each plot. The optical light curves are shown by opened triangles, if they are
available. They are fitted by a smooth broken power law or a simple power law as
displayed by dashed lines.}\label{XRT_LC}
\end{figure*}
\begin{figure*}
\includegraphics[angle=0,scale=0.40]{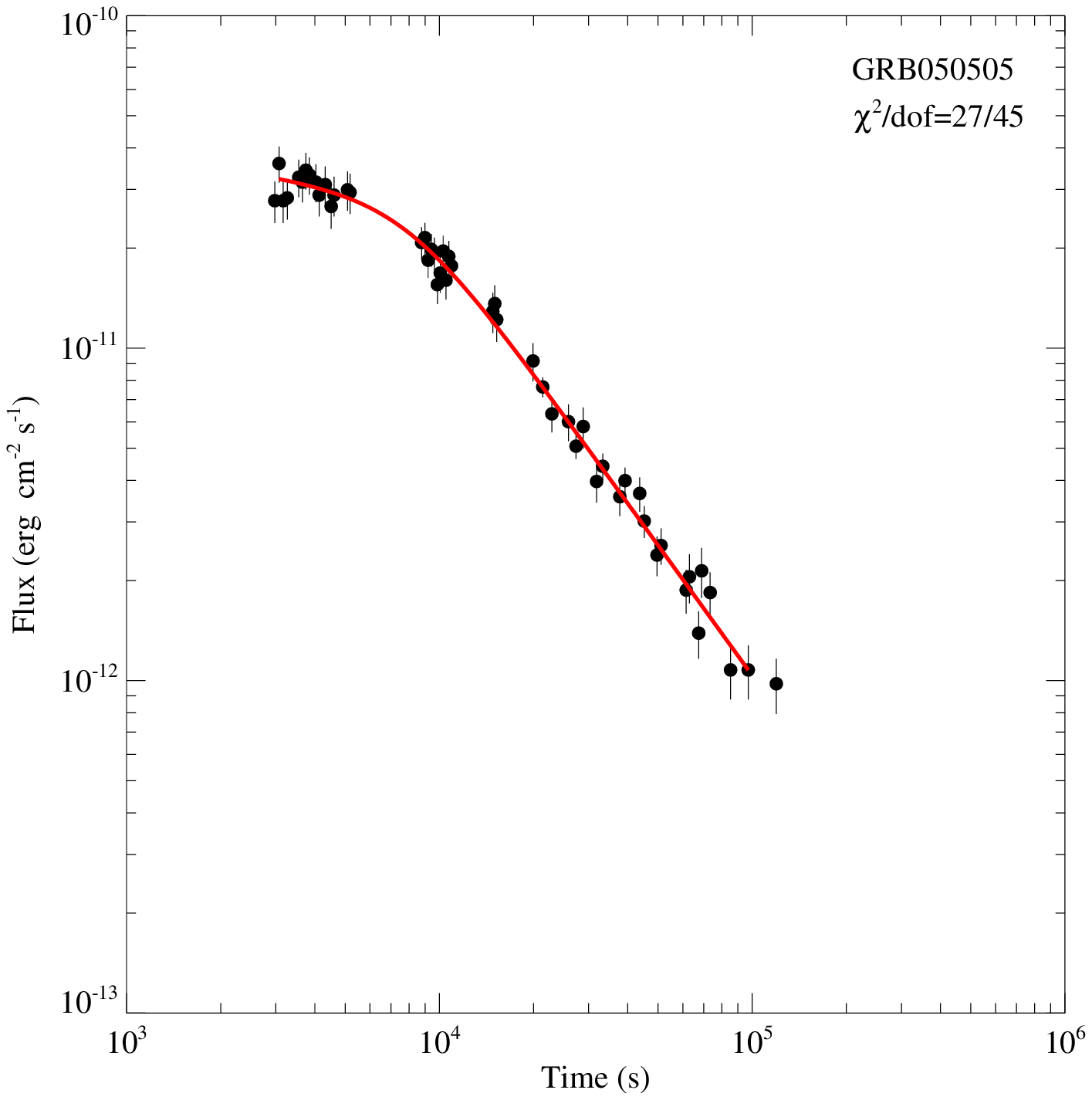}
\includegraphics[angle=0,scale=0.40]{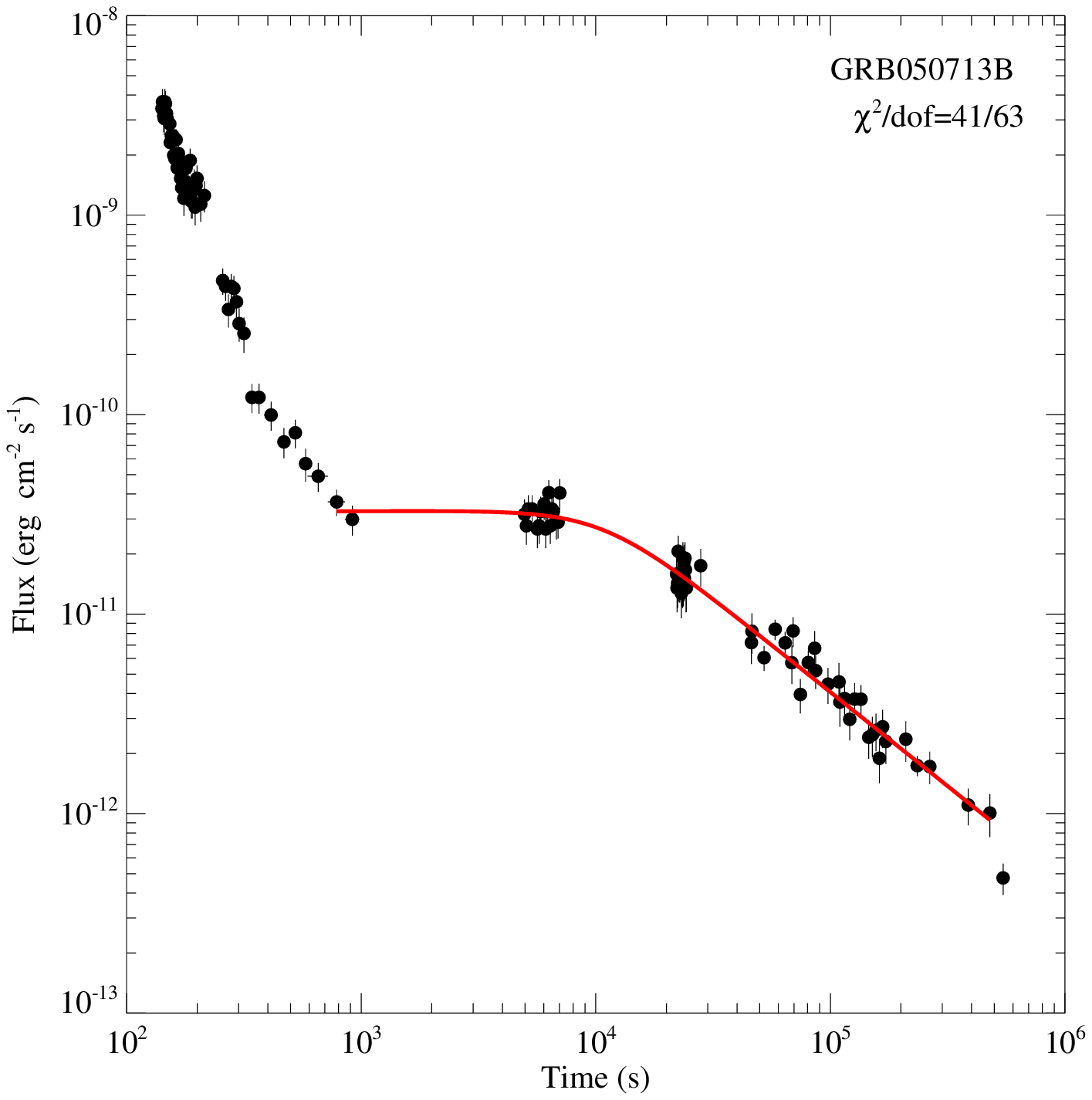}
\includegraphics[angle=0,scale=0.40]{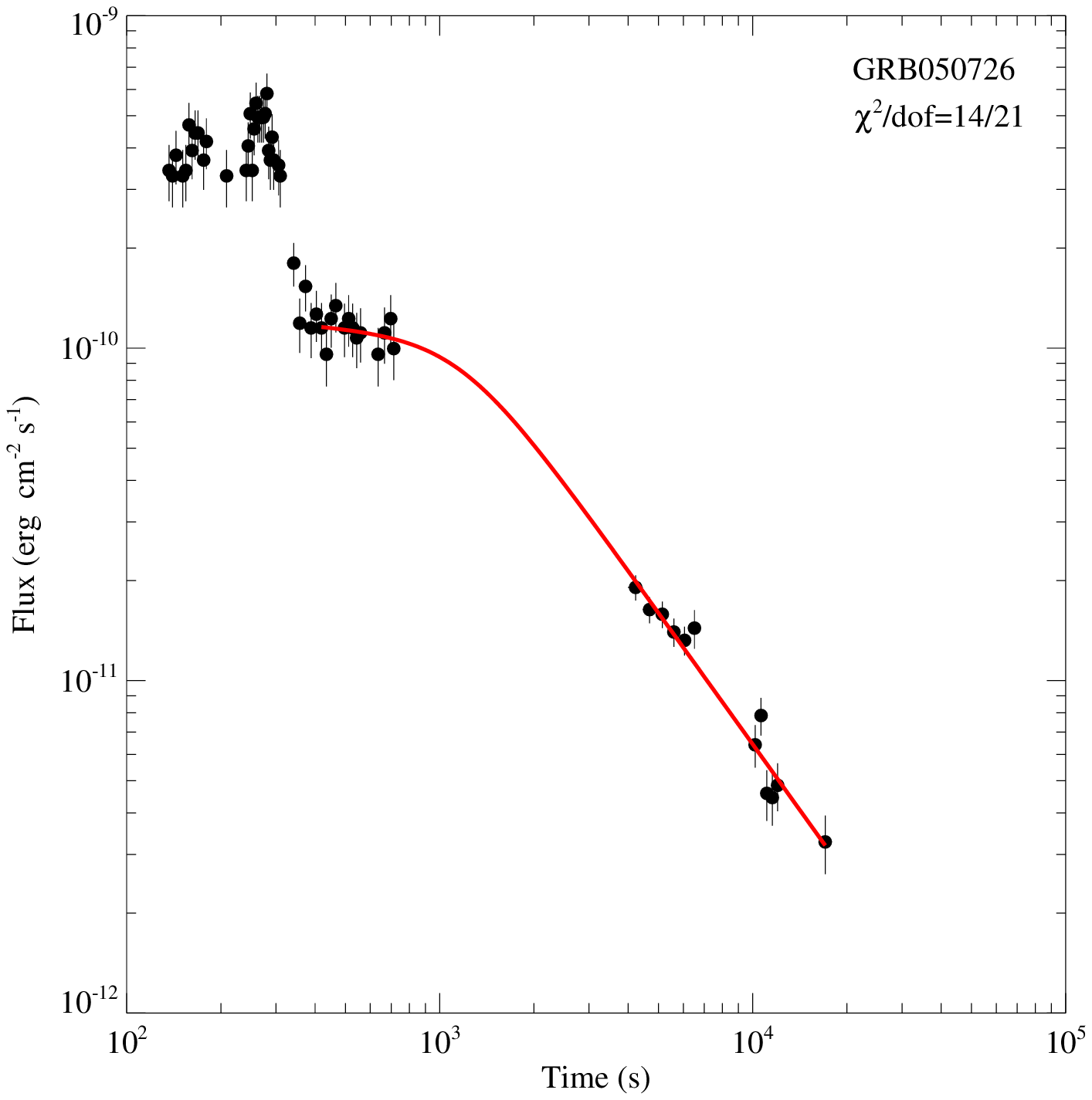}
\includegraphics[angle=0,scale=0.40]{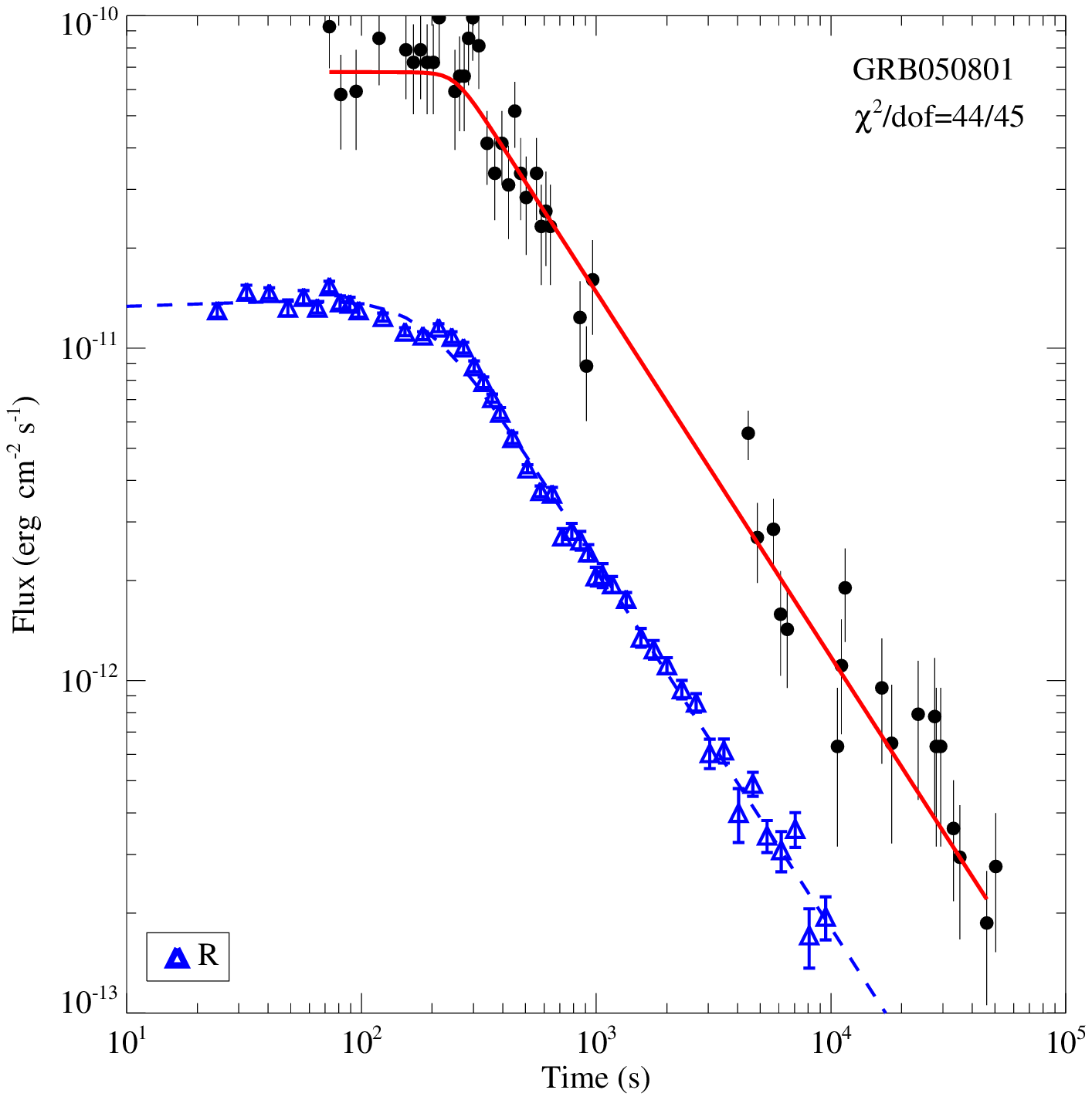}
\includegraphics[angle=0,scale=0.40]{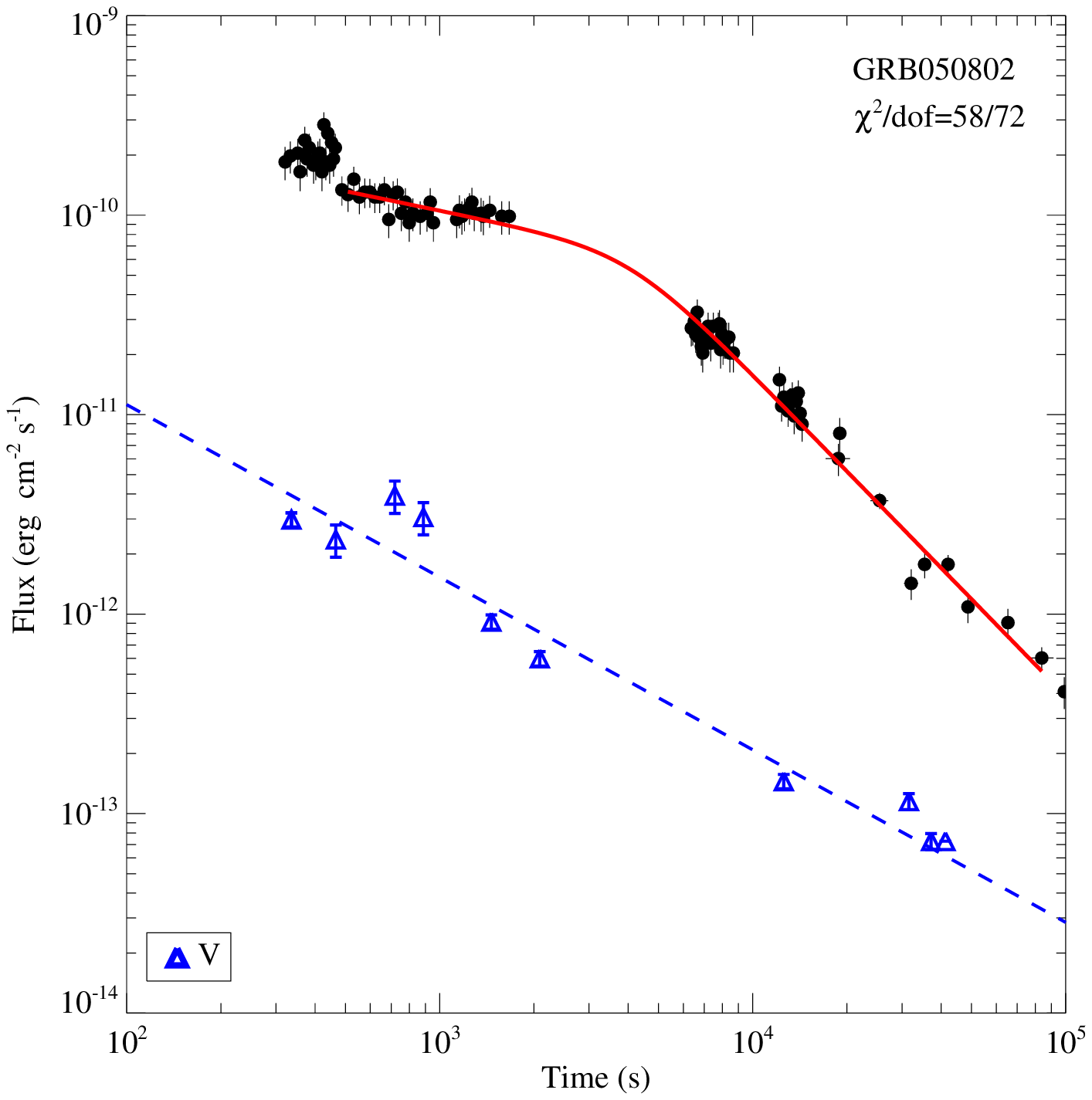}
\hfill
\includegraphics[angle=0,scale=0.40]{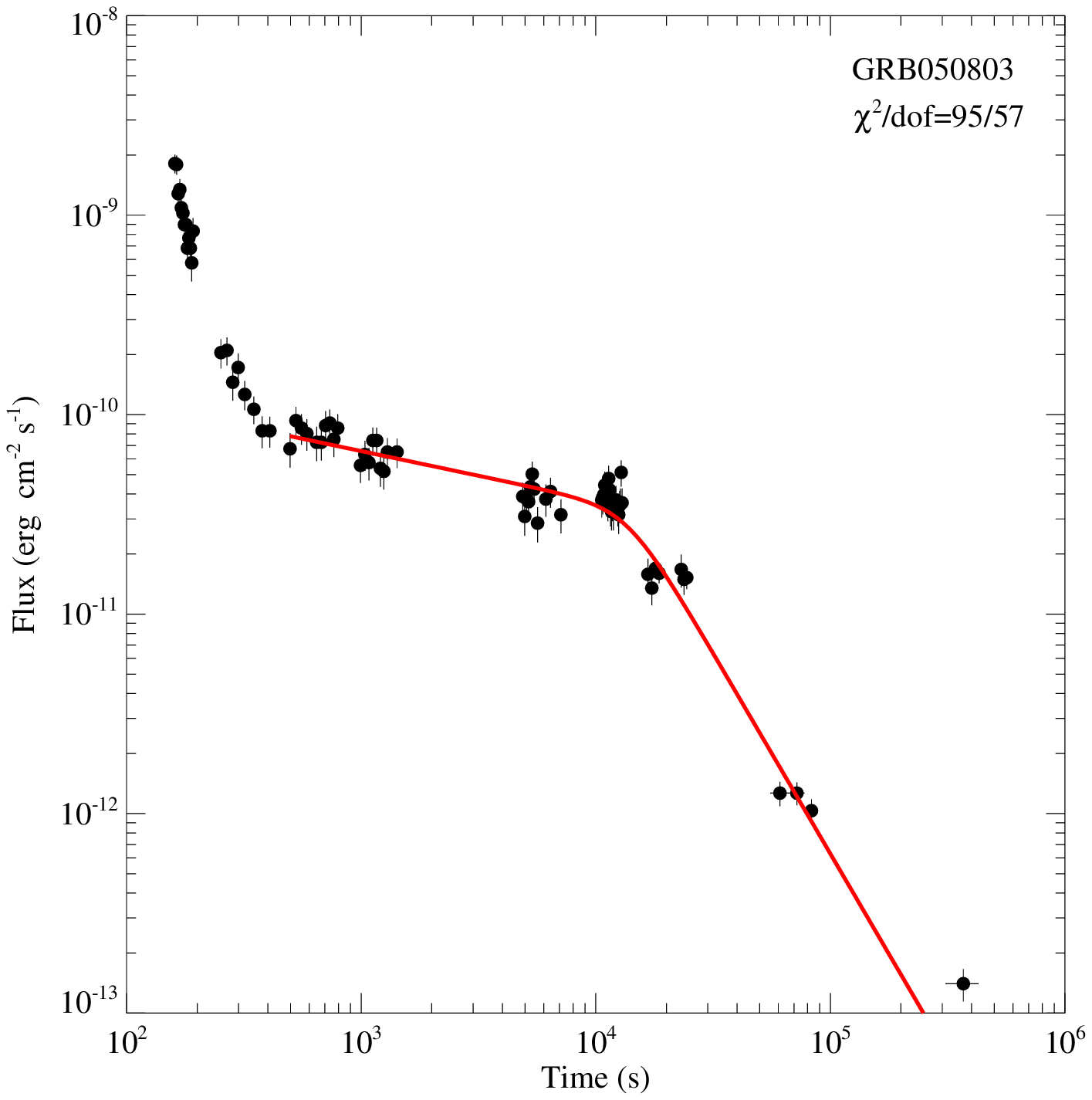}
\center{Fig.2  (continued)}
\end{figure*}
\begin{figure*}
\includegraphics[angle=0,scale=0.40]{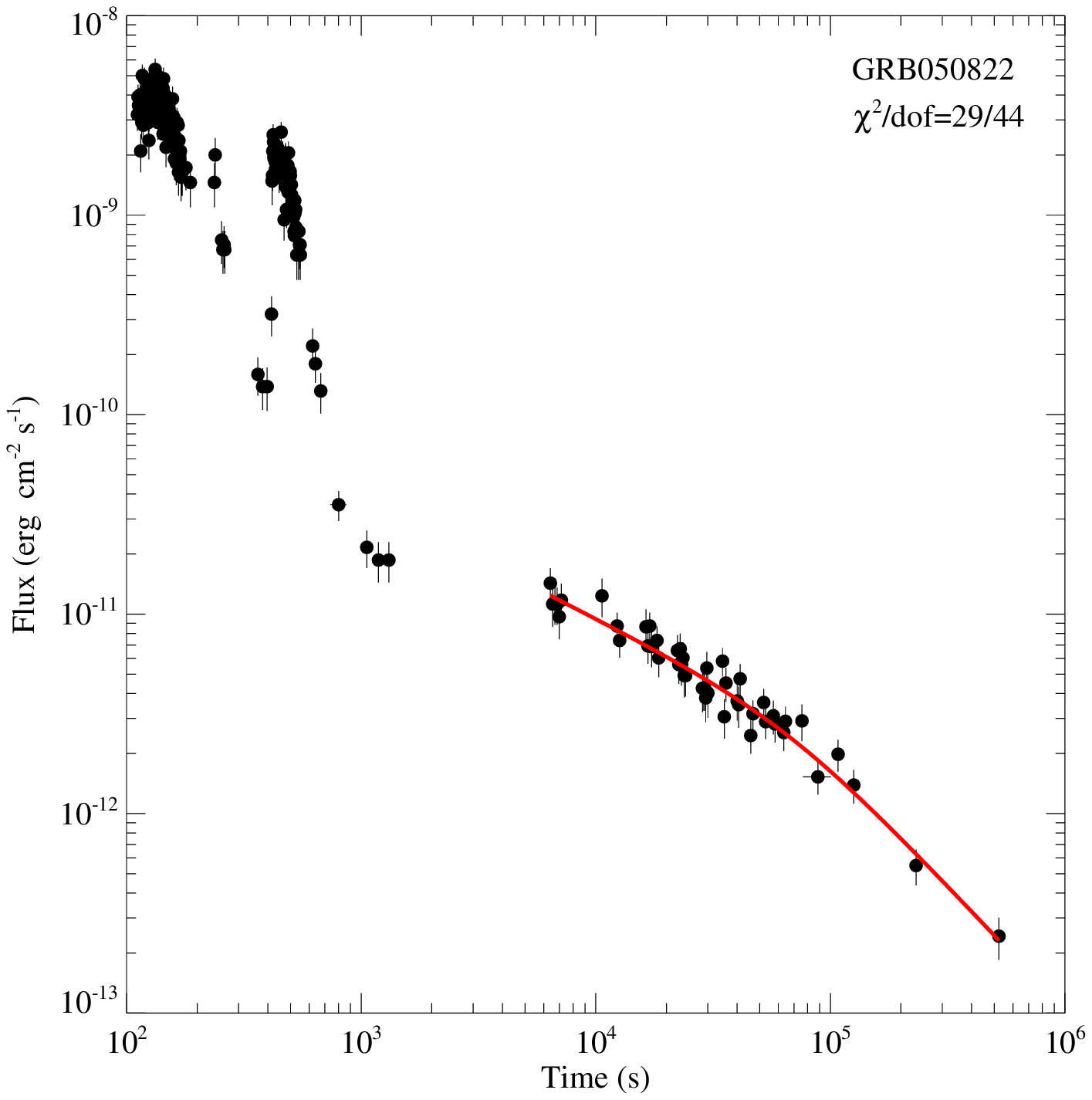}
\includegraphics[angle=0,scale=0.40]{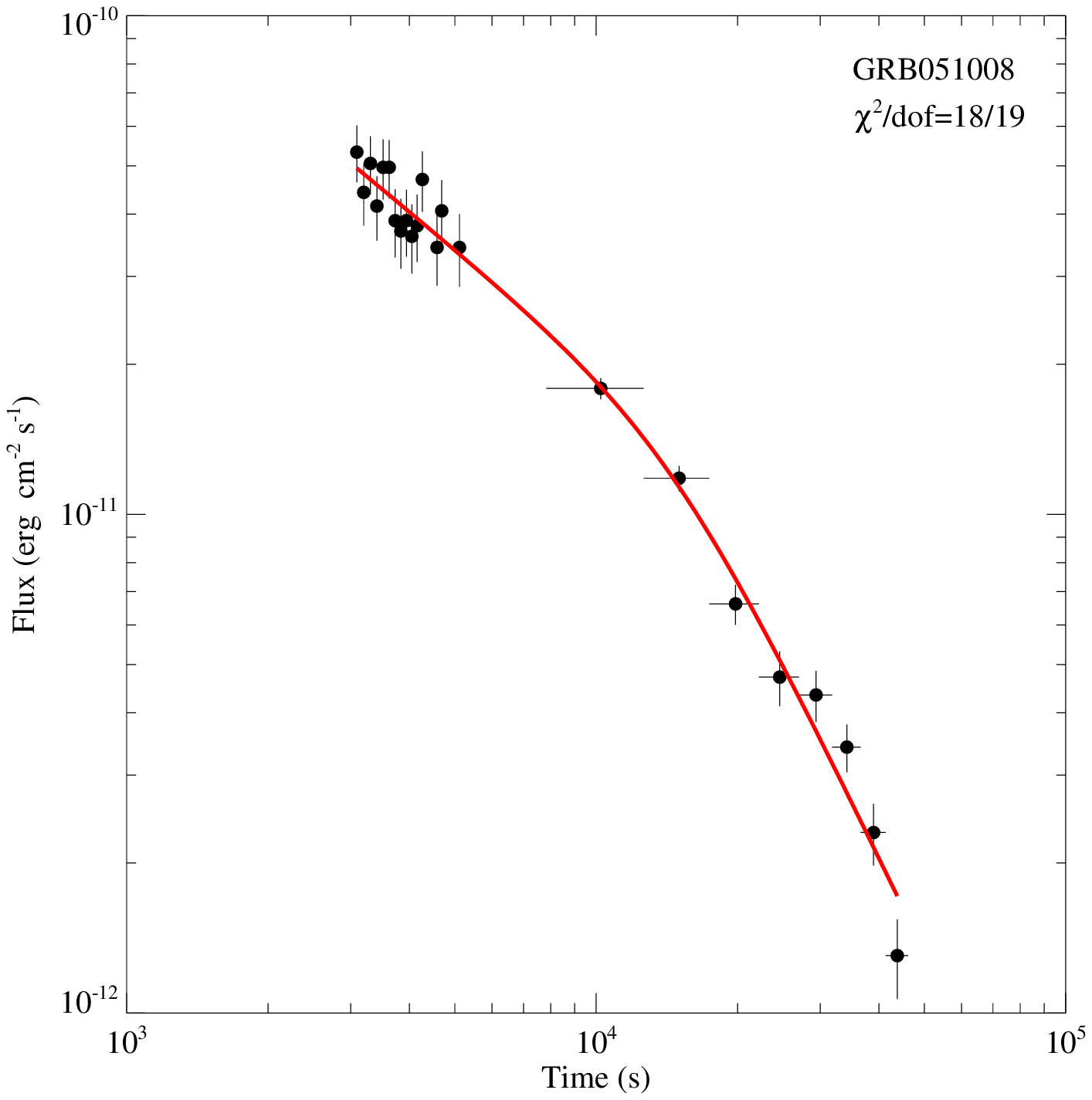}
\includegraphics[angle=0,scale=0.40]{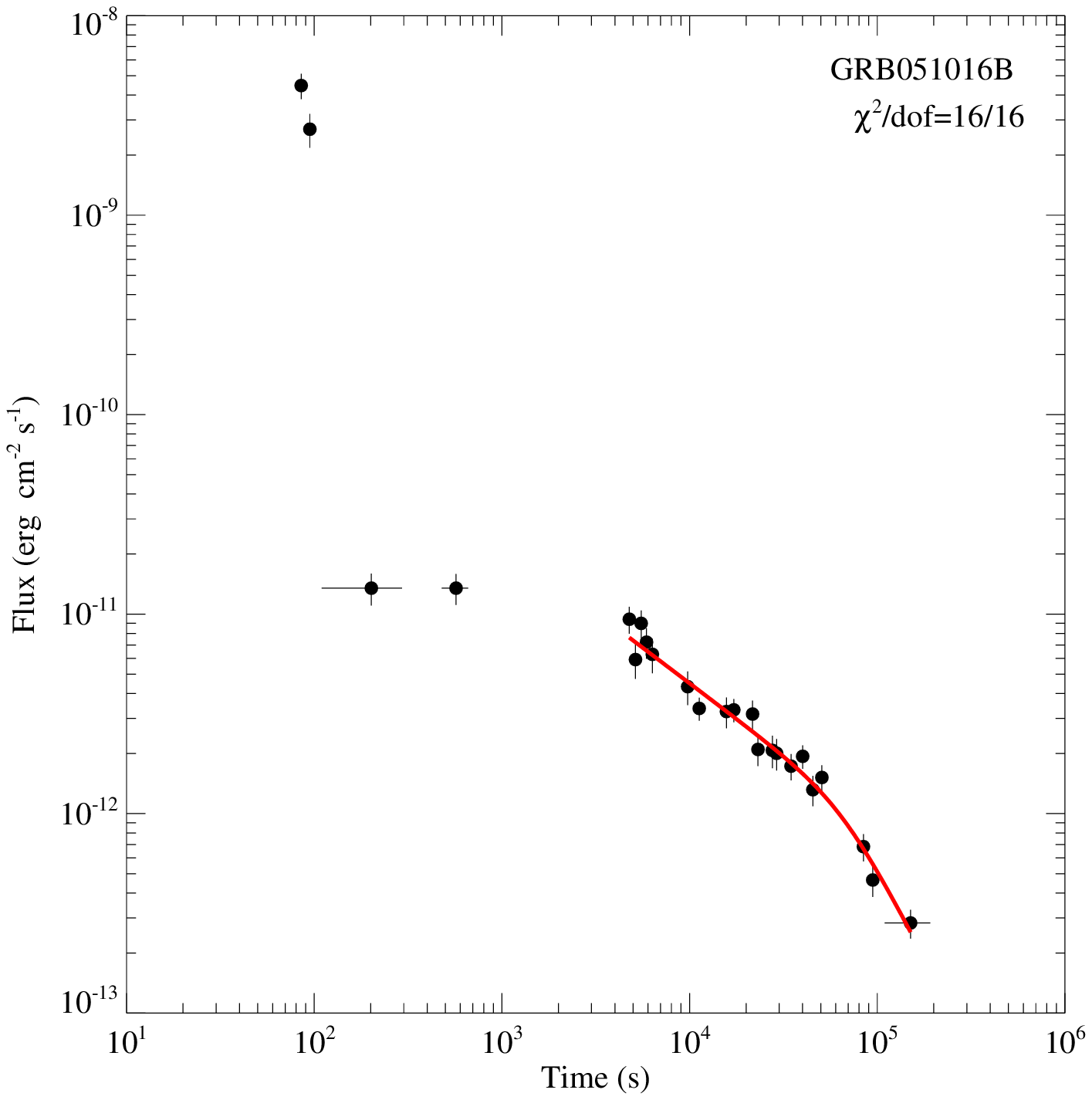}
\includegraphics[angle=0,scale=0.40]{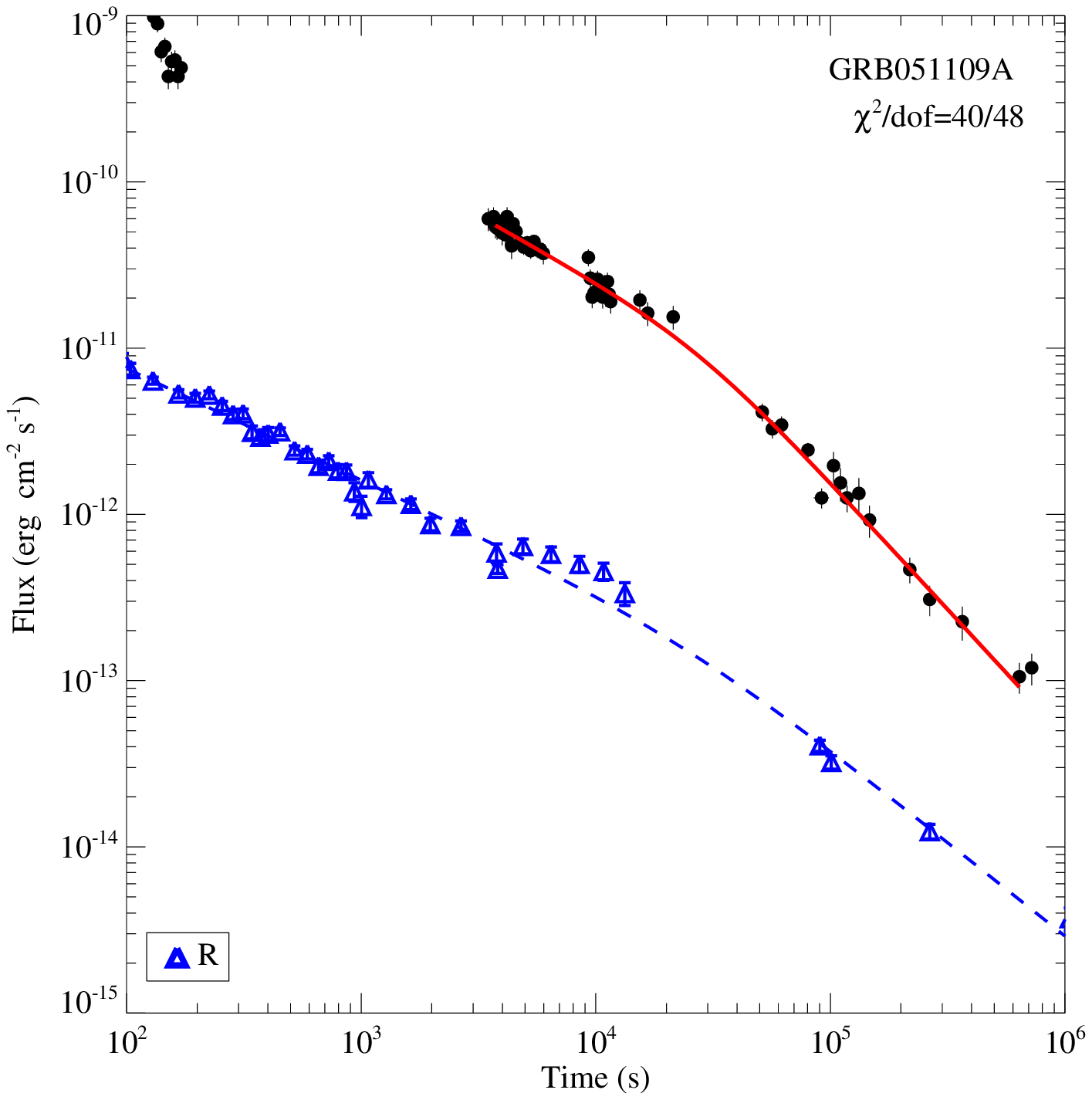}
\includegraphics[angle=0,scale=0.40]{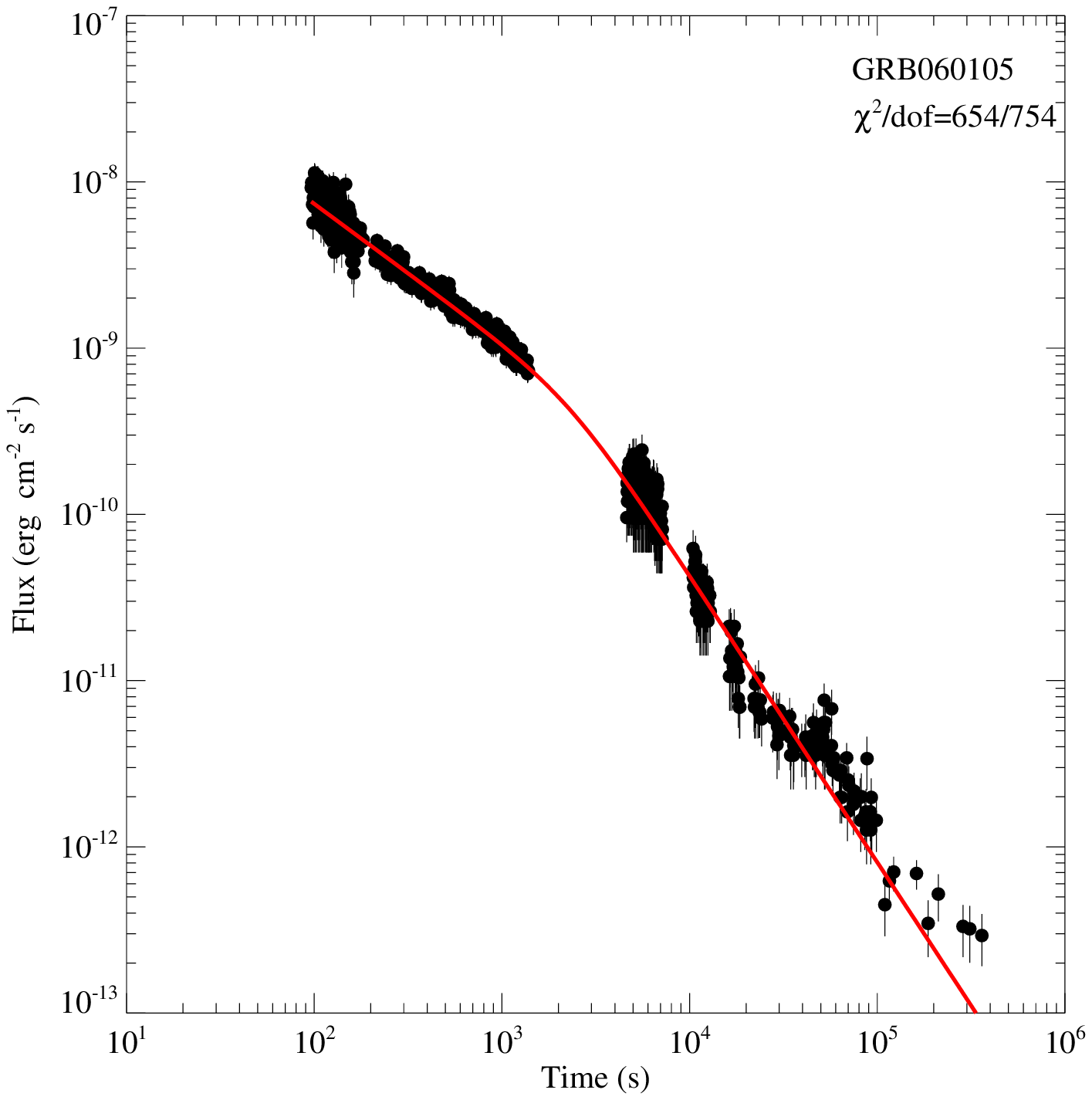}
\hfill
\includegraphics[angle=0,scale=0.40]{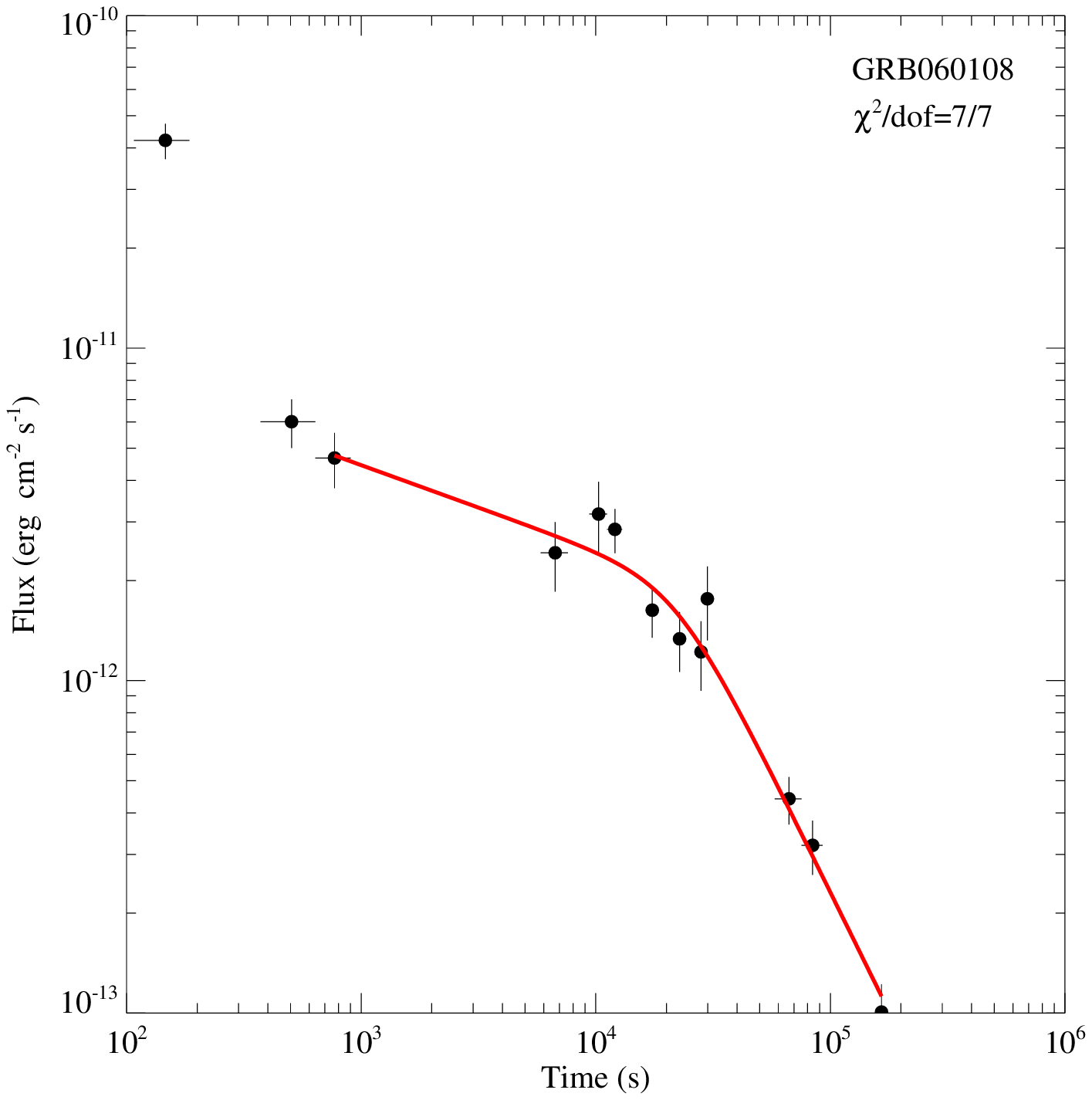}
\center{Fig.2  (continued)}
\end{figure*}
\begin{figure*}
\includegraphics[angle=0,scale=0.40]{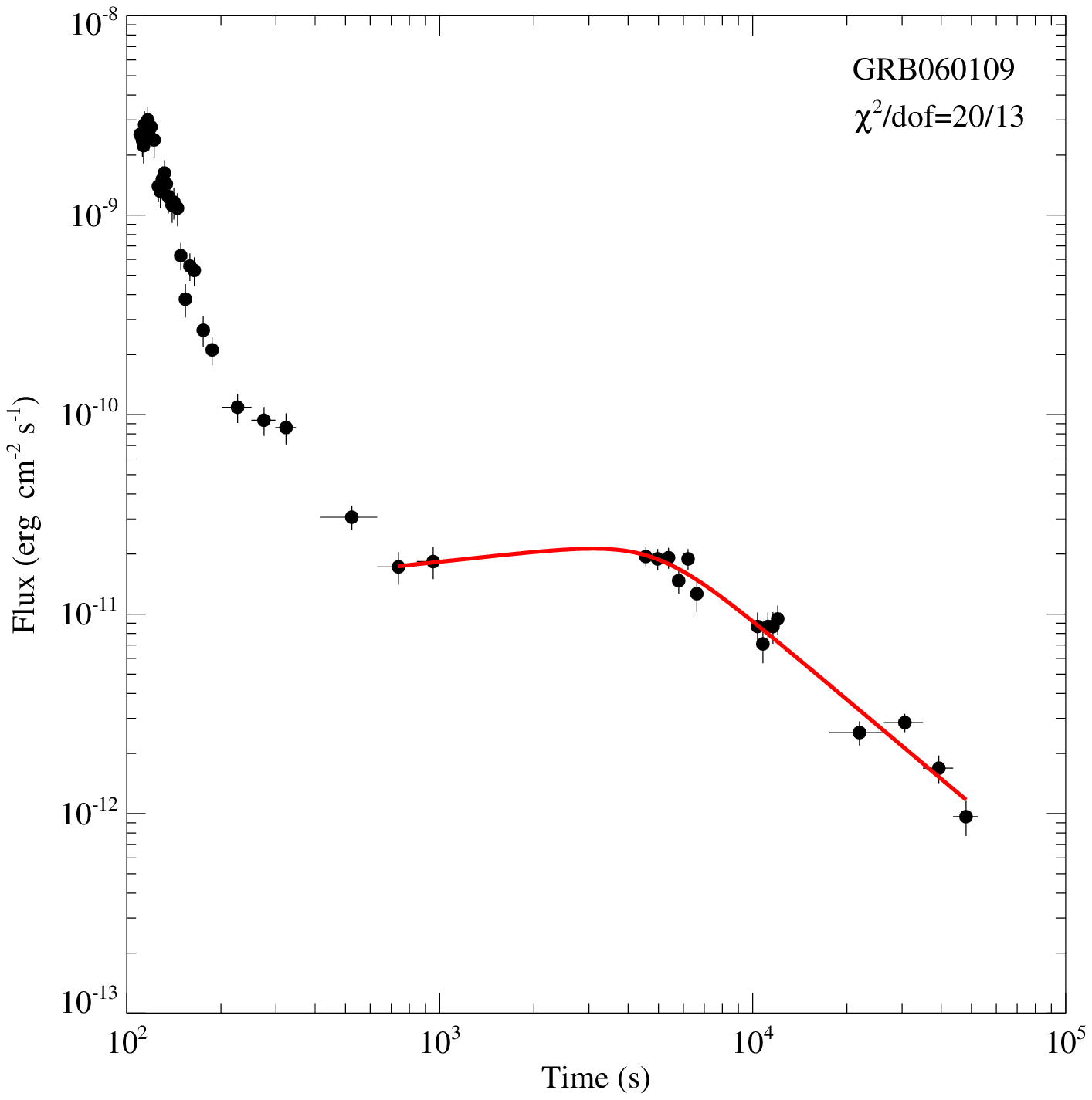}
\includegraphics[angle=0,scale=0.40]{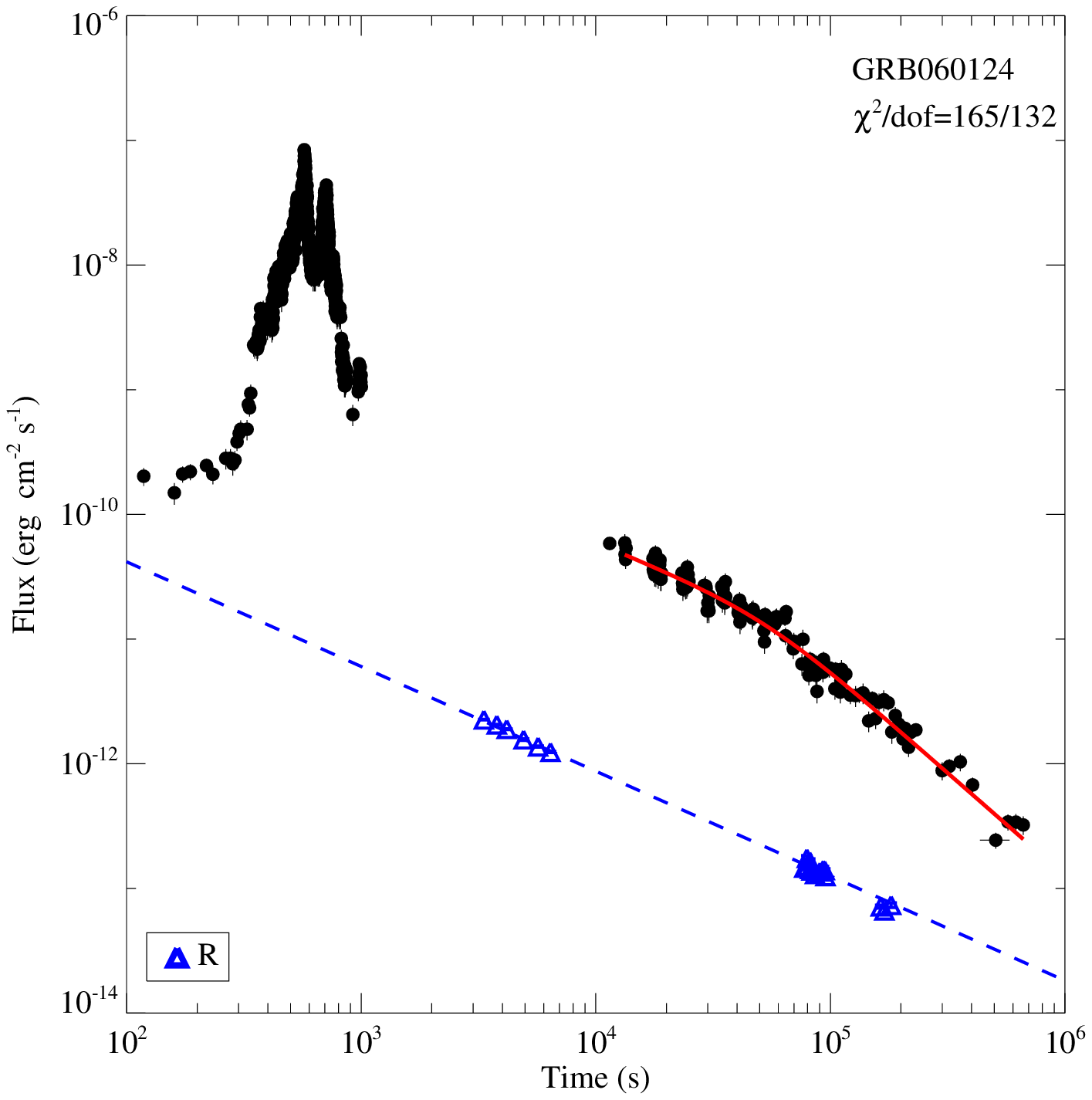}
\includegraphics[angle=0,scale=0.40]{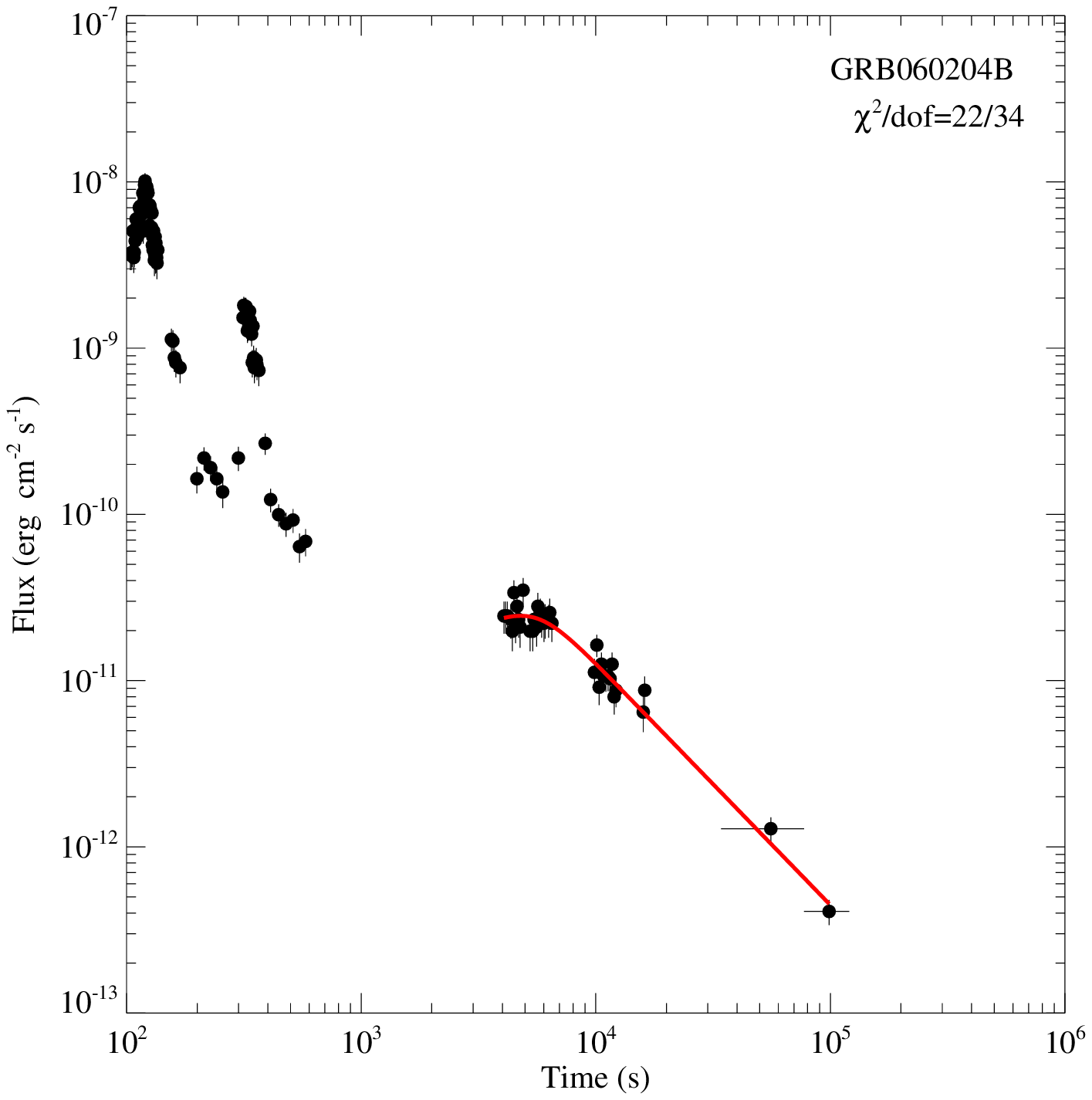}
\includegraphics[angle=0,scale=0.40]{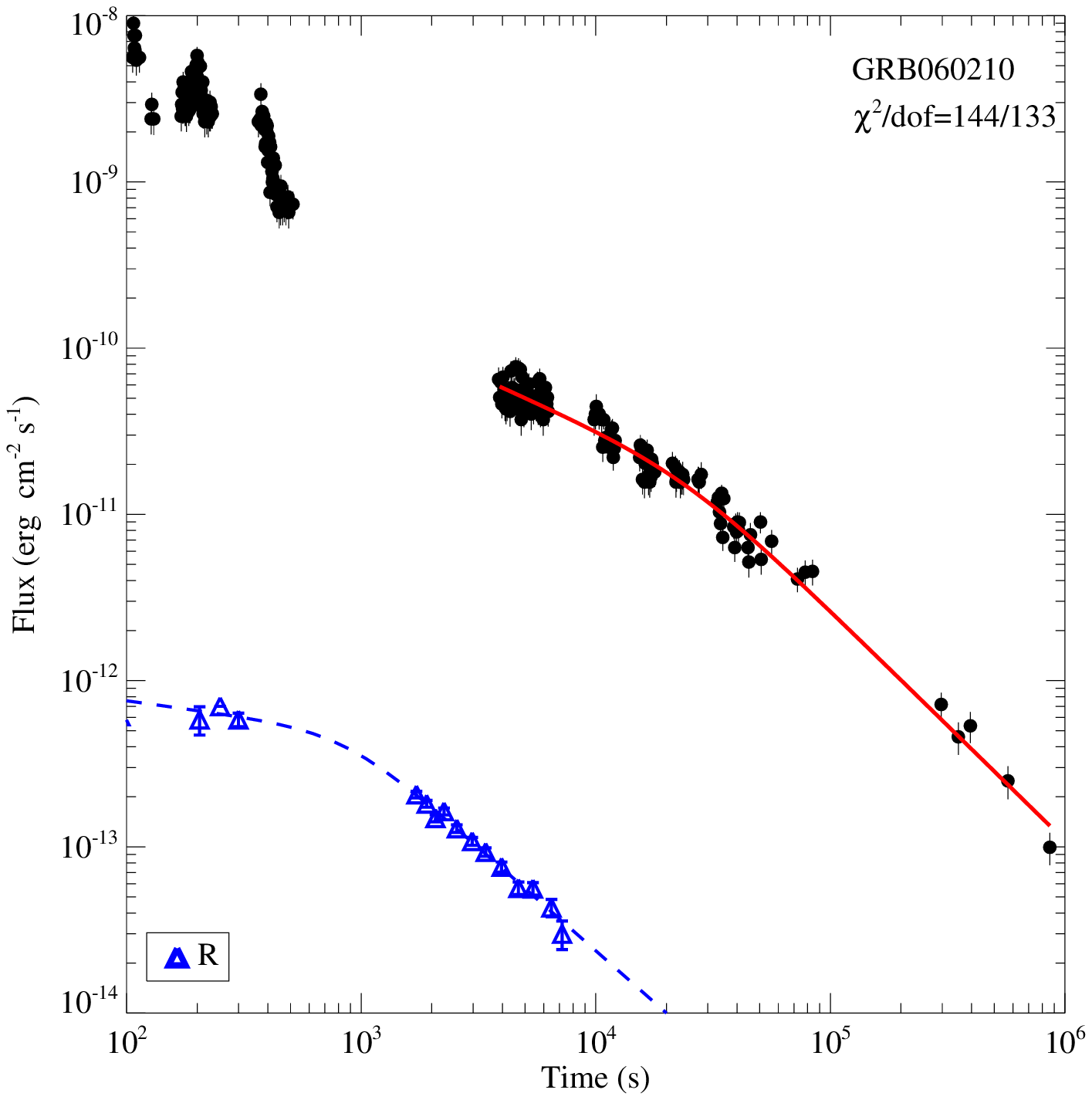}
\includegraphics[angle=0,scale=0.40]{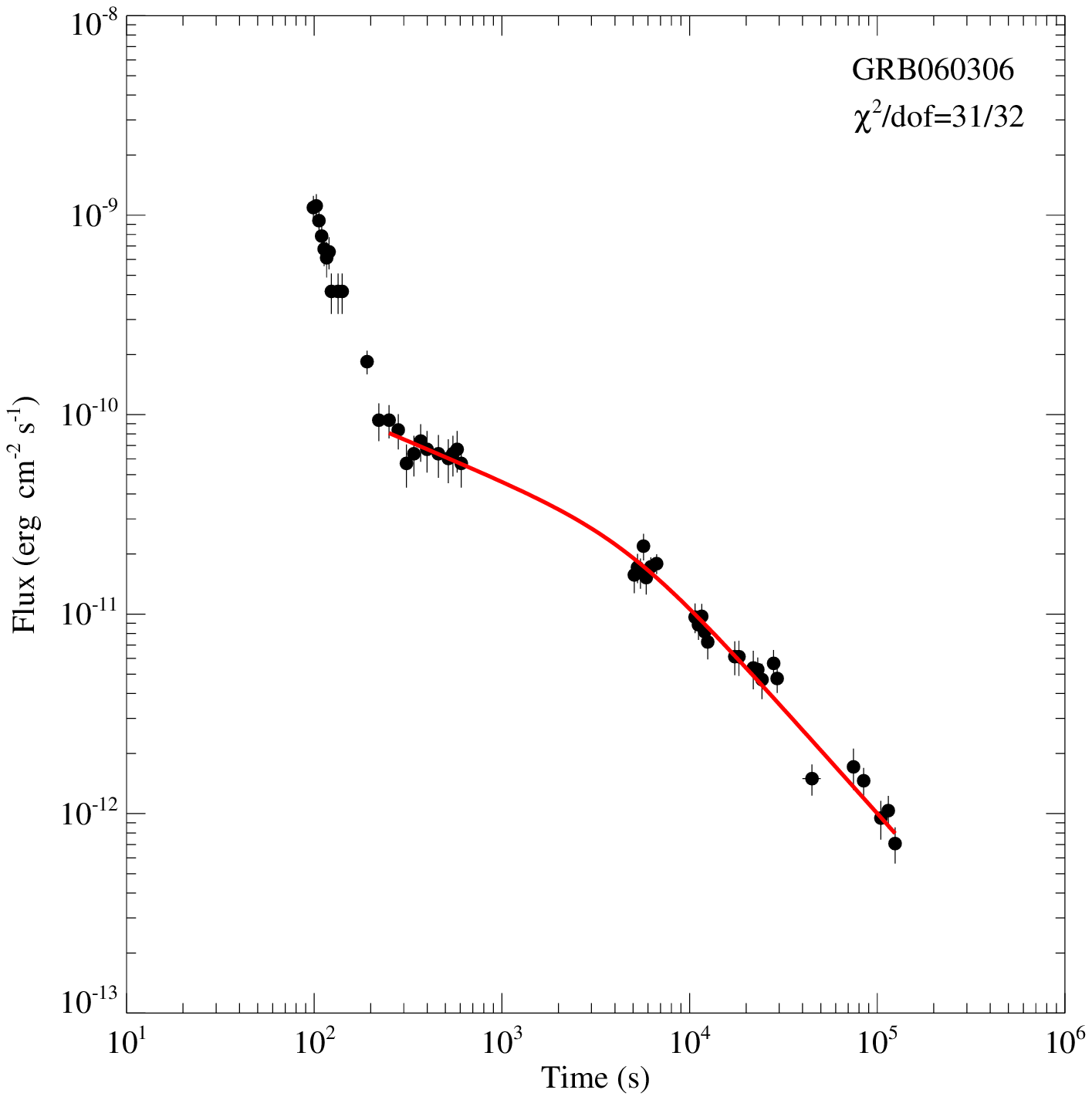}
\hfill
\includegraphics[angle=0,scale=0.40]{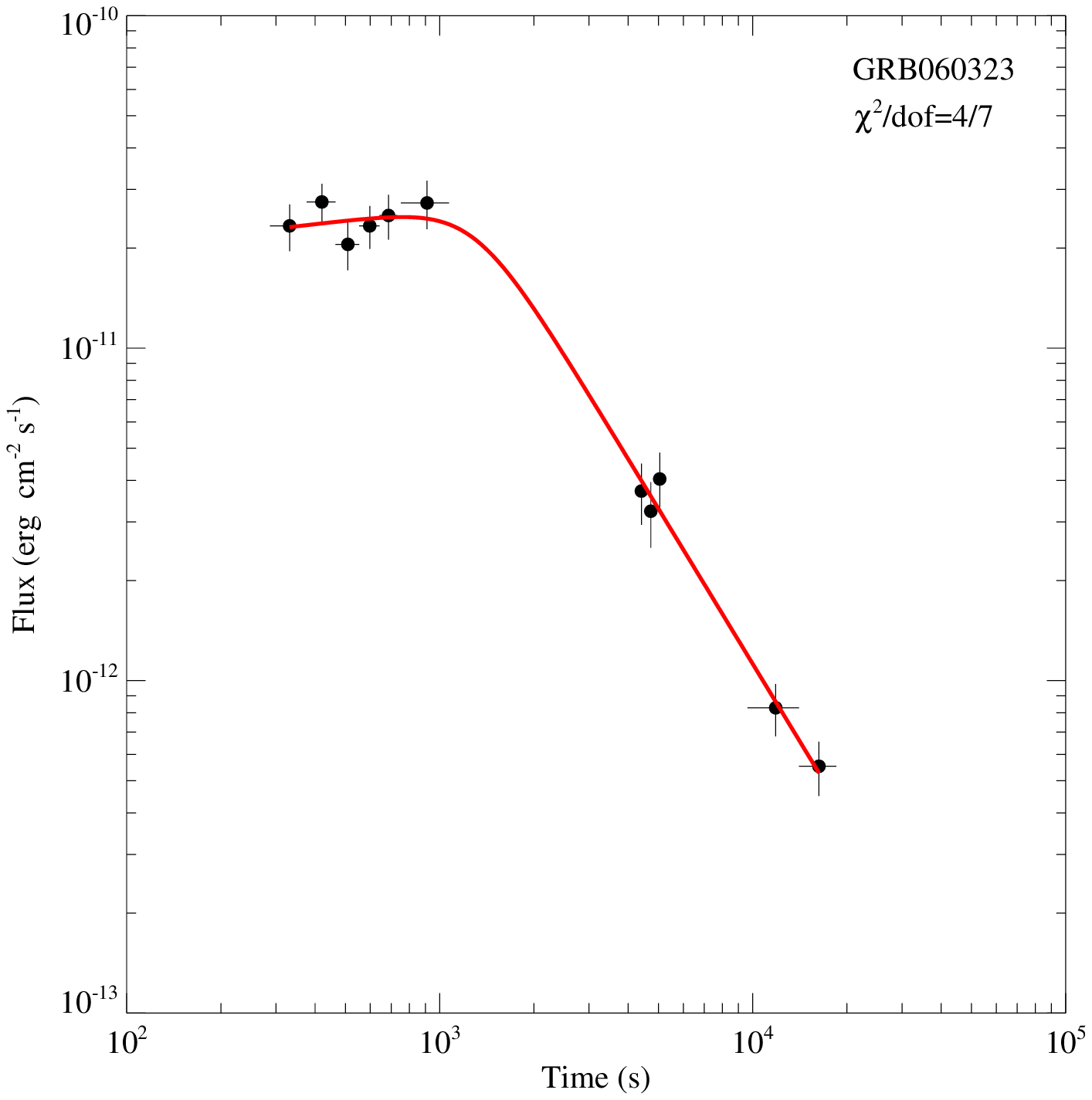}
\center{Fig.2  (continued)}
\end{figure*}
\begin{figure*}
\includegraphics[angle=0,scale=0.40]{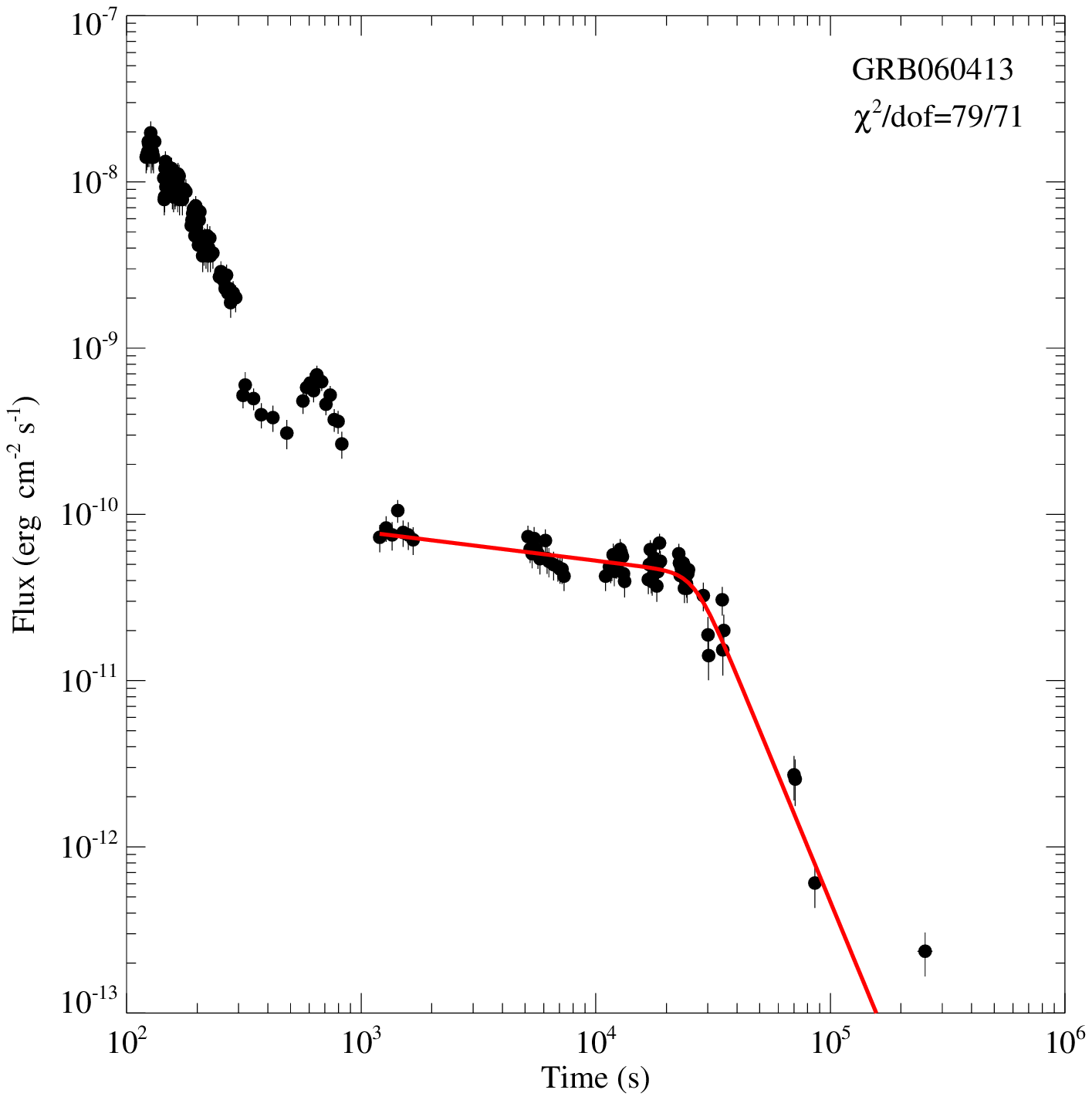}
\includegraphics[angle=0,scale=0.40]{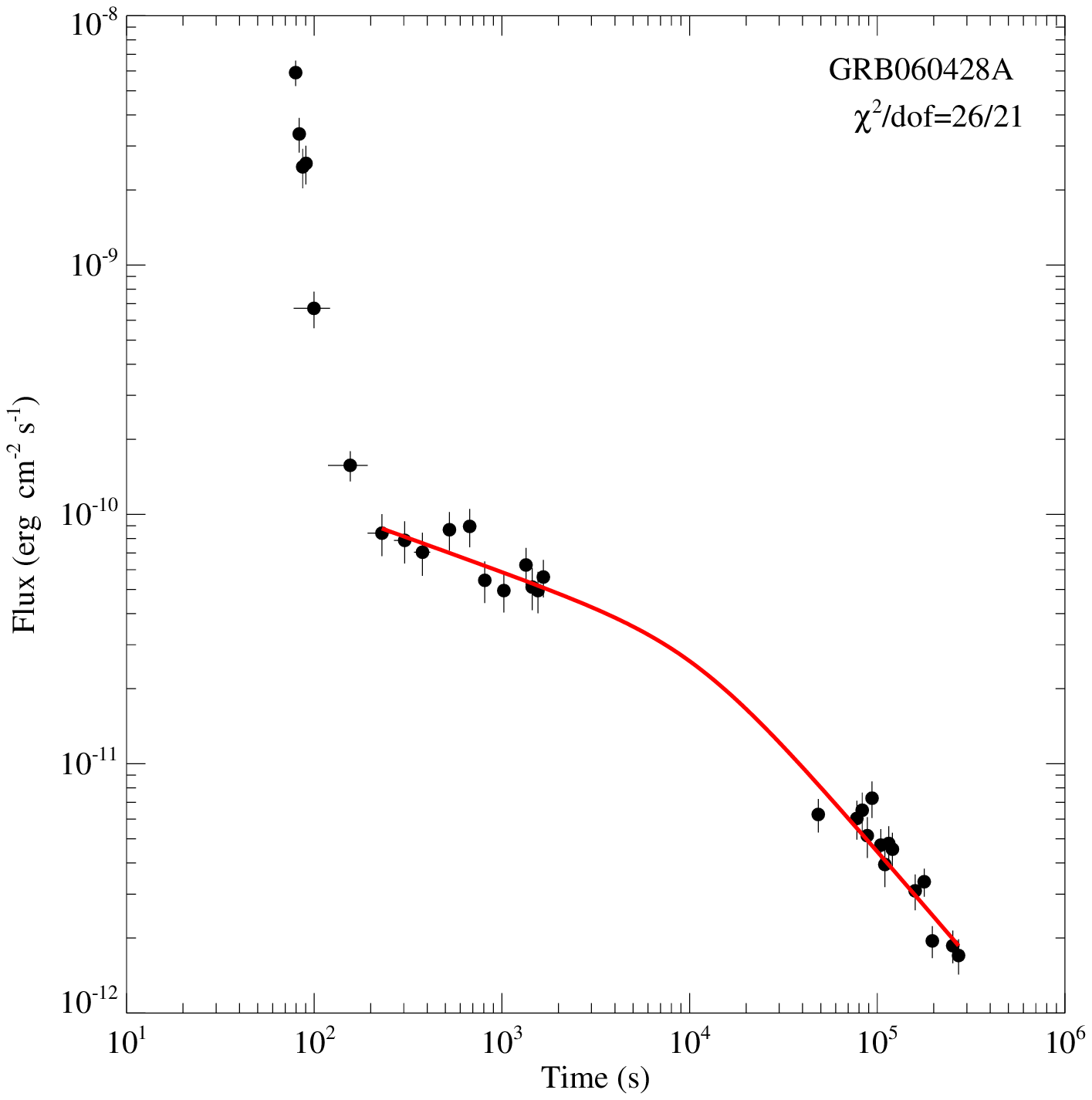}
\includegraphics[angle=0,scale=0.40]{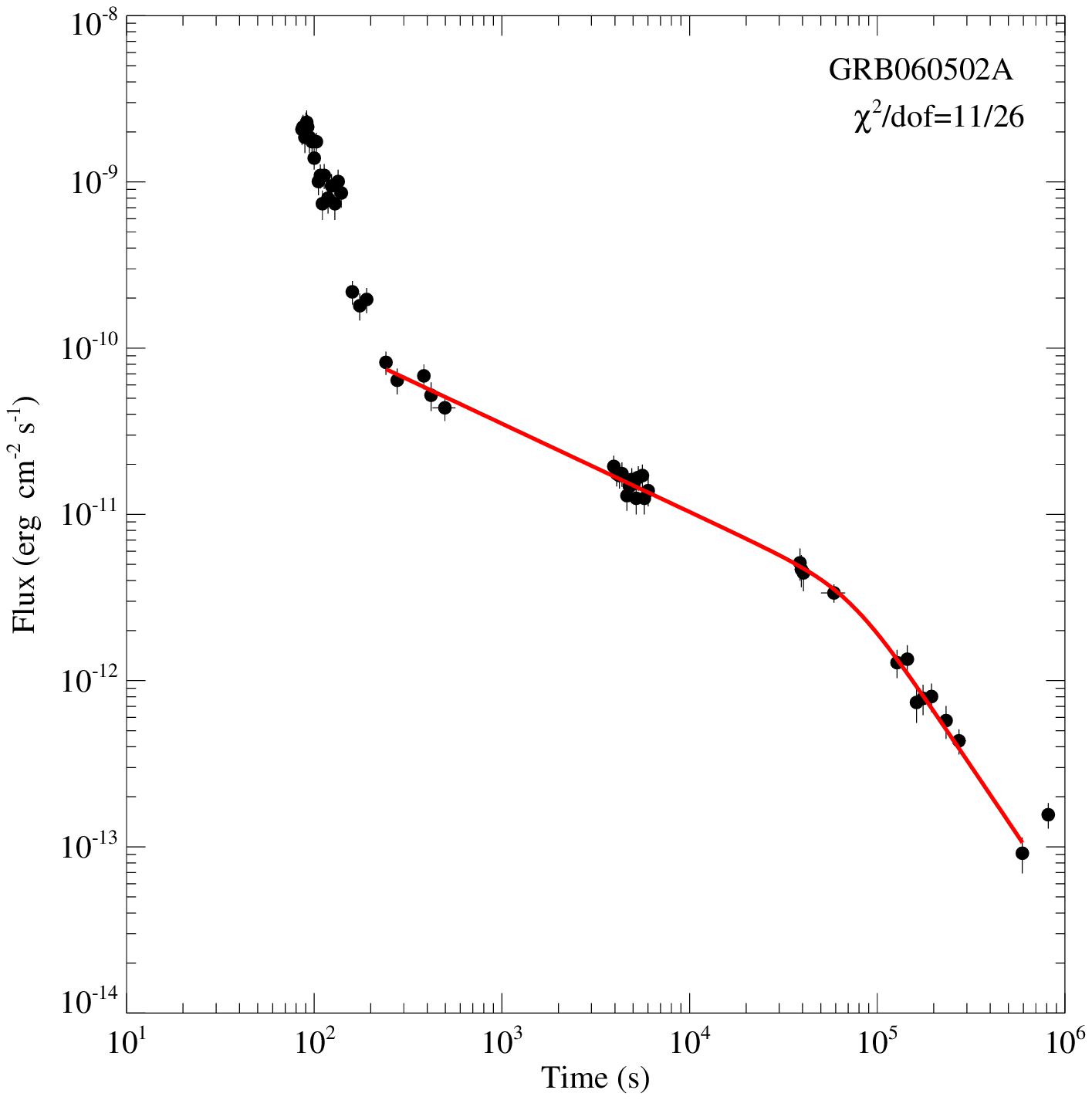}
\includegraphics[angle=0,scale=0.40]{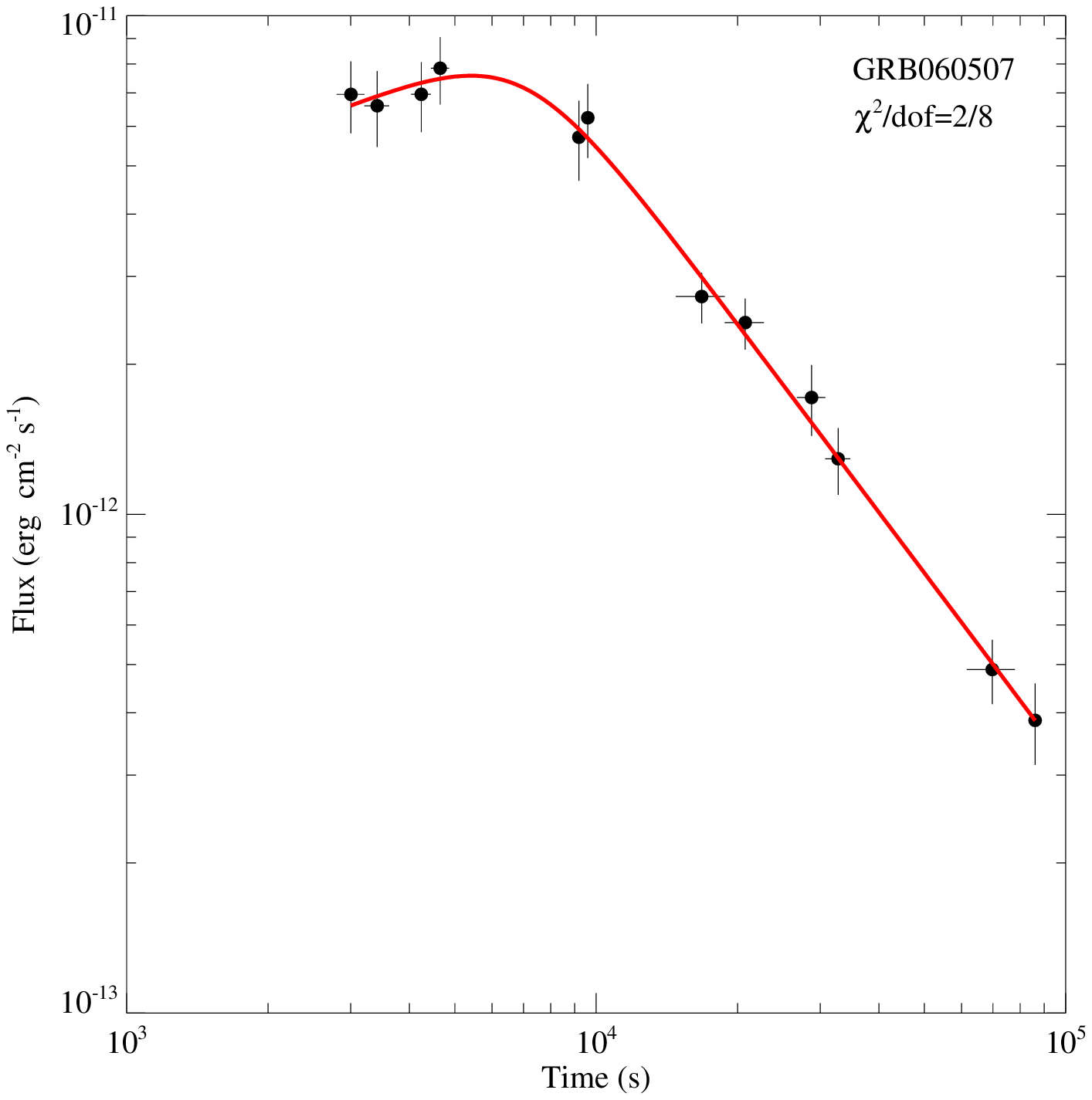}
\includegraphics[angle=0,scale=0.40]{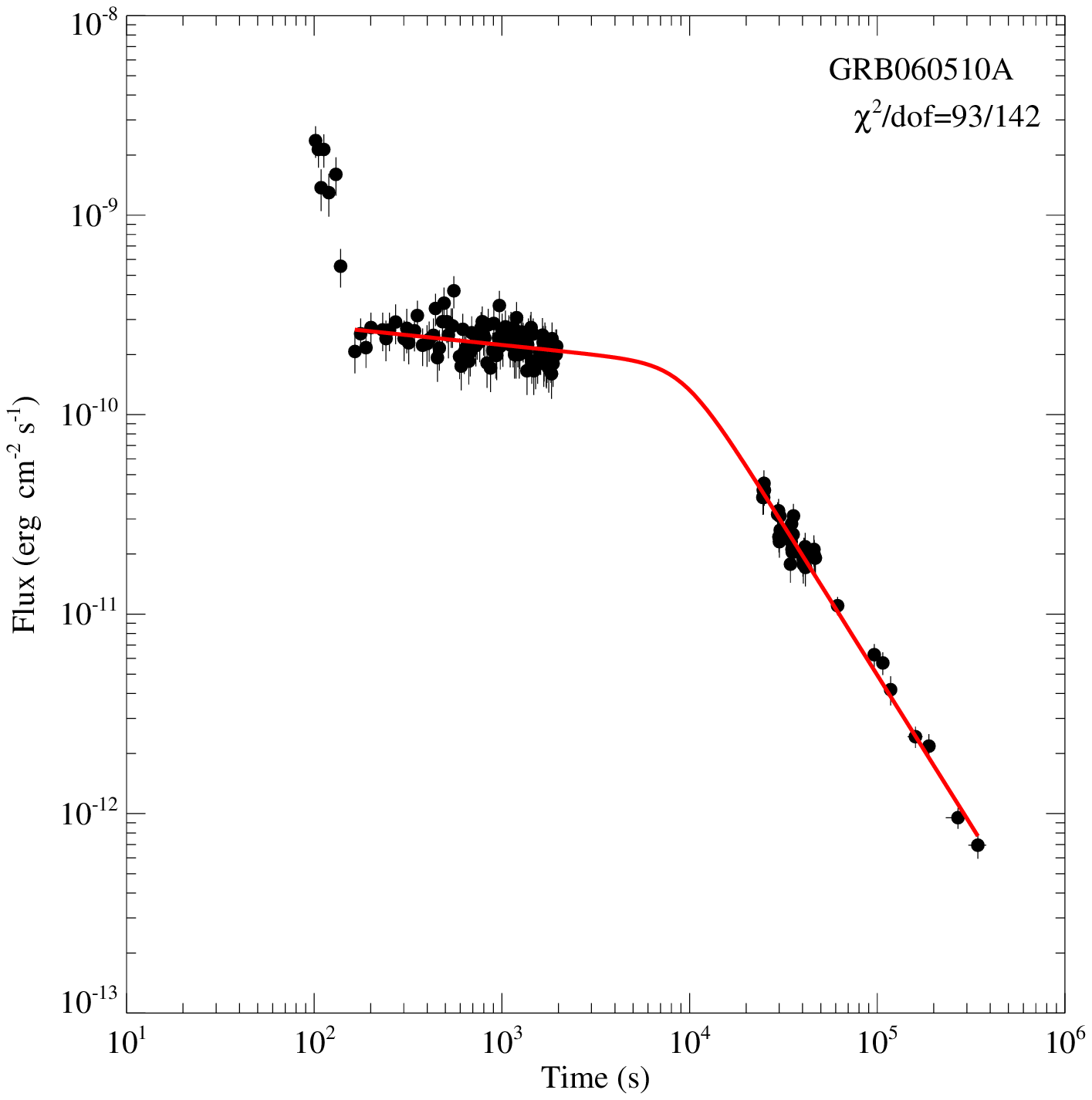}
\hfill
\includegraphics[angle=0,scale=0.40]{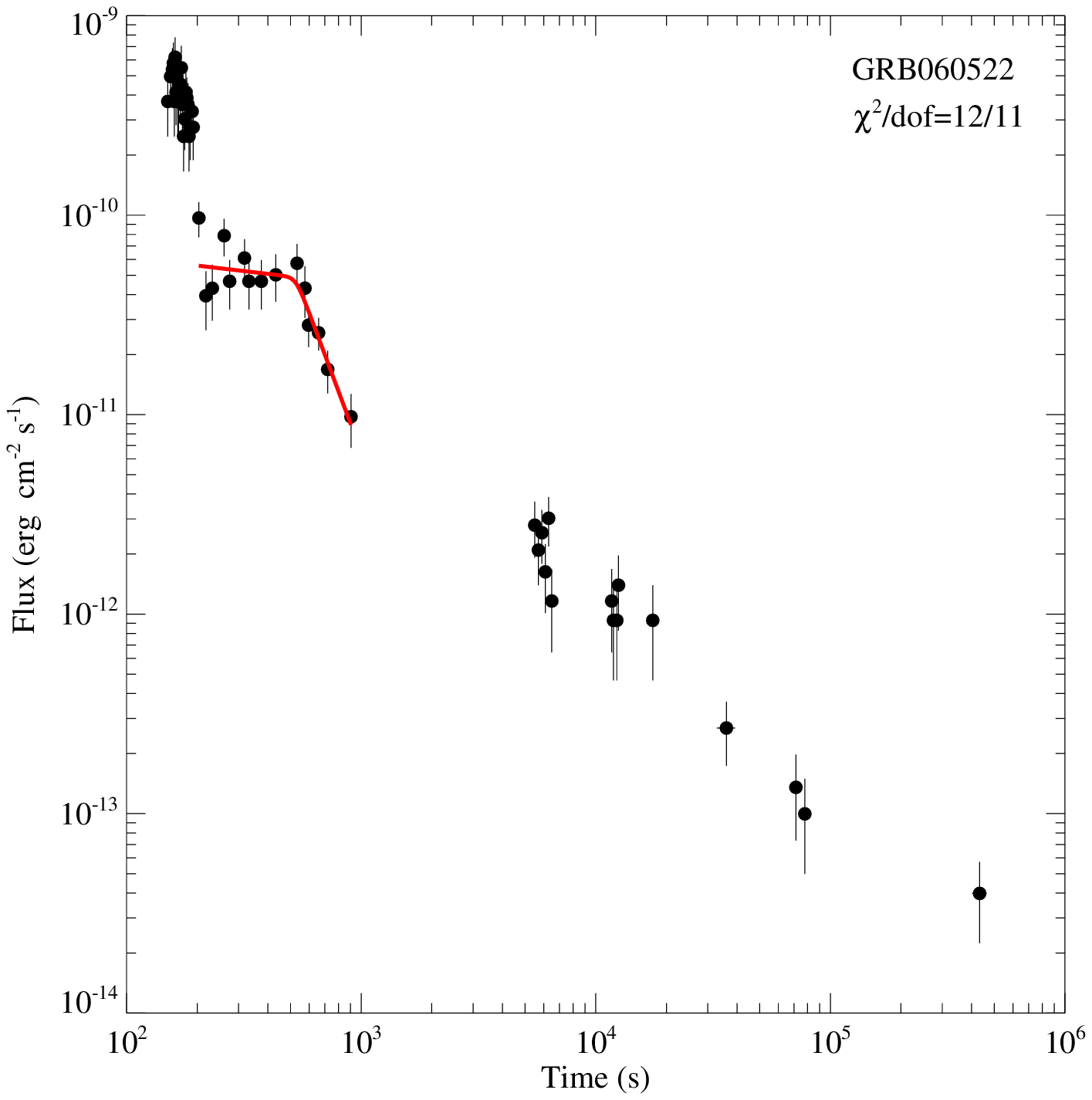}
\center{Fig.2  (continued)}
\end{figure*}
\begin{figure*}
\includegraphics[angle=0,scale=0.40]{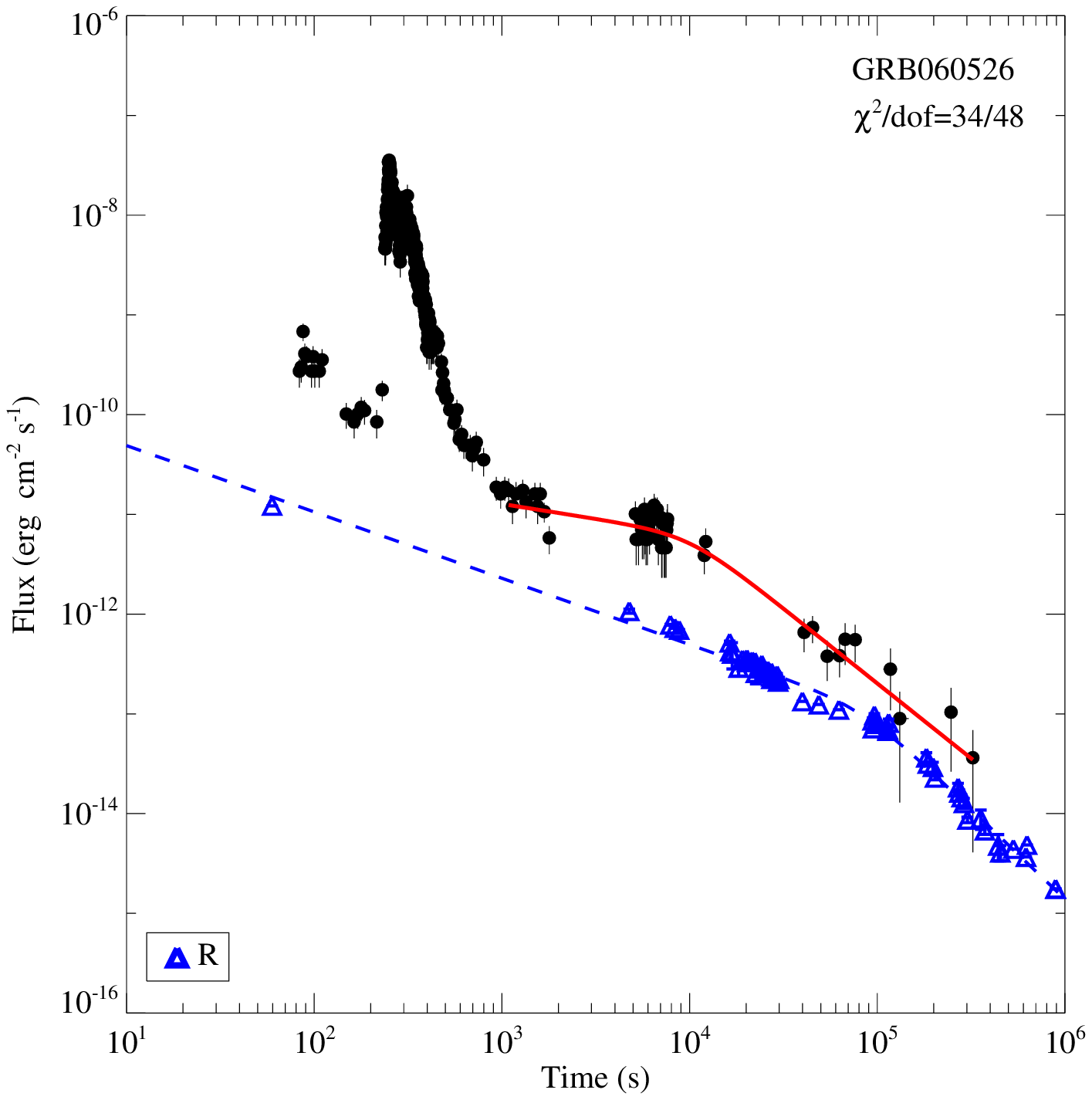}
\includegraphics[angle=0,scale=0.40]{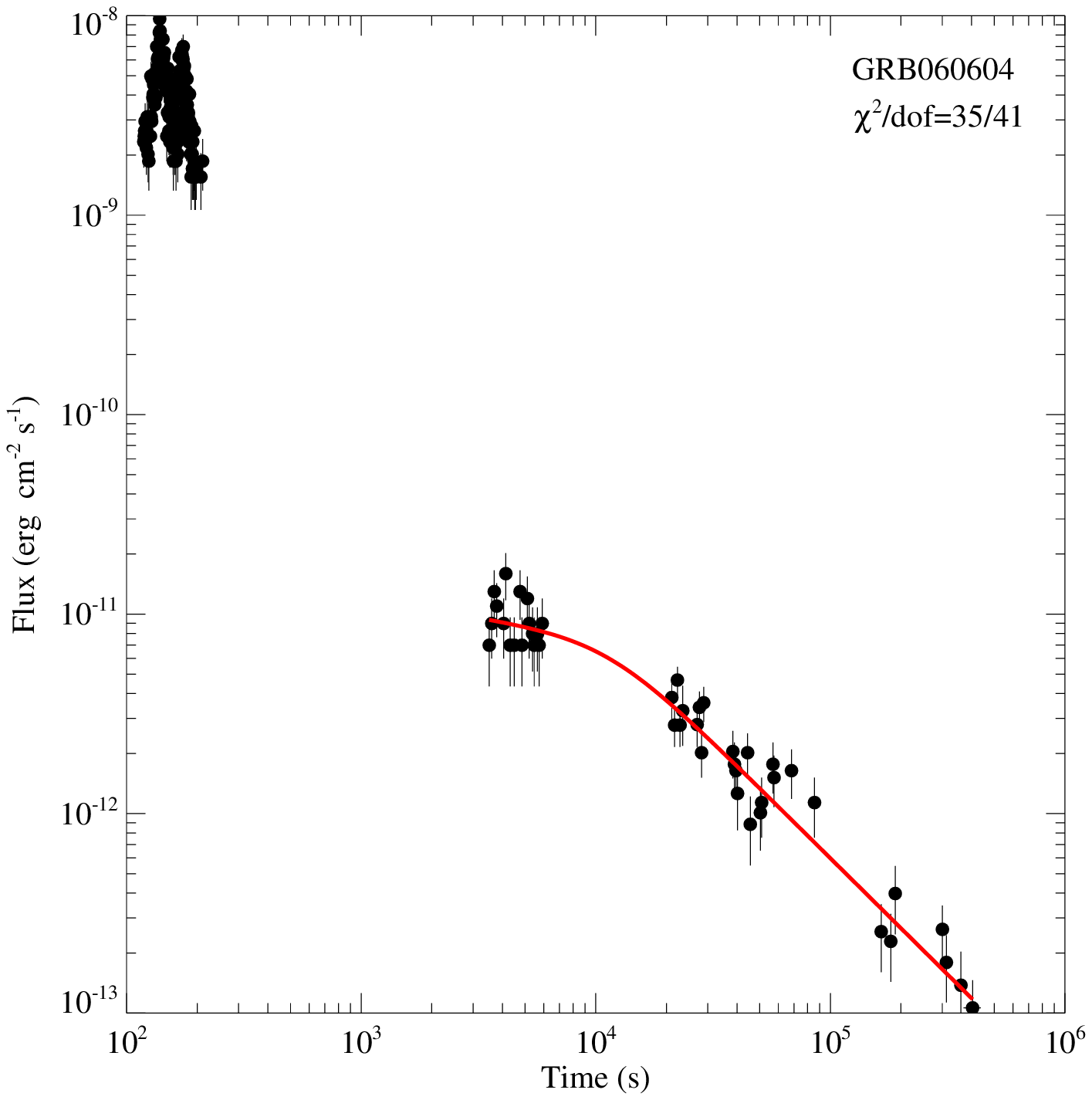}
\includegraphics[angle=0,scale=0.40]{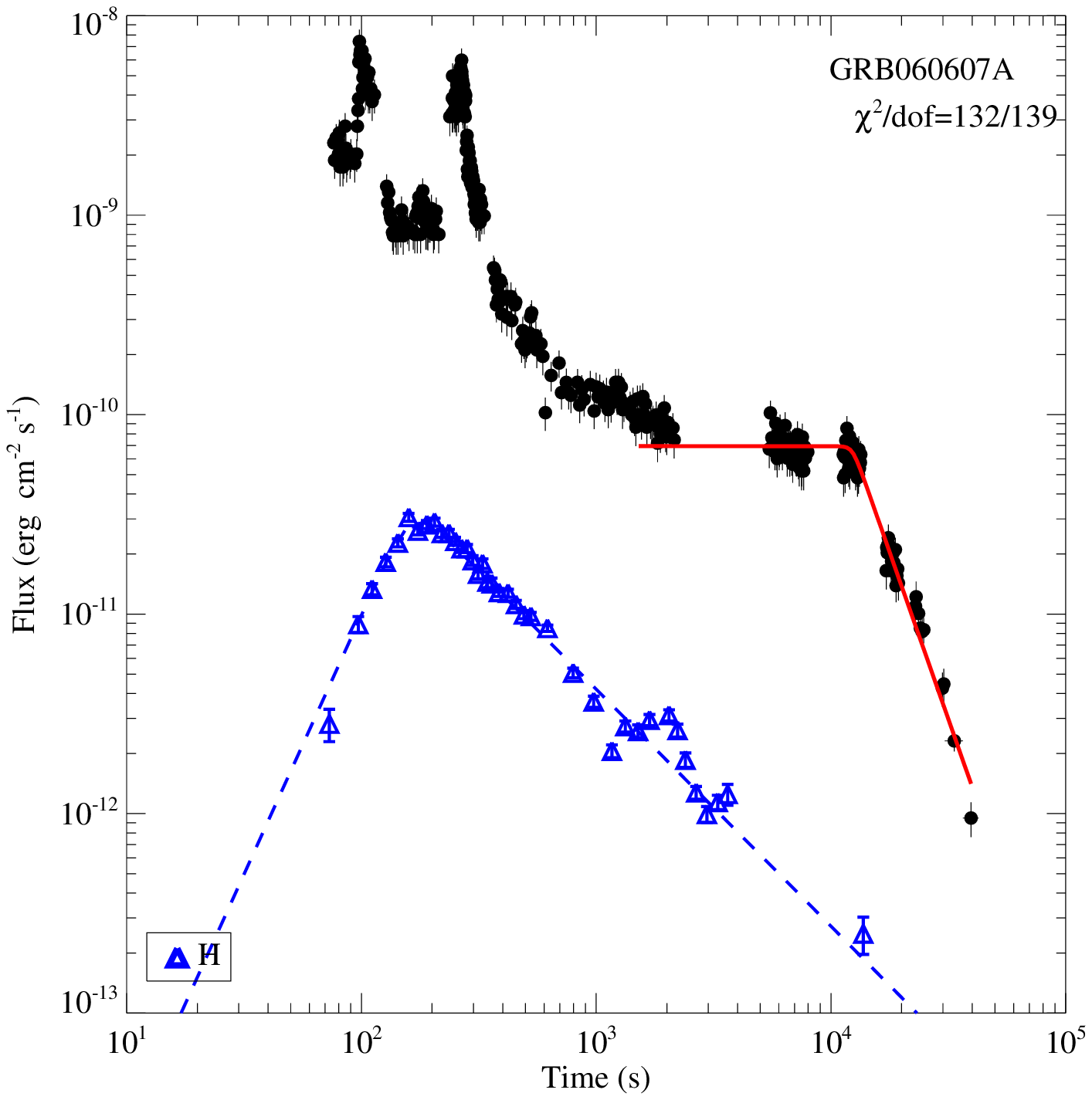}
\includegraphics[angle=0,scale=0.40]{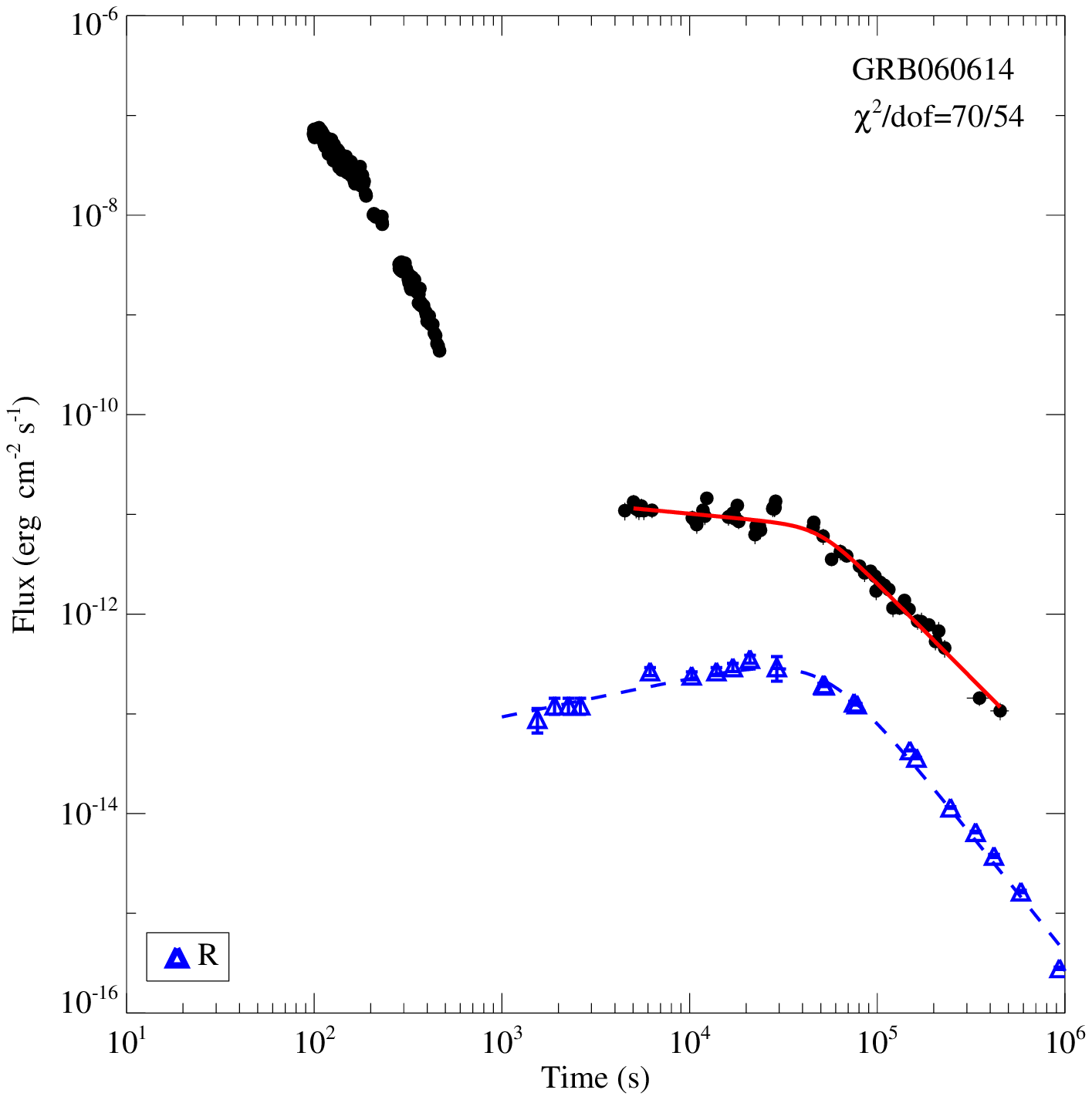}
\includegraphics[angle=0,scale=0.40]{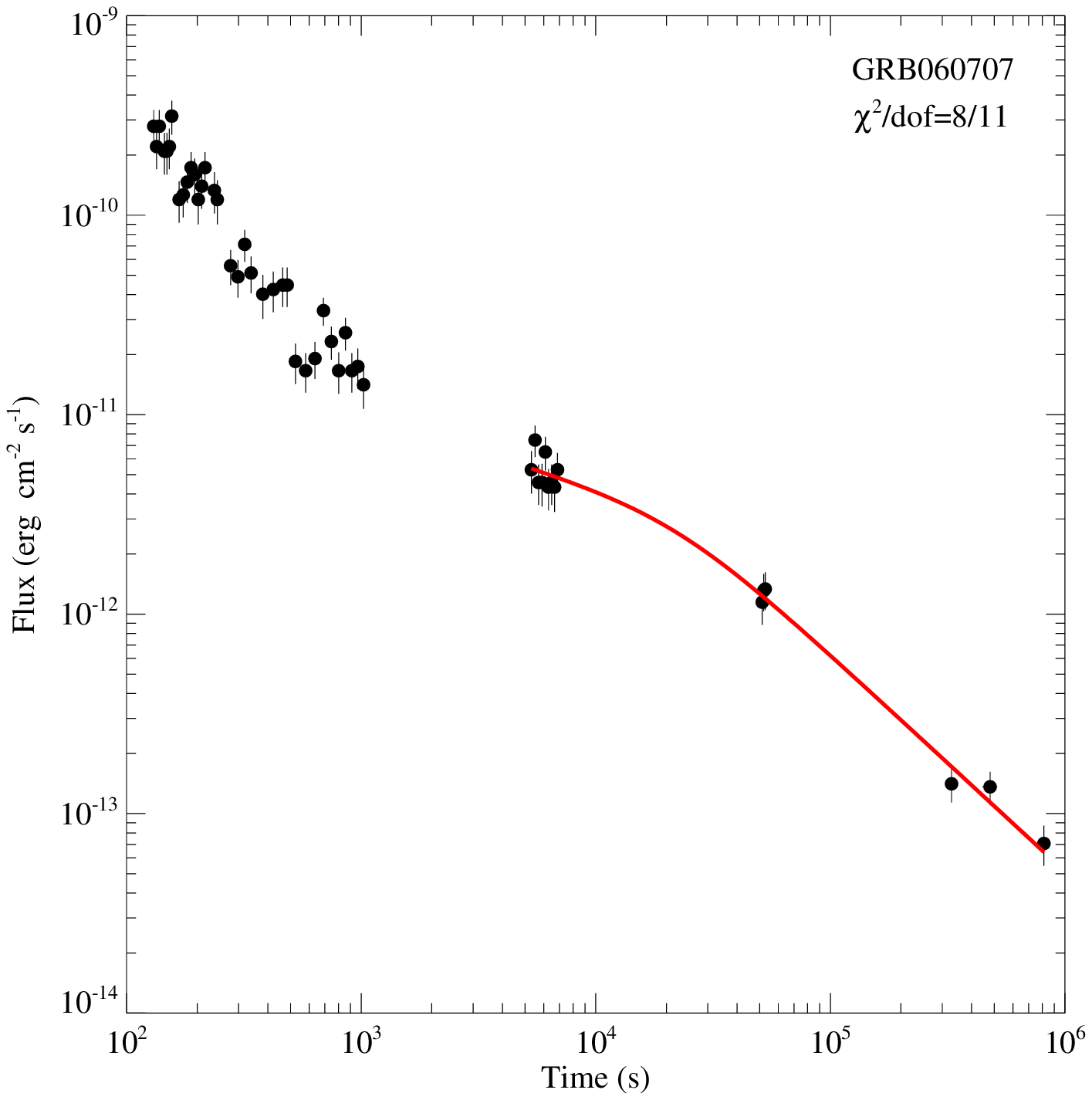}
\hfill
\includegraphics[angle=0,scale=0.40]{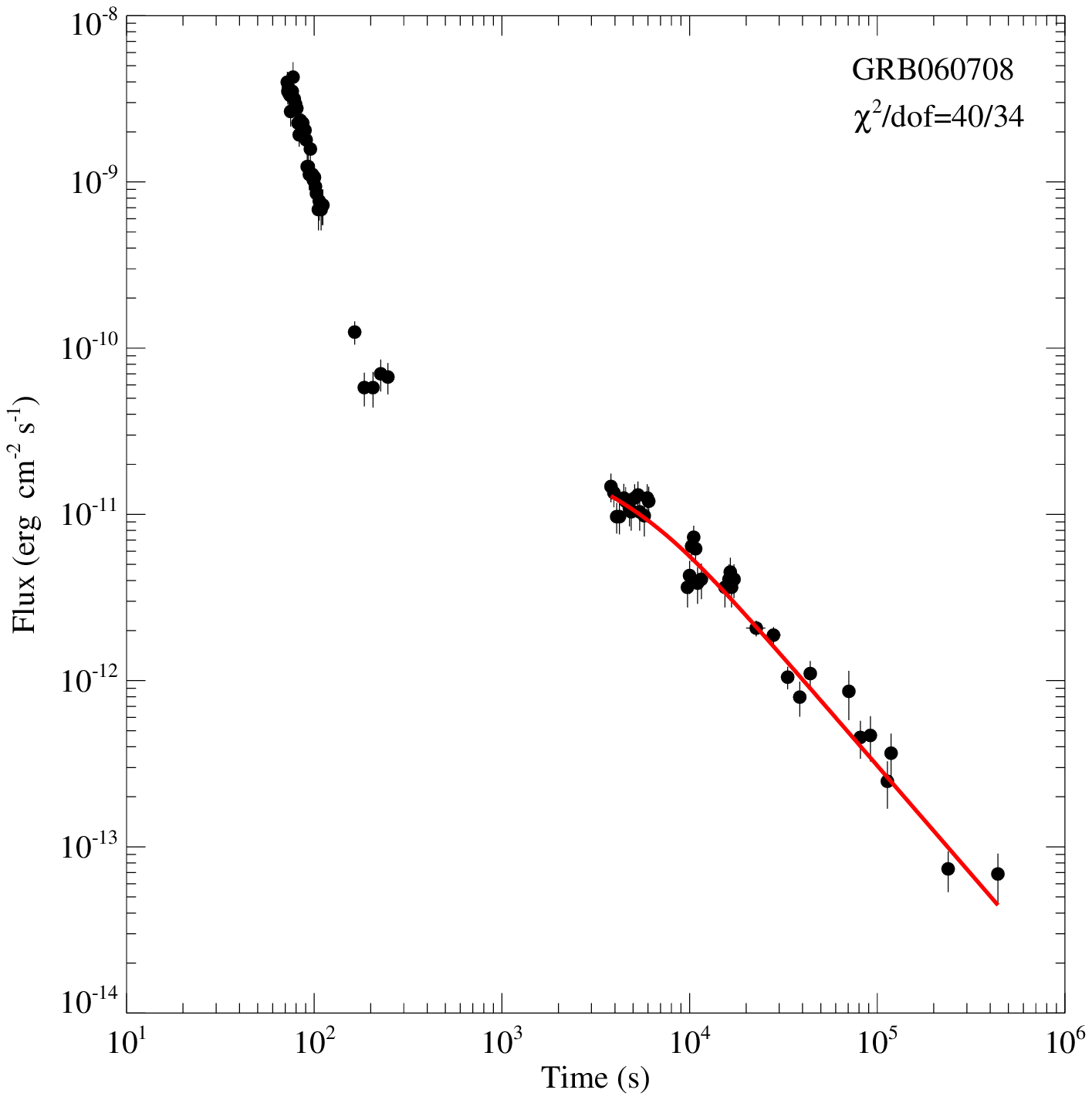}
\center{Fig.2  (continued)}
\end{figure*}
\begin{figure*}
\includegraphics[angle=0,scale=0.40]{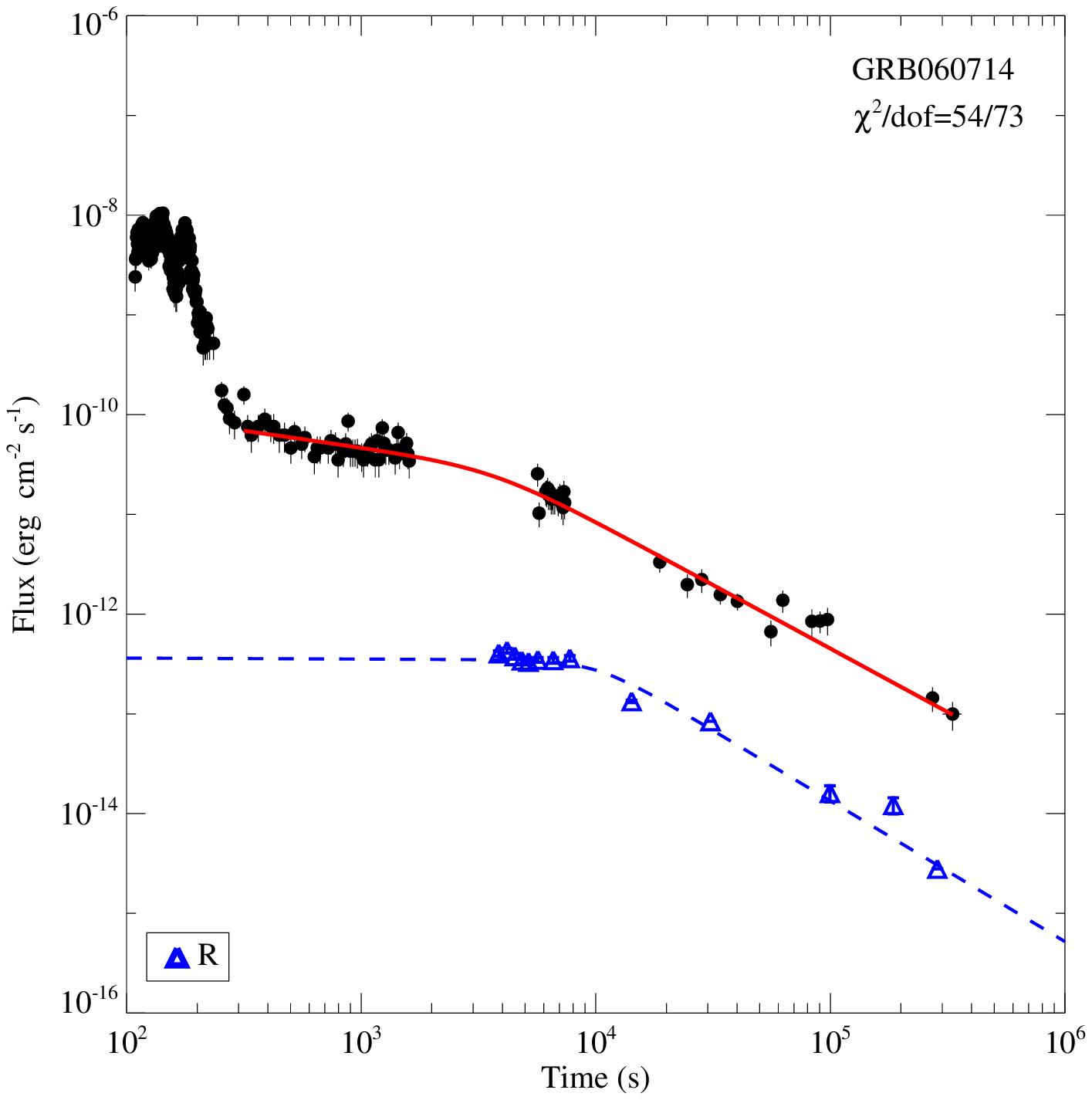}
\includegraphics[angle=0,scale=0.40]{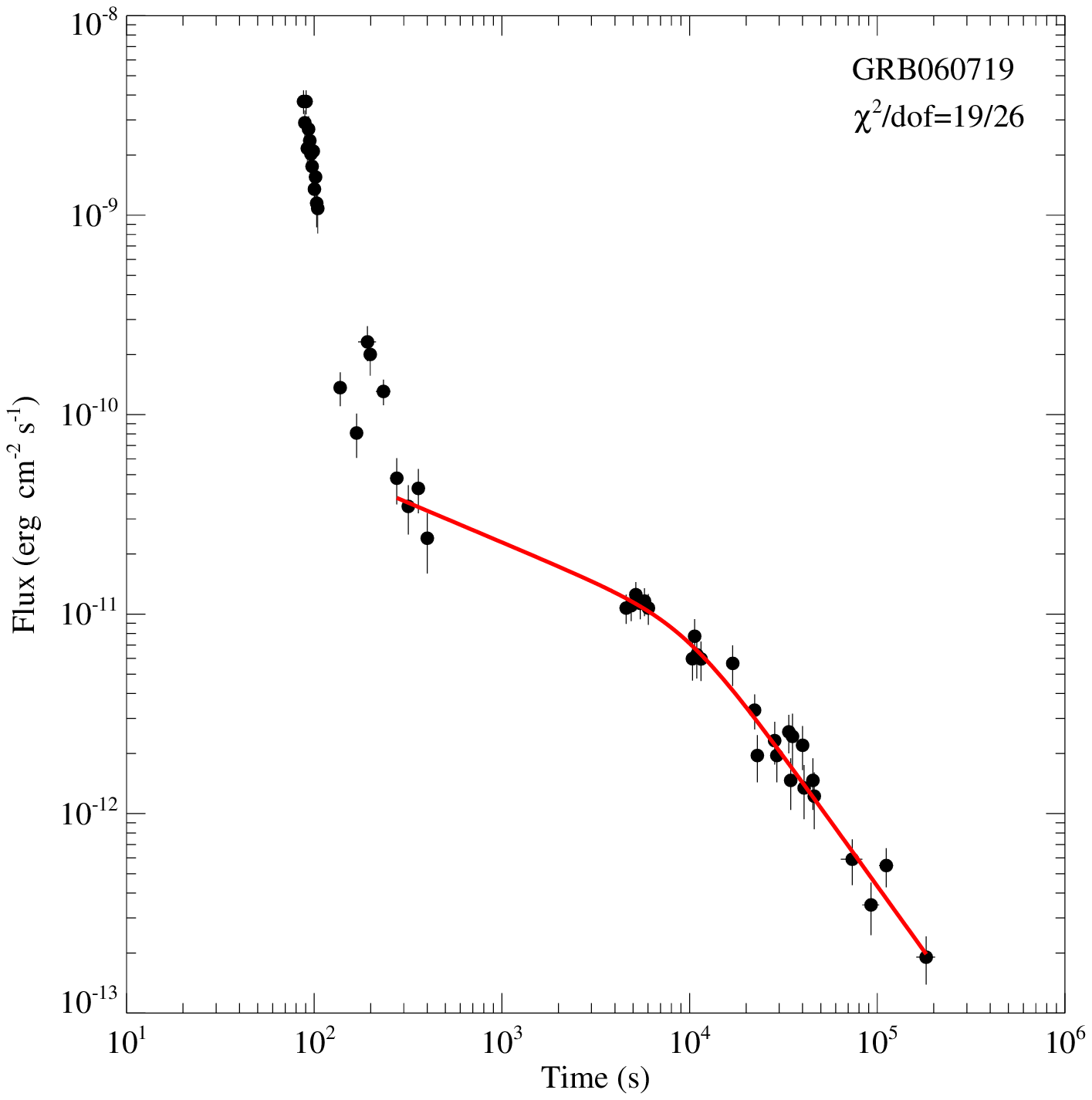}
\includegraphics[angle=0,scale=0.40]{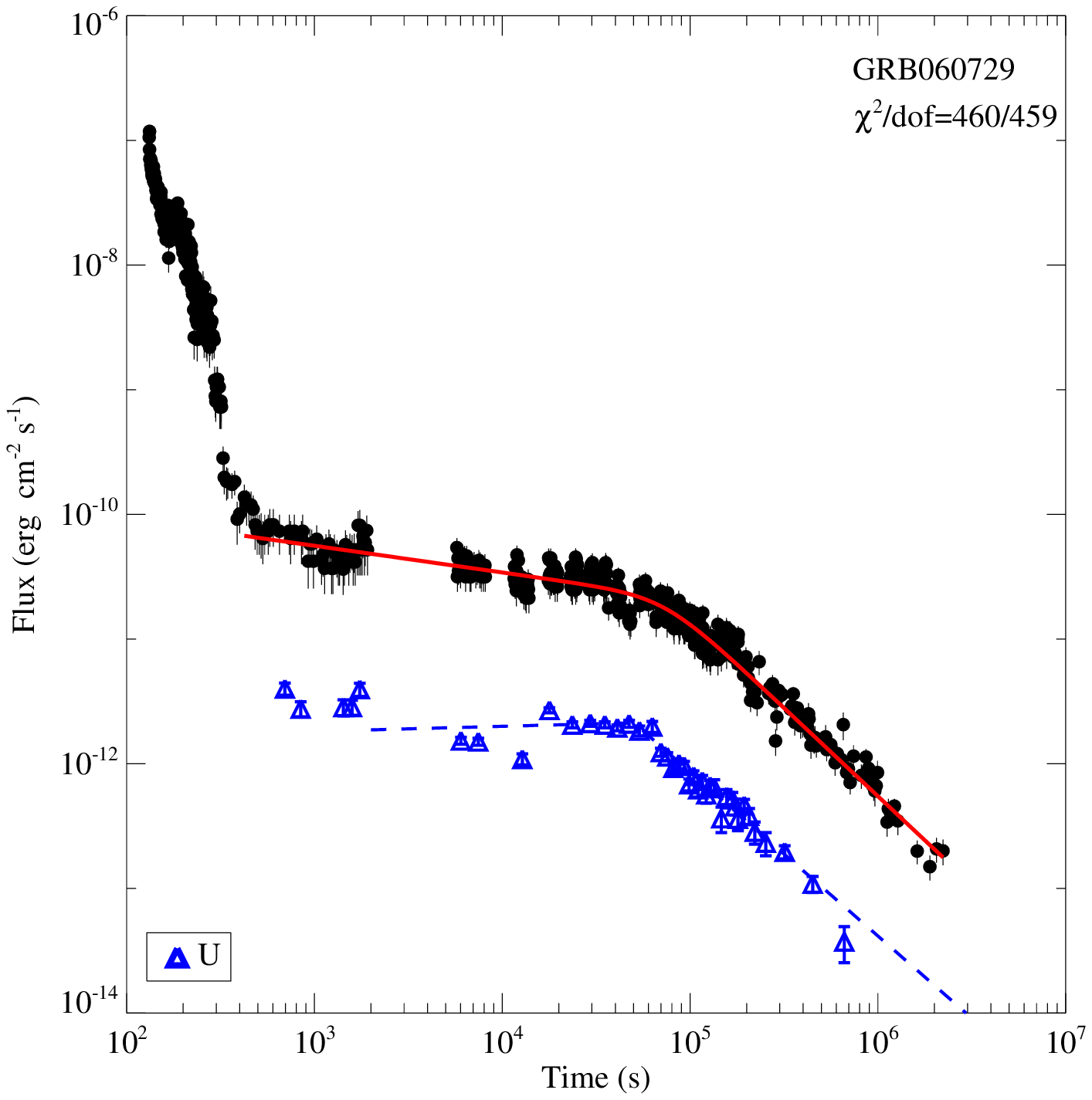}
\includegraphics[angle=0,scale=0.40]{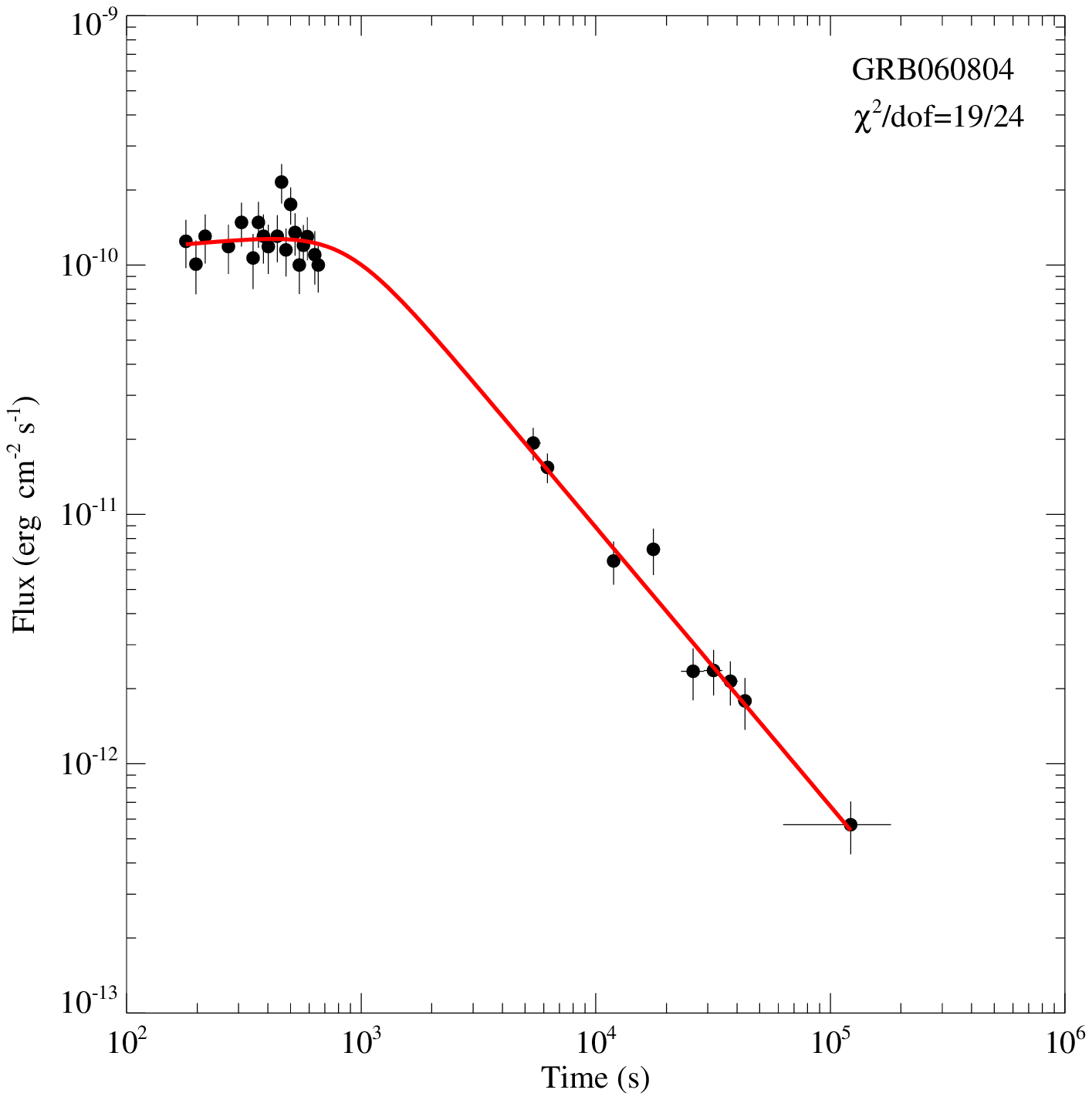}
\includegraphics[angle=0,scale=0.40]{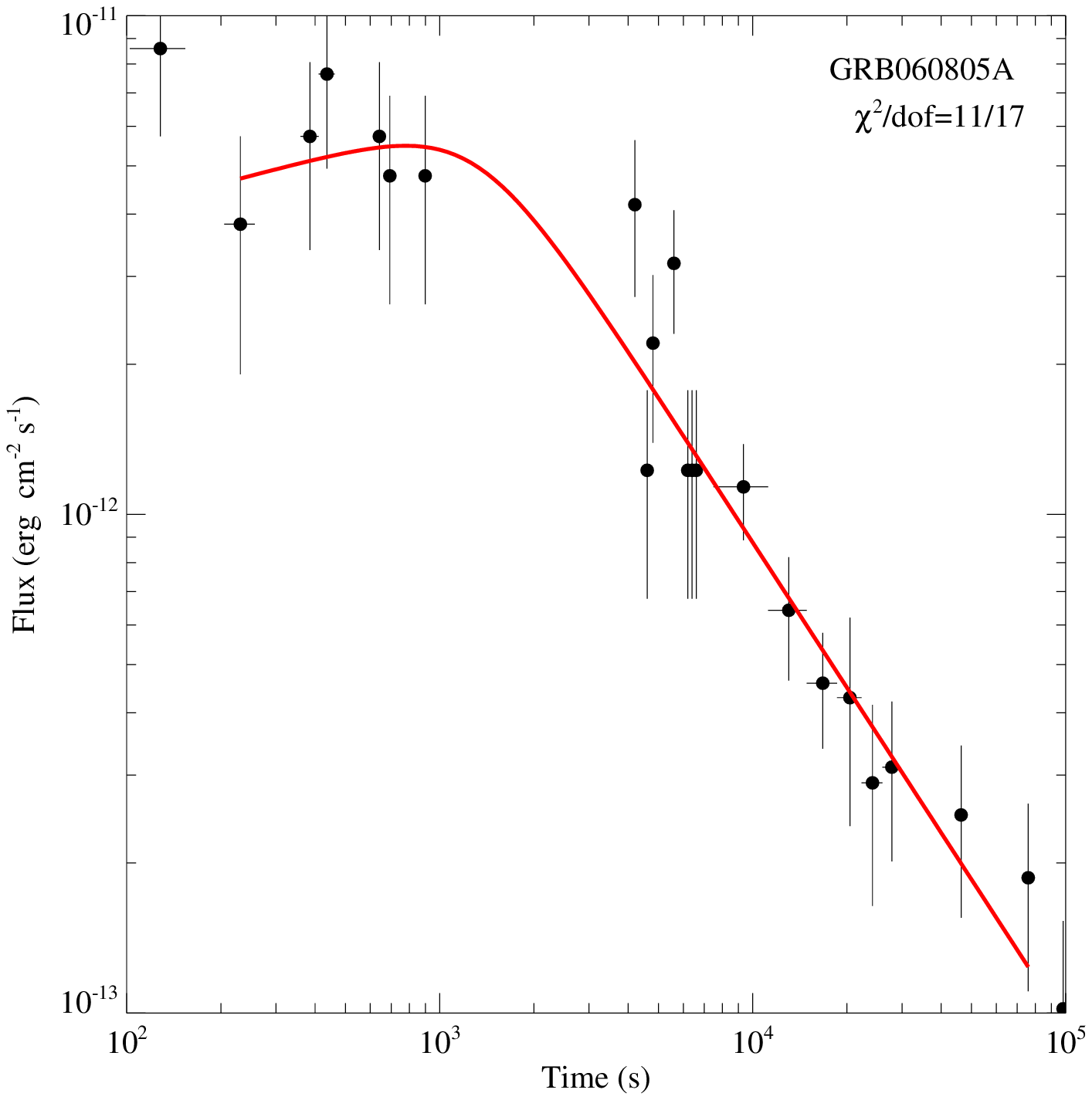}
\hfill
\includegraphics[angle=0,scale=0.40]{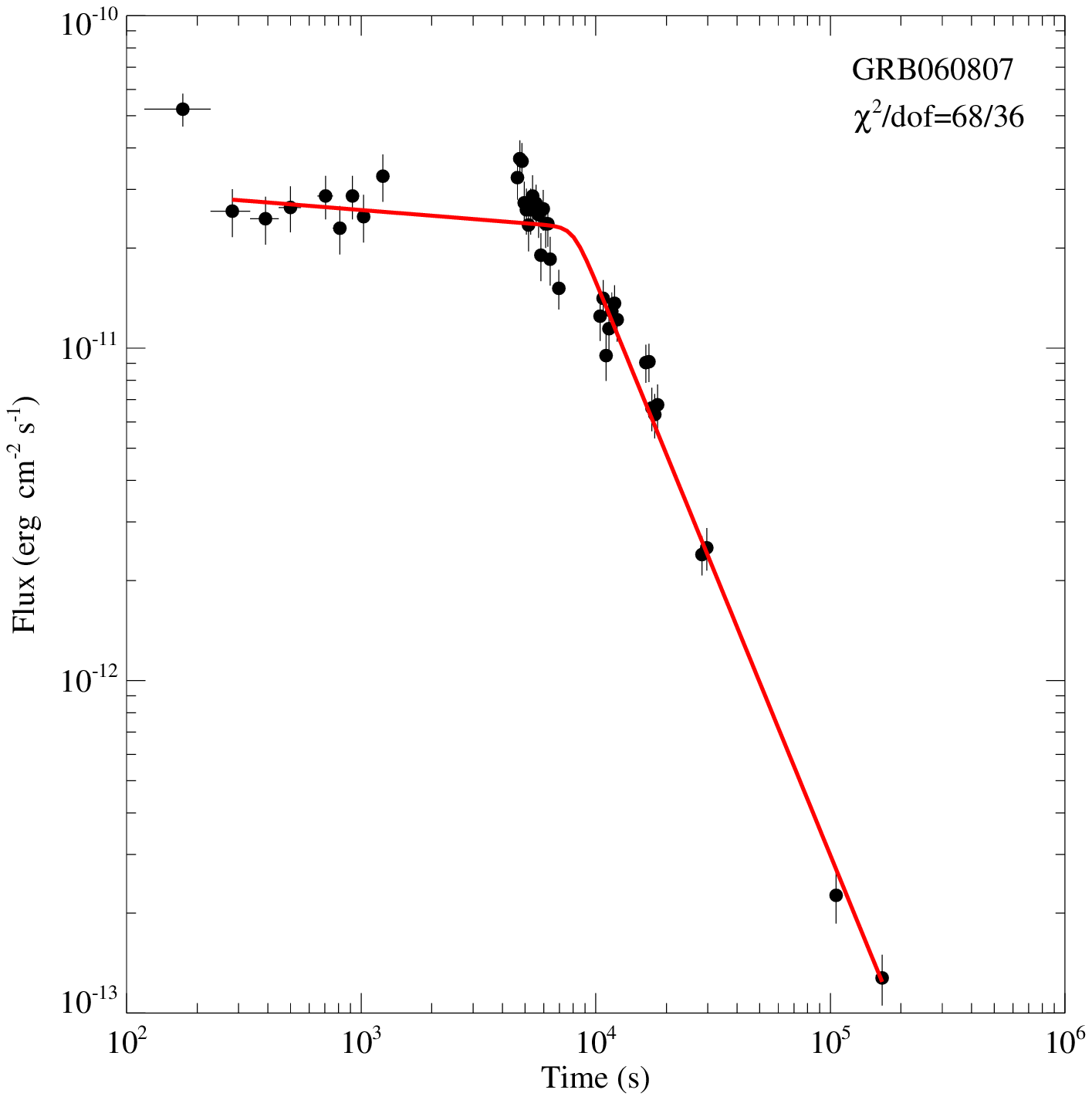}
\center{Fig.2  (continued)}
\end{figure*}
\begin{figure*}
\includegraphics[angle=0,scale=0.40]{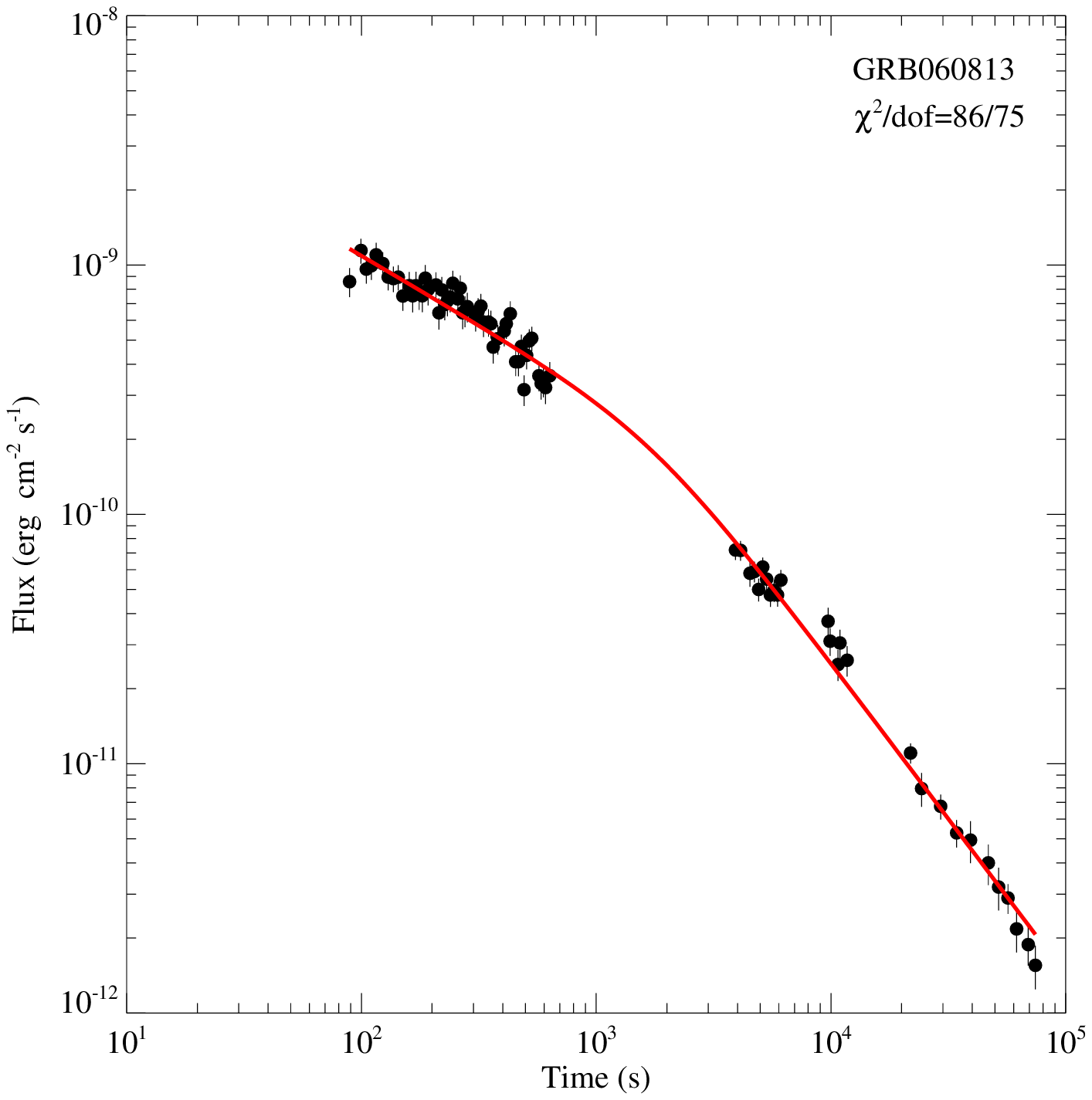}
\includegraphics[angle=0,scale=0.40]{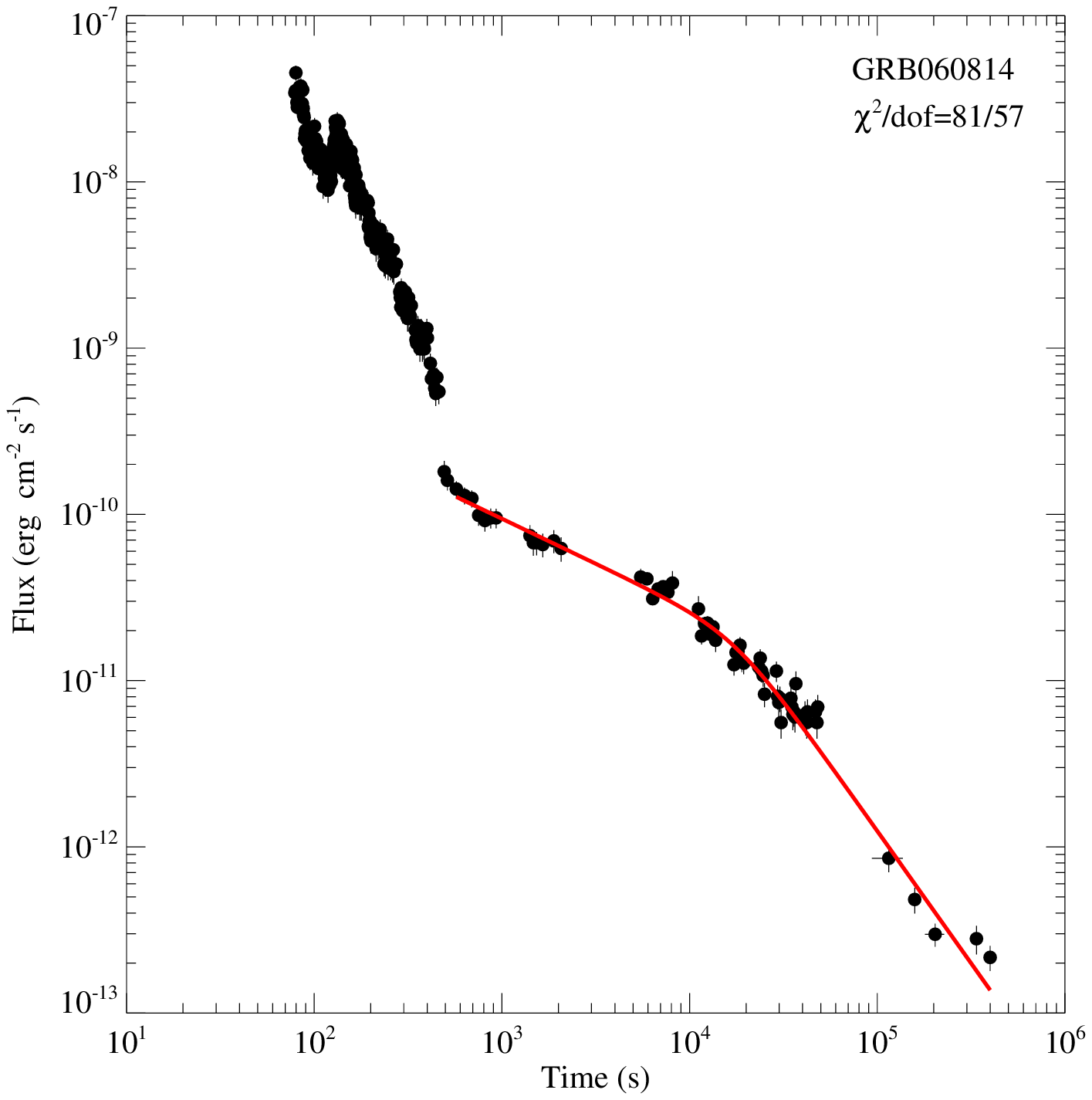}
\includegraphics[angle=0,scale=0.40]{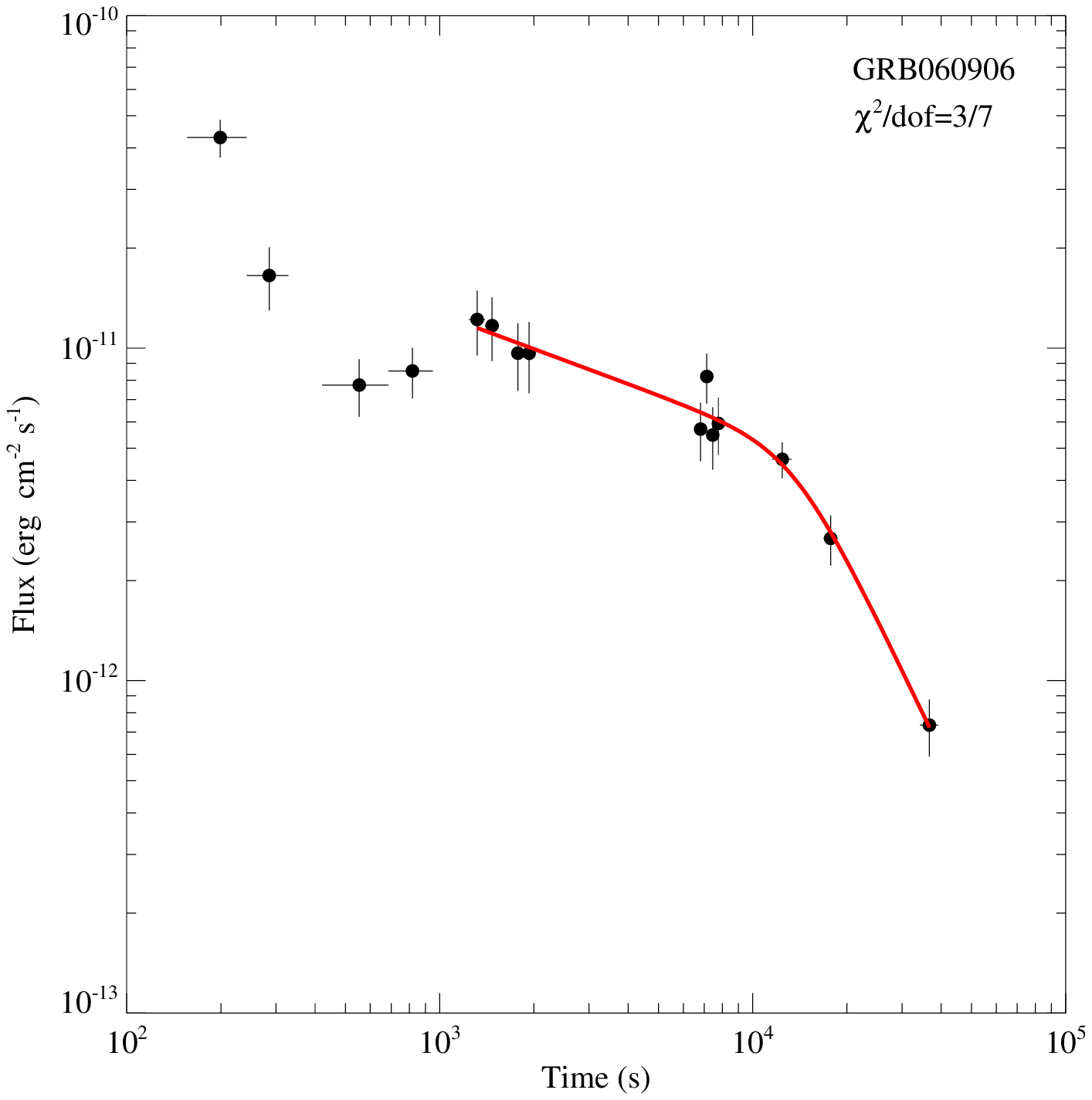}
\includegraphics[angle=0,scale=0.40]{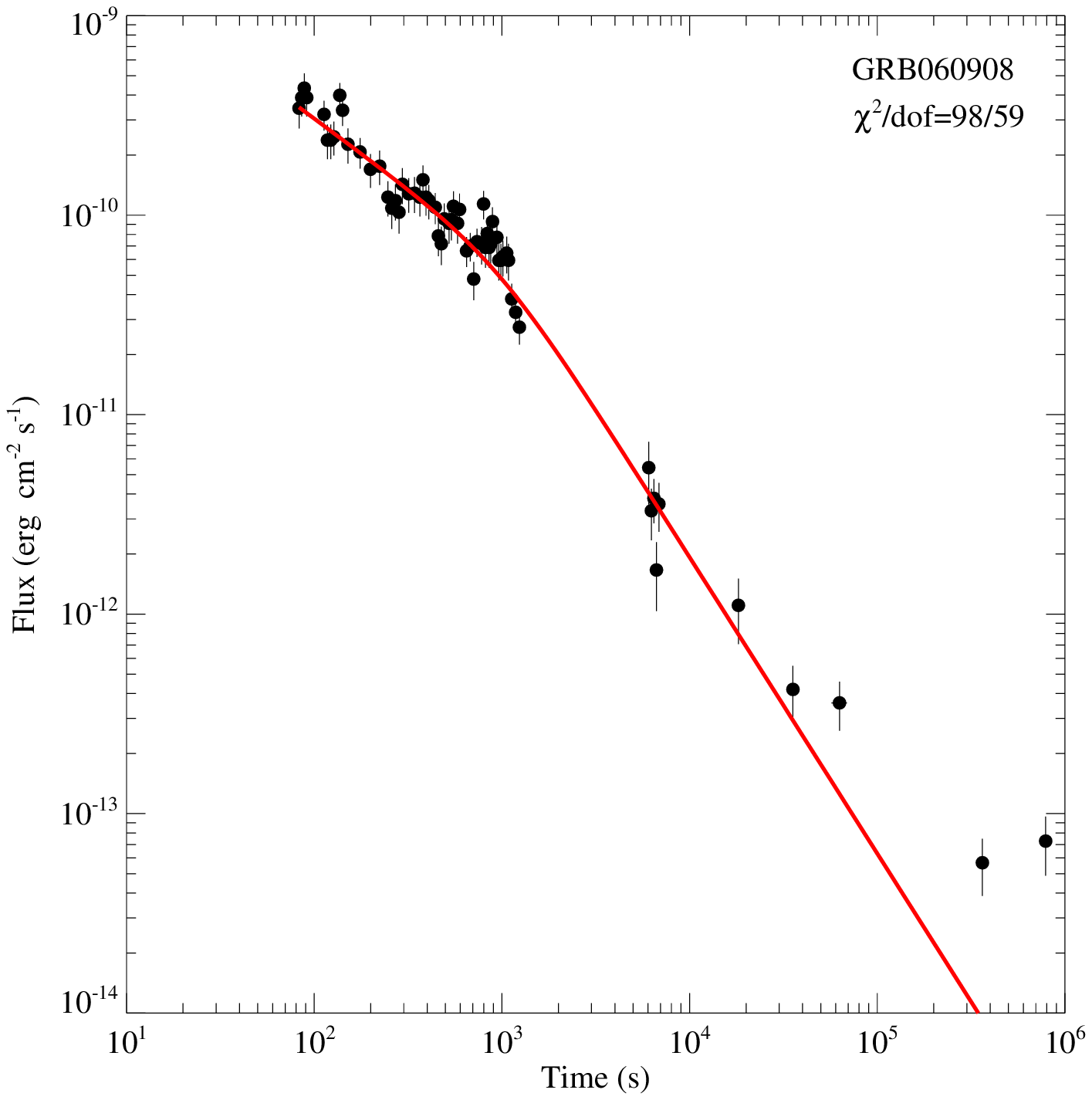}
\includegraphics[angle=0,scale=0.40]{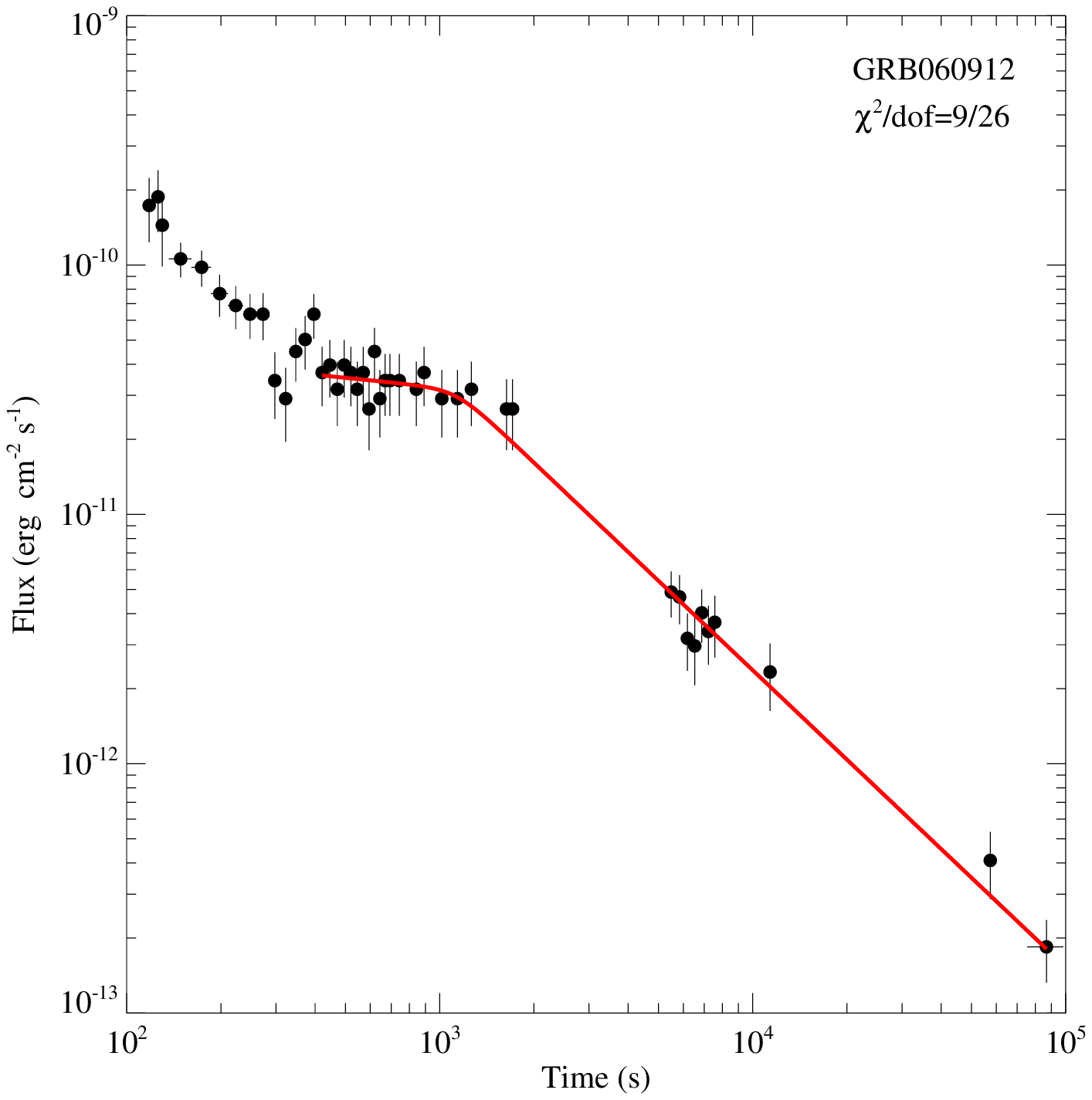}
\hfill
\includegraphics[angle=0,scale=0.40]{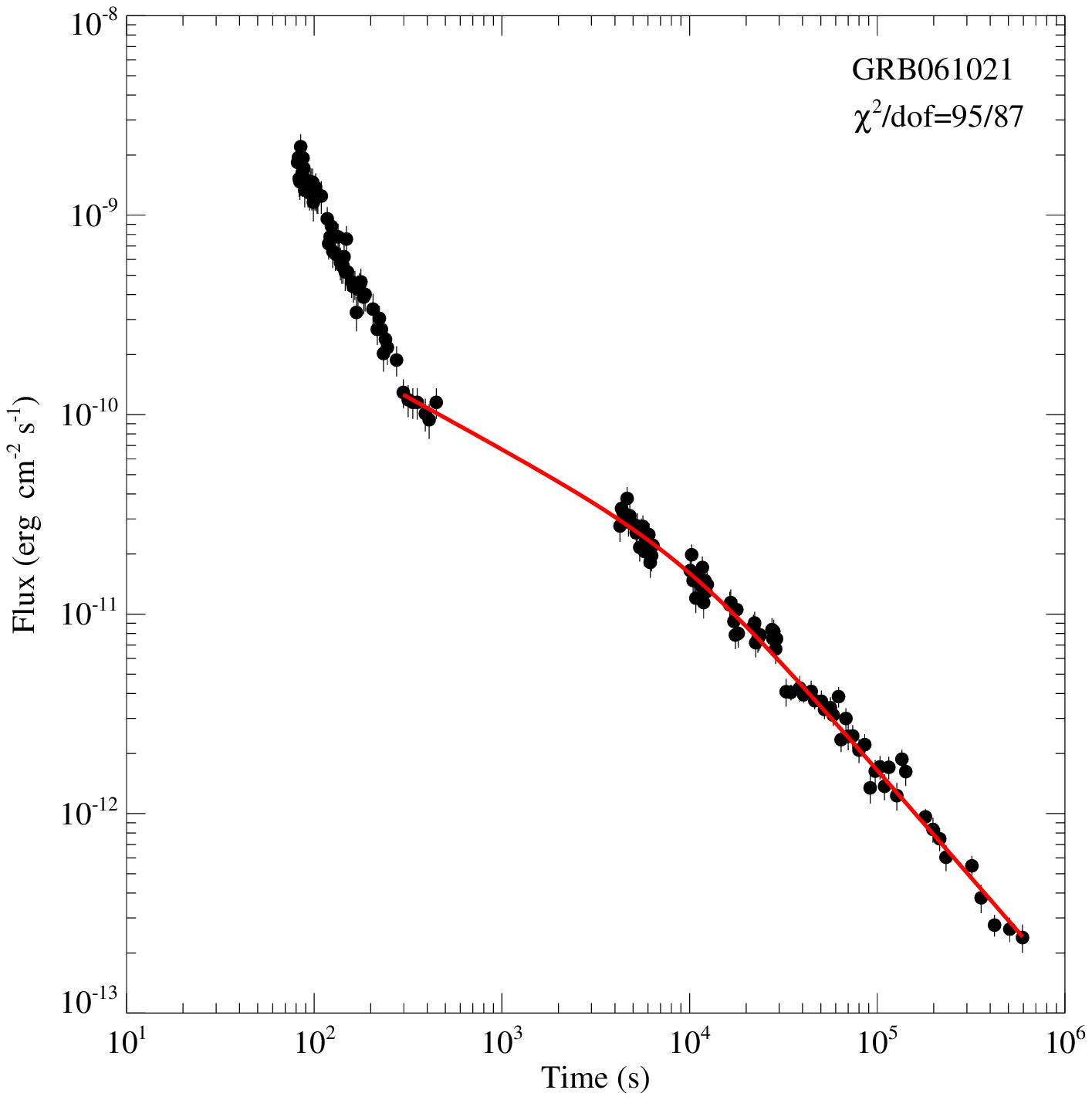}
\center{Fig.2 (continued)}

\end{figure*}
\begin{figure*}
\includegraphics[angle=0,scale=0.40]{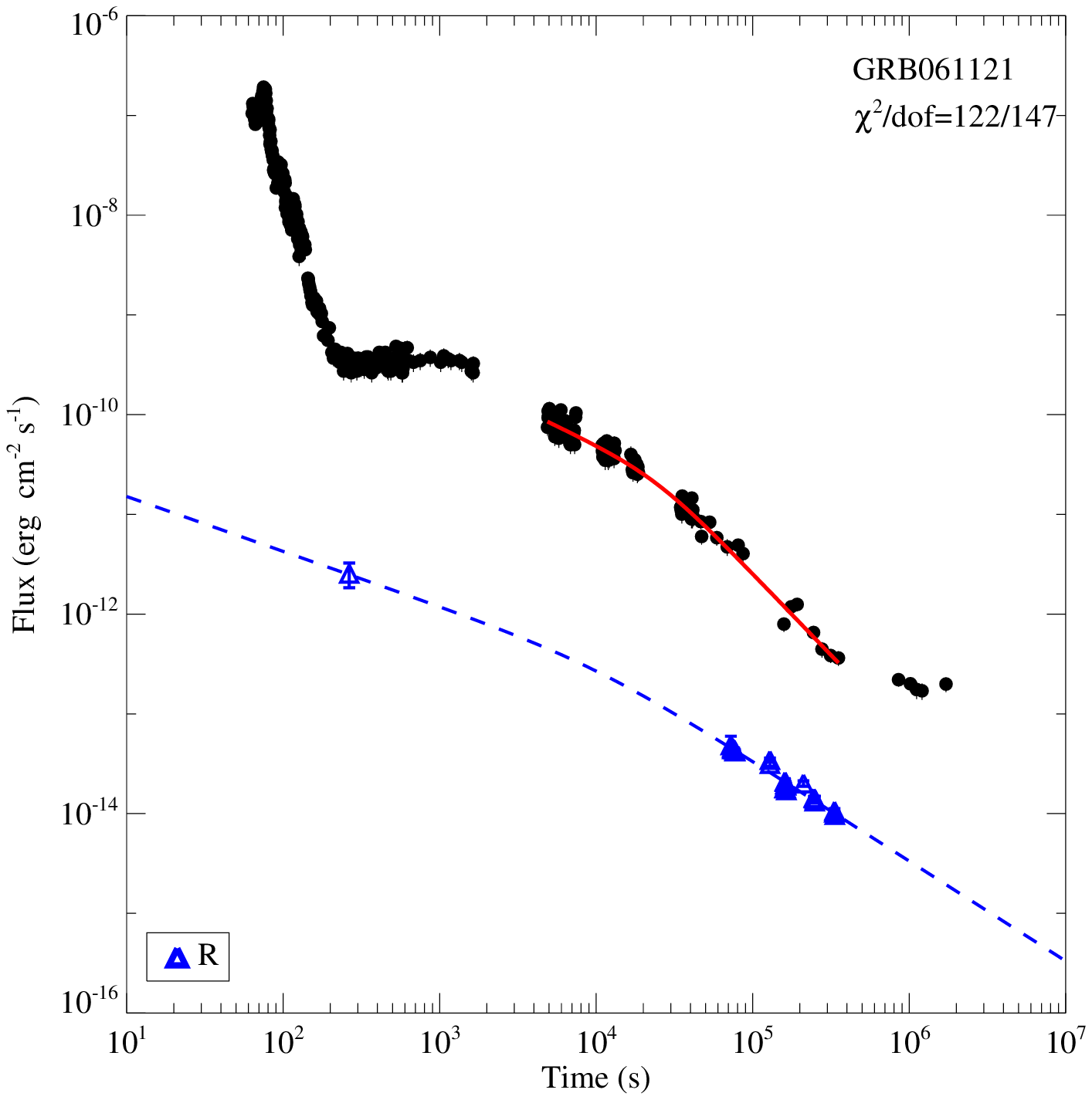}
\includegraphics[angle=0,scale=0.40]{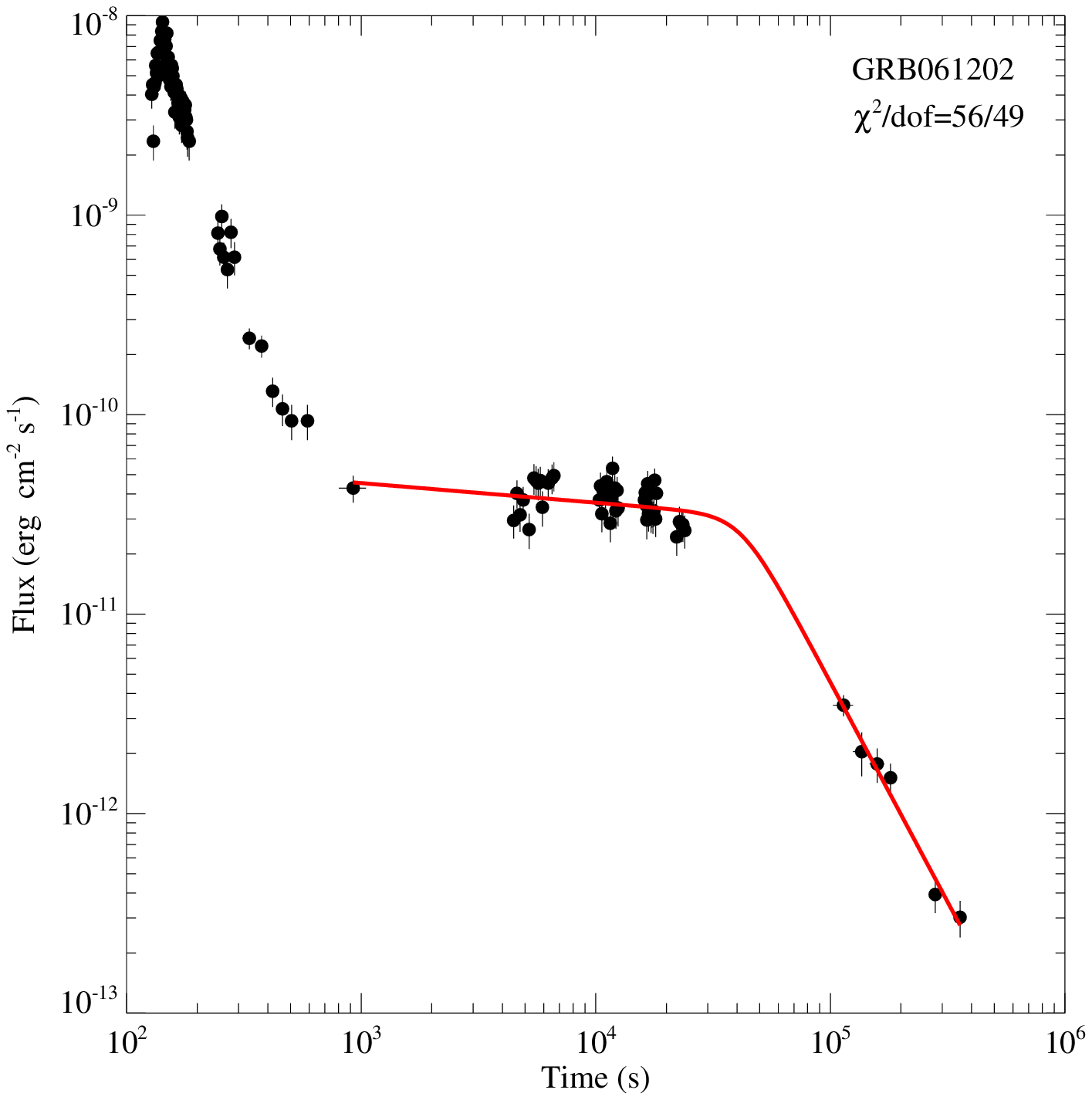}
\includegraphics[angle=0,scale=0.40]{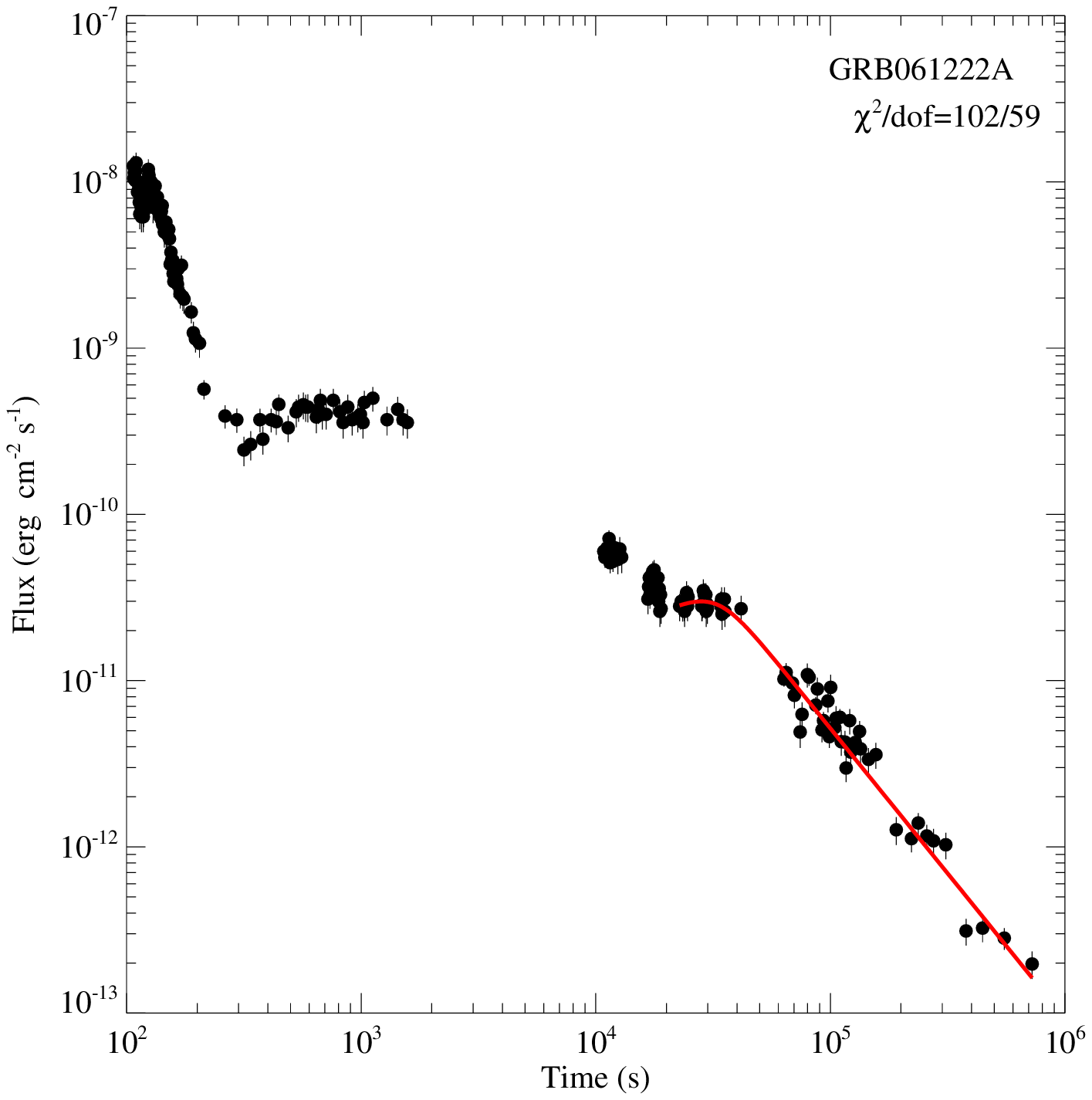}
\includegraphics[angle=0,scale=0.40]{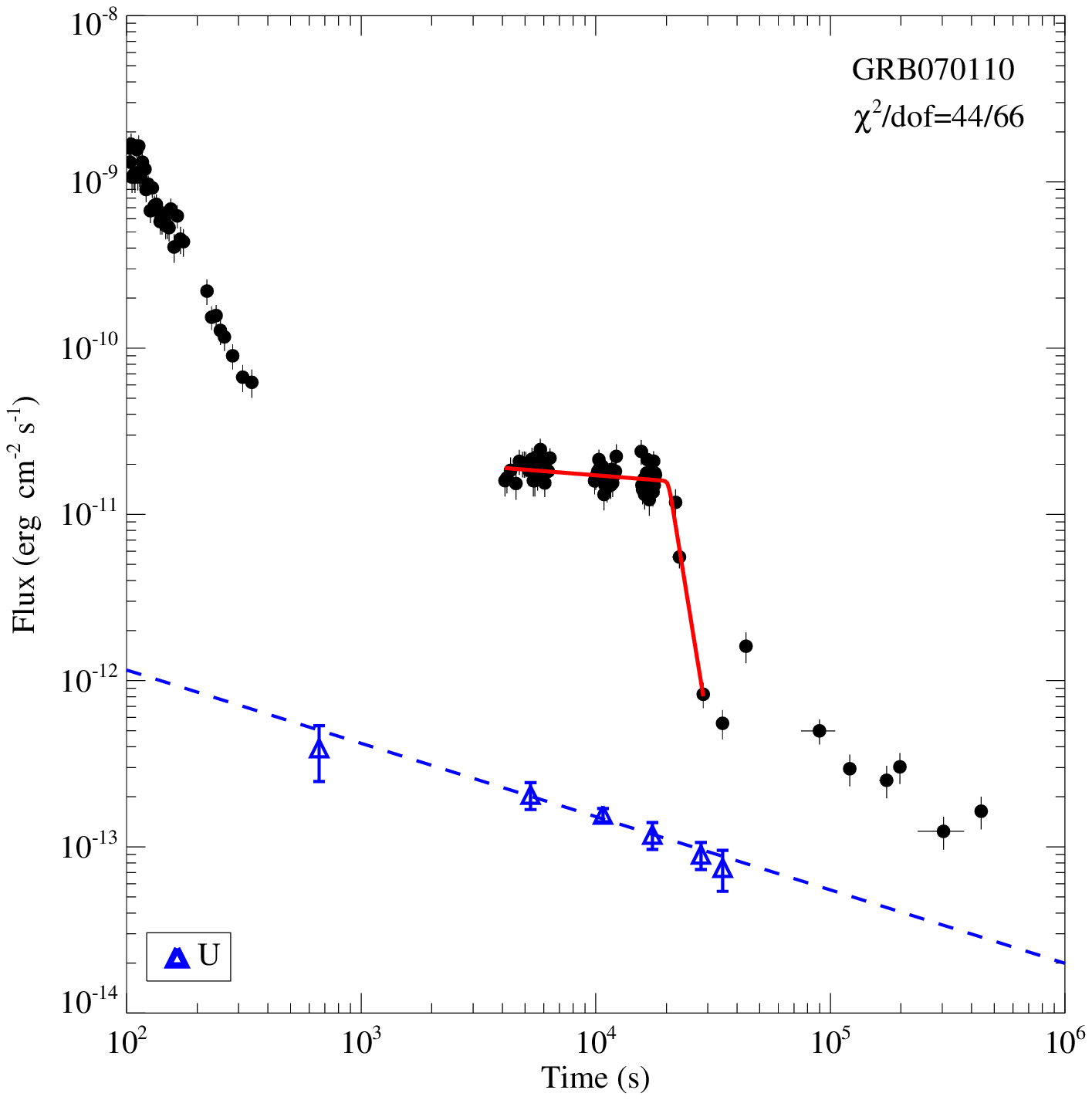}
\includegraphics[angle=0,scale=0.40]{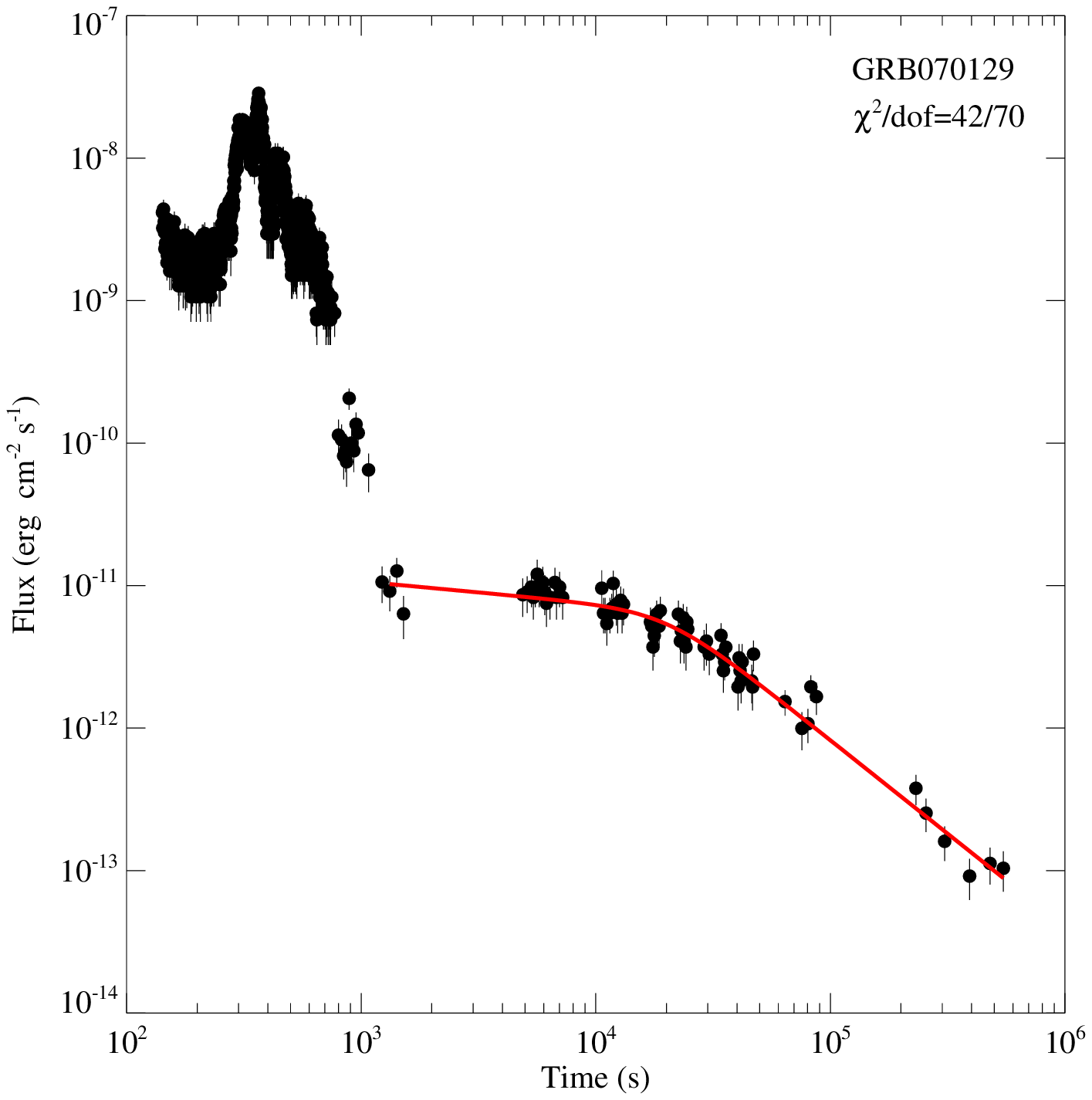}
\center{Fig.2  (continued)}
\end{figure*}

\begin{figure}
\epsscale{0.8} \plotone{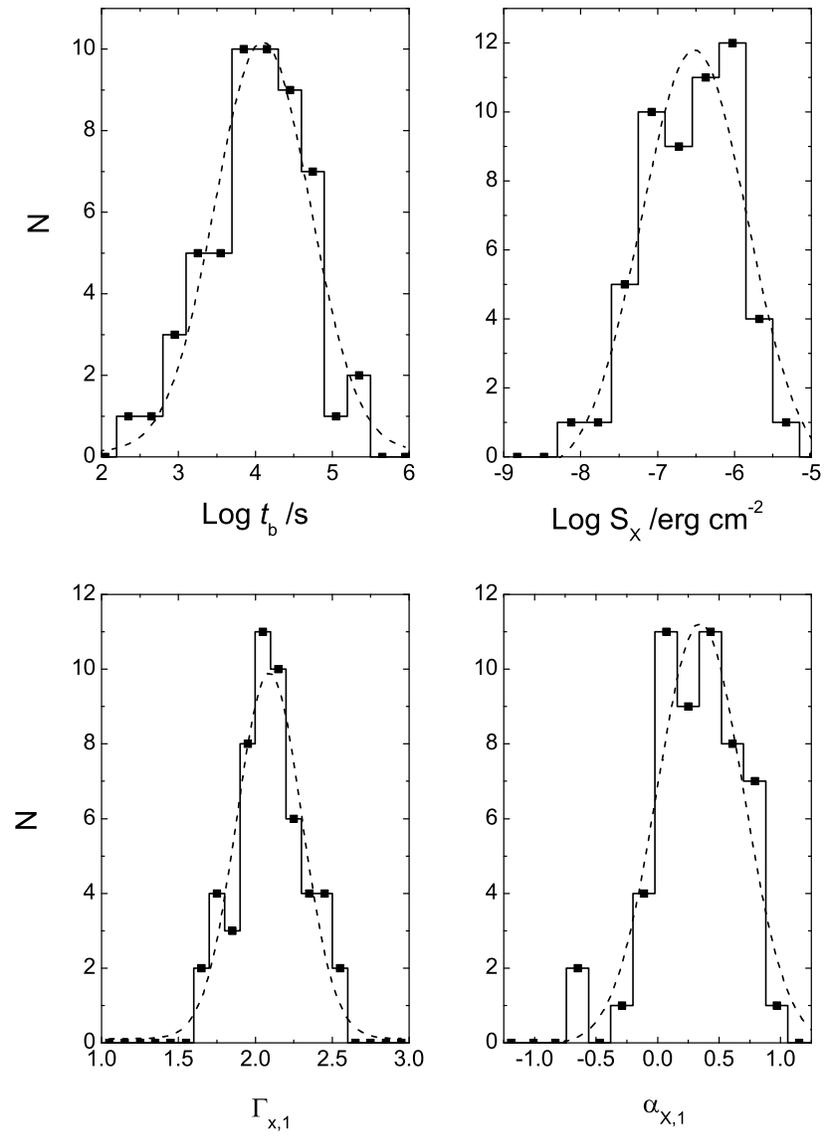} \caption{Distributions of the characteristics of
the shallow decay segment for the bursts in our sample. The dashed lines are the
fitting results with Gaussian functions.} \label{Fig_Distribution}
\end{figure}

\begin{figure}
\epsscale{0.8} \plotone{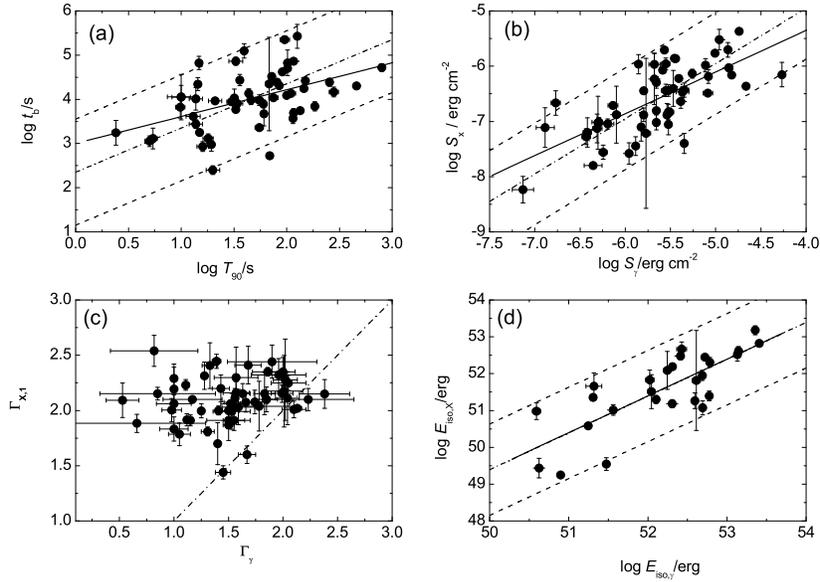} \caption{The correlations between the data of the
shallow decay phase and the prompt gamma-ray phase. The {\em solid} line in each
panel is the best fit. The {\em dashed} lines mark a $2\sigma$ region defined as
$y=x+(A\pm 2\times \sigma_A)$, where $y$ and $x$ are the quantities in the $y$
and $x$-axes, respectively, and $A$ and $\sigma_A$ are the mean and its $1\sigma$
standard error of $y-x$, respectively. The {\em dash-dotted} line is $y=x$. }
\label{Fig_Correlation}

\end{figure}

\begin{figure}
\epsscale{0.8} \plotone{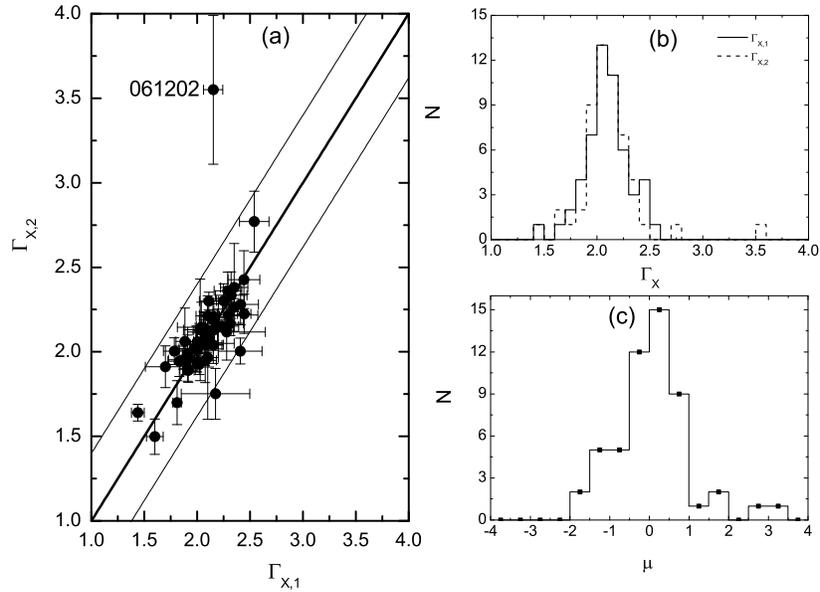} \caption{The comparison between $\Gamma_{\rm
X,1}$ and $\Gamma_{\rm X,2}$: (a) $\Gamma_{\rm X,1}$ vs. $\Gamma_{\rm X,2}$. The
{\em solid} lines marks the equality line $\Gamma_{\rm X,2}=\Gamma_{\rm X,1}$ and
its 2$\sigma$ region; (b) The histograms of $\Gamma_{\rm X,1}$ and $\Gamma_{\rm
X,2}$; (c) The distribution of $\mu$.} \label{Fig_Comparison}
\end{figure}

\begin{figure}
\epsscale{0.8} \plotone{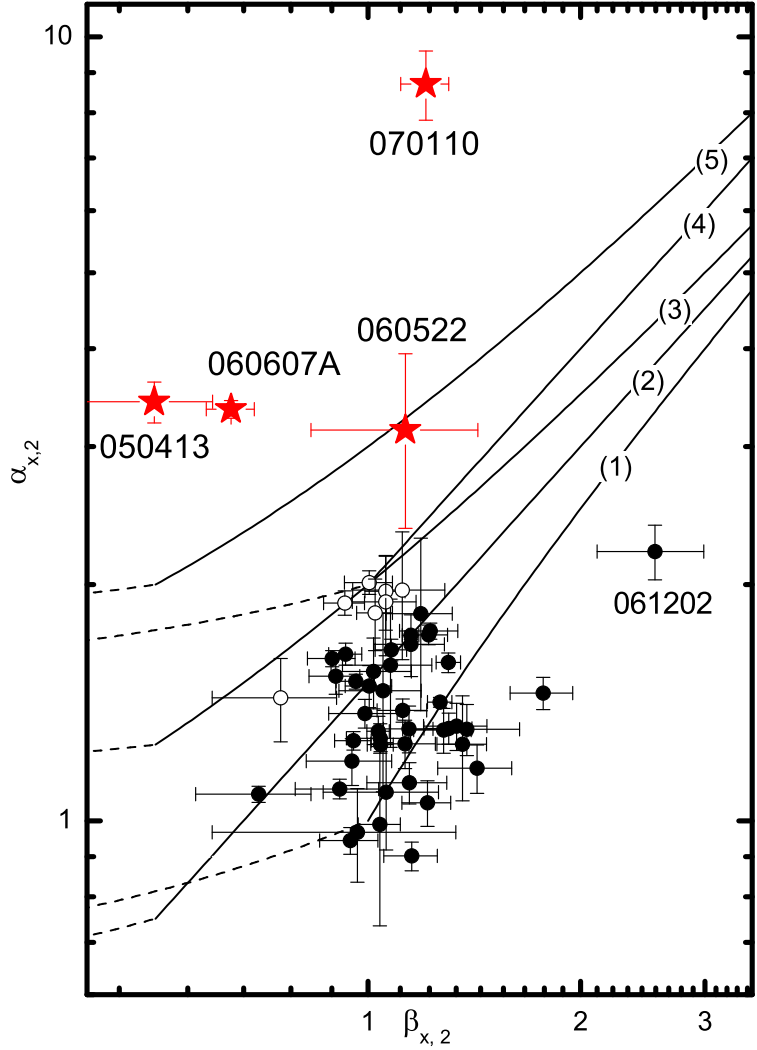} \caption{The temporal decay index $\alpha_{\rm
X,2}$ as a function of the spectral index $\beta_{\rm X,2}$ for the post-break
segment as compared with the closure correlations of various external shock
afterglow models: (1) $\nu>\max(\nu_c, \nu_m)$; (2) $\nu_m<\nu<\nu_c$ (ISM, slow
cooling); (3) $\nu_m<\nu<\nu_c$ (Wind, slow cooling) (4) $\nu>\nu_c$ (Jet, slow
cooling) (5) $\nu_m<\nu<\nu_c$ (Jet, slow cooling). The solid lines are those for
electron distribution index $p>2$, and the dashed lines are for $p<2$. The {\em
solid} dots represent the bursts whose $\alpha_{\rm X,2}$ and $\beta_{\rm X,2}$
satisfy the models (1) and (2), and the open dots represent those bursts can be
explained with the model (3). The stars are those bursts that significantly
deviate from the external shock afterglow models including 060522 (see discussion
in the text).}.\label{Fig_Model}
\end{figure}

\begin{figure}
\epsscale{0.8} \plotone{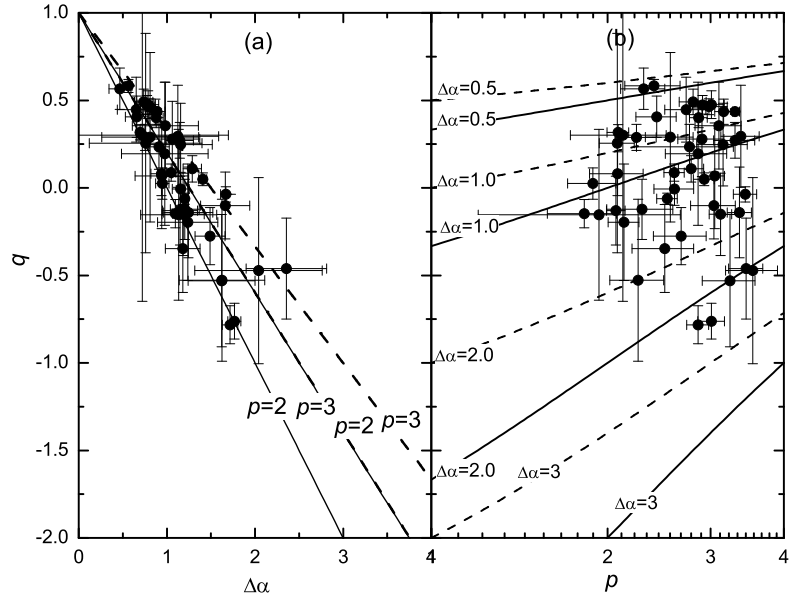} \caption{The distributions of the 49 GRBs whose
normal decay phases are consistent with the external shock models in the
$q-\Delta \alpha$ ({\em panel a}) and $q-p$ ({\em panel b}) planes along with the
model predictions for $\nu>\max(\nu_c,\nu_m)$ (solid lines) and $\nu_m<\nu<\nu_c$
(dashed lines).}.\label{Fig_Model}
\end{figure}

\begin{figure}
\epsscale{0.8} \plotone{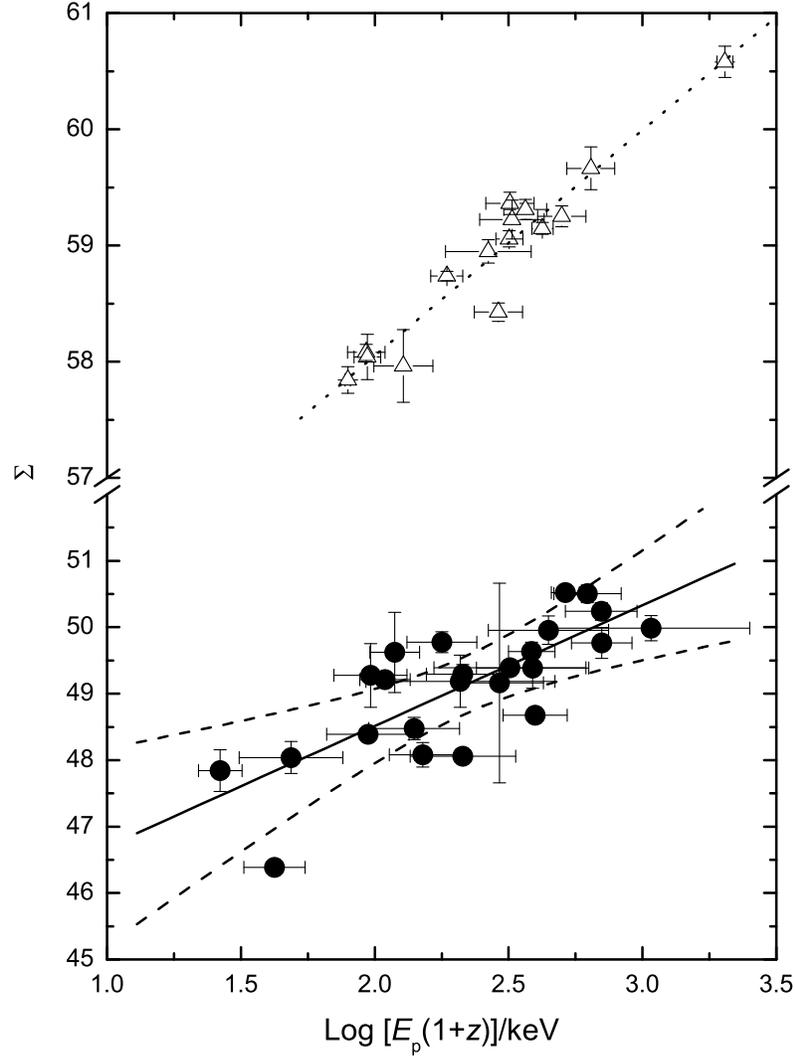} \caption{Comparison of the $E_{\rm iso, X}-E_{\rm
p}^{'}-t_{\rm b}^{'}$ relation  with the bursts in our sample (the solid dots;
the solid and dashed lines mark the best fit in $3\sigma$ level) with the
Liang-Zhang relation derived with pre-{\em Swift} bursts (the open triangles and
the dotted line is the best fit; Liang \& Zhang 2005), where $\Sigma_X\equiv \log
E_{\rm iso, X}-\kappa_2 t_{\rm b}^{'}$ and $\Sigma_\gamma\equiv \log E_{\rm iso,
\gamma}-\kappa_2 t_{\rm b, O}^{'}$.} \label{Liang_Zhang}
\end{figure}

\end{document}